\documentclass[twocolumn,longbib]{aastex7}

\submitjournal{The Planetary Science Journal}
\accepted{March 7, 2025}

\usepackage{amsmath} 
\usepackage{tikz}
\newcommand\wye[1][]{%
    \tikz\draw[thick, line cap=round,x=1ex,y=1ex,#1]
    (0,0) -- ++(90:1)
    (0,0) -- ++(-30:1)
    (0,0) -- ++(-150:1);
}

\newcounter{movie}

\counterwithin{equation}{section}

\begin{document}
%
%
\title{Planetesimal Impact Vapor Plumes and Nebular Shocks form Chondritic Mixtures}

\shorttitle{Impact vapor and nebular shocks}

%
%
\author[0000-0001-9606-1593]{Sarah T. Stewart}
\affiliation{Arizona State University}
\email{sstewa56@asu.edu}

\author[0000-0001-5365-9616]{Simon J. Lock}
\affiliation{University of Bristol}
\email{s.lock@bristol.ac.uk}
\author[0000-0001-5065-4625]{Philip J. Carter}
\affiliation{University of Bristol}
\email{p.carter@bristol.ac.uk}

\author[0000-0002-2140-3605]{Erik J. Davies}
\altaffiliation{Now at California Air Resources Board}
\affiliation{University of California Davis}
\email{erikjdavies@gmail.com}

\author[0000-0003-1209-3569]{Michail I. Petaev}
\affiliation{Harvard University}
\email{mpetaev@fas.harvard.edu}

\author[0000-0003-2164-0720]{Stein B. Jacobsen} 
\affiliation{Harvard University}
\email{jacobsen@neodymium.harvard.edu}

\shortauthors{Stewart, Lock, Carter, et al.}
\correspondingauthor{Sarah T. Stewart, sstewa56@asu.edu}

\begin{abstract}
The origin of chondrules, and the chondritic sedimentary rocks that dominate the meteoritic record, is a long-standing problem in planetary science. Here, we develop a physical model for the formation of chondritic mixtures as an outcome of vaporizing collisions between planetesimals that were dynamically excited by the growth and migration of planets. We present calculations of nebular shock waves generated by impact vapor plumes and focus on aspects of the plume interaction with the nebular gas and dust that have been neglected in previous studies of impact ejecta. We find that, when water dominates the vapor, the plumes are relatively cool. However, the plume expansion is supersonic and can drive strong shock waves in the dusty nebular gas. Portions of these nebular shock fronts initially melt nebular dust, forming chondrules that are coupled to the moving front. As the shock front expands and cools, the chondrules solidify while the shock front entrains additional dust. Eventually, the plume expansion stalls and then hydrodynamically collapses, turbulently mixing variably processed dust and size-sorted chondrules. For probable impact parameters and nebular conditions during giant planet growth and migration, the impact-generated mixtures have characteristics that span the range observed in chondritic meteorites, providing an environment for rapid formation of chondritic assemblages after chondrule formation. Our impact vapor and nebular shocks (IVANS) model links chondrule formation to the overall context of planet formation and provides a framework for interpreting the detailed chronological and geochemical record contained in chondritic meteorites.
\end{abstract}

\keywords{\uat{Impact phenomena}{779} --- \uat{Planetesimals}{1259} --- \uat{Shocks}{2086} --- \uat{Chondrules}{229} --- \uat{Chondrites}{228} --- \uat{Planetary migration}{2206}}

\section{Introduction}

In standard models of terrestrial planet formation, protoplanets grew by collisional accretion from a population of planetesimals, and the final planets were assembled by a period of stochastic giant impacts between protoplanets \citep[e.g.,][]{Chambers_2004}. Thus, in this framework, planetesimals are considered the building blocks of terrestrial planets. These building blocks are traditionally thought to be represented by the composition of undifferentiated meteorites, called chondrites, which are samples from small bodies in the solar system today that never went through a stage of bulk melting. However, it has become clear that the genetic relationship between chondrites and planets is not as straightforward as once thought.

Chondrites are comprised of various proportions of chondrules and grains of Fe-Ni metal embedded in a much finer matrix of silicate and organic materials. Chondrules are sub-mm-sized silicate spherules that were transiently heated to partial or complete melting. They are evidence for widespread thermal processing of material in the solar nebula during the formation of planetesimals. To date, there is no consensus on the principal physical mechanism that formed chondrules because no single process has been able to explain the diverse properties of chondrules and chondrites. For a full discussion of this problem, the reader is referred to one of the comprehensive reviews: \citet[][]{Hewins_Jones_Scott_1996,Rubin_2000,Scott2014,Connolly_Jones_2016,russell2018chondrules,Marrocchi_2024}. Here, we summarize the key observational constraints on the origin of chondrules and chondrite formation.

The main period of chondrule formation ranges from the age of calcium-aluminum-rich inclusions (CAIs), the oldest solids that condensed in the solar system, to about 4 Myr after the formation of CAIs \citep[e.g.,][]{Connelly_2012,Zhu_2019,Zhu_2020}. These ages overlap with the lifetime of the hydrogen-rich gas of the solar nebula, whose presence is required for the gentle assembly of chondritic mixtures. Based on the observed petrologic and chemical differences between chondrite groups, assembly of chondrules into chondrites was a local process in spatially and temporally separated reservoirs \citep{Jones2012,Ruzicka2012}. Most meteorites we have in our collections are samples from bodies in the present-day asteroid belt, and chondrites were originally thought to have formed within or near the main belt. However, the isotopic differences between two major groupings of meteorites \citep{Warren2011,Budde2016,Scott2018} have motivated the hypothesis that one group (isotopically carbonaceous chondrites or CC group) originated from the outer solar system while the other group (isotopically non-carbonaceous chondrites or NC group) formed in the inner solar system. Jupiter, whose rapid growth and large mass opened a gap in the protosolar disk, is often invoked as the spatial divider between the NC and CC meteorite groups \citep[e.g.,][]{Scott2018}; however, the details of the evolution of the protoplanetary disk is still an active area of research. Overall the chemical, petrologic, and isotopic data suggest that chondrites must have formed in both the inner and outer solar system and over an extended period during the lifetime of the nebular gas.

The historical idea that chondrites were the pre-accretionary building blocks of planets was confounded by the discovery that differentiation of both small and large bodies occurred contemporaneously with chondrule formation. The parent bodies of iron meteorites differentiated within the first Myr after CAIs \citep[e.g.,][]{Kleine_2009}, and Mars reached about two thirds of its mass within a few Myr of CAIs \citep{Kleine2017,Tang_Dauphas_2014}. In addition, Jupiter and Saturn must have accreted their hydrogen rich atmospheres before the nebular gas dissipated on time scales of 3 to 6 Myr \citep{Ribas2014}. Thus the formation time of the planetary embryos that seeded the gas giants must have overlapped with chondrule formation. Since chondrite formation must have happened in parallel to planet growth, how do chondrites fit into the overall history of planet formation?

Without a clear answer to this question, the valuable information contained in the meteorite record cannot be completely integrated into our understanding of planet formation. The diverse set of environmental conditions indicated by the observations has prevented a general consensus from being reached as to the origin of chondrites. For example, chondrules require multiple localized heating events \citep[e.g.,][]{Hewins_Connolly_Libourel_2005}. The chemical composition of chondrules implies that they formed in transient nebular environments that were substantially enhanced in solids compared to the expected mean conditions in the solar nebula \citep[e.g.,][]{Alexander2008}. In addition, the diversity of chemical and isotopic properties amongst the different chondrite groups indicates that the mechanism(s) of chondrule formation and chondrite assembly must have occurred over wide regions of the solar nebula \citep{Scott2018}. Finally, the meteoritic data imply that some chondrites were assembled quickly after the formation of their enclosed chondrules \citep[e.g.,][]{Ruzicka2012,Metzler_2012,Ruzicka_Hugo_Friedrich_Ream_2024}, requiring a local mechanism to bring together materials with vastly different thermal histories; however, simultaneously, other data indicate that some chondrites contain chondrules that formed at distinctly different times \citep{Connelly_2012} or locations \citep{Williams_2020} in the nebula.

The voluminous data from meteorites have thus provided a set of perplexing requirements on chondrule and chondrite formation. In response, a remarkable diversity of physical processes have been proposed to explain the origin of chondrules. These works have been grouped broadly into processes that act within the protoplanetary disk (referred to as nebular or canonical models) and processes that act on planetary bodies (planetary or non-canonical models) \citep{Connolly_Jones_2016}. These two classes of models are each favored to explain different aspects of chondrule and chondrite chemistry. Broadly, a nebular environment for chondrules is required for the assembly of the primitive chondritic mixture and to explain the rarity of signatures of planetary silicate differentiation in chondrites; however, planetary contributions have been invoked to explain the necessary elevated dust-to-gas ratios and varying oxygen fugacities required by aspects of chondrule chemistry. 

In this work, we present a hybrid model for chondrule formation in nebular shock fronts driven by vapor plumes from collisions between planetesimals. We study the impact generation of nebular shocks in one, two, and three-dimensional geometries and develop scaling relationships for the nebular disturbance from impact events. These nebular shock fronts form transient physicochemical environments in the solar nebula that expand, stall, and collapse to form mixtures of materials with different thermal and chemical histories. We examine how free-floating dust in the nebula influences the plume-nebula interaction and calculate the size of both melted and unmelted particles that are coupled to the shock front environment. Our study builds upon the planetary context provided by previous works that demonstrated that vaporizing collisions between planetesimals were a common event in the solar system during the growth and migration of the giant planets \citep{Davies2020,Carter_2020,Carter_Stewart_2022}. We identify phenomena during impact events that have not been deeply investigated in previous works and relate these physical processes to the inferred environmental conditions for the formation of chondrules and chondritic mixtures.

\subsection{Planetary context for vaporizing collisions between planetesimals}

Collisions are a fundamental process for planet formation. However, in most $N$-body studies of planet accretion where the focus was on the collisional growth of protoplanets, the mutual collisions between planetesimals have been neglected for computational efficiency. In addition, previous studies of the outcomes of planetesimal-planetesimal collisions have largely neglected the presence of a surrounding gas. Such collision studies have primarily focused on (i) the early growth phase prior to significant dynamical excitation by protoplanets when mutual impact velocities were small \citep[e.g.,][]{Leinhardt_Richardson_2002} or (ii) impact erosion and disruption for application to the collisional evolution of the asteroid belt \citep[e.g.,][]{Holsapple_Housen_2019}. There is little previous work focused on mutual planetesimal collisions in the presence of the solar nebula gas \citep[e.g.,][]{Hood2009} or collisions that produced significant degrees of vaporization of the colliding material. The interactions between the impact vapor plume created by vaporizing collisions and nebula is the central topic of our study.

\begin{figure*}
\plotone{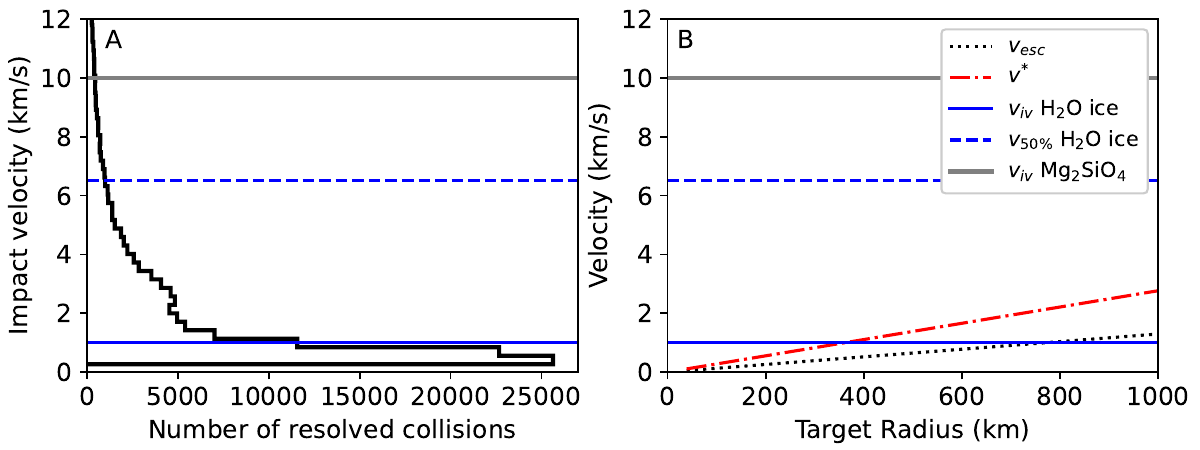}
\caption{{\bf Many planetesimal collisions are disruptive and vaporizing.} A. Histogram of mutual planetesimal collision velocities (corrected for particle size inflation) from a Grand Tack $N$-body simulation (022GTJf6hgas) in \citet{Carter_Stewart_2022}. In this example, 57\% of the collisions are faster than 1 km~s$^{-1}$ (see Fig.~\ref{fig:impact-pres-prob} in Appendix~\ref{sec:imp-prob}). B. Escape velocities ($v_\mathrm{esc}$, dotted black line) and catastrophic disruption velocities ($v^*$, red dot-dash line) compared to the critical velocities for shock-induced vaporization: incipient vaporization ($v_{iv}$) for H$_2$O ice at an initial temperature $T_0=150$~K (about 1~km~s$^{-1}$, solid blue line) and forsterite (Mg$_2$SiO$_4$) at $T_0=300$~K (about 10~km~s$^{-1}$, grey line) and 50\% vaporization ($v_{50\%}$) for H$_2$O ice (about 6.5~km~s$^{-1}$, blue dashed line). Giant planet migration leads to erosive or disruptive collisions between 100s km sized planetesimals. Many events also partially vaporize water ice, but impact-vaporization of silicates is much less common. }
\label{fig:disrupt}
\end{figure*}

The growth and migration of the giant planets strongly affects the dynamics of the entire solar system \citep[e.g.,][]{Walsh2011,Carter2015,Raymond2017a,Clement2018,Carter_2020}. \citet{Carter2015} and \citet{Carter_Stewart_2022} included calculations of the mutual collisions between planetesimals in studies of the growth of the terrestrial planets with different giant planet migration histories. They found that gravitational stirring by giant planets leads to high velocity collision between planetesimals, as shown in  Figure~\ref{fig:disrupt}A for the Grand Tack scenario \citep{Walsh2011}. The collision velocities exceed that needed for the onset of shock-induced vaporization of planetary materials. The critical impact velocities for the onset of shock-induced vaporization (also referred to as the threshold for incipient vaporization) of ice, metals, and silicates have been determined experimentally \citep{Stewart2008,Kraus_2015,Davies2020}. The velocity required to reach different vaporization criteria depend on the peak pressures generated by the collisions (initial internal pressures are negligible for planetesimals). In solar system formation, a large fraction of impacts exceed the impact velocity required for incipient vaporization of water ice (solid blue line in Figure~\ref{fig:disrupt}A) but generally fall below the conditions for incipient vaporization of silicates (dashed grey line in Figure~\ref{fig:disrupt}A). 

Giant planet perturbations of planetesimals also produce erosive conditions. The size distribution of fragments depends on the masses of the colliding bodies \citep[e.g.,][]{Housen_Holsapple_1990,Leinhardt2012}. The key parameters for predicting collision outcomes are the escape velocity, $v_{\rm esc}$, and the critical velocity for catastrophic disruption, $v^*$ (Figure~\ref{fig:disrupt}B). Catastrophic disruption is defined as a collision where the largest remnant is less than half of the original total mass, and super-catastrophic events have a largest remnant less than 10\% of the original total mass. The critical velocity corresponds to the required specific impact energy for removal of exactly half the total mass \citep[$v^*$ is the equal mass collision case from][]{Leinhardt2012}. The $N$-body simulations demonstrate that, during a Grand Tack event, most collisions between 100-km scale planetesimals are in the catastrophic or super-catastrophic regimes \citep{Carter2015,Carter_Stewart_2022}.  The exact numbers, timings and locations of vaporizing and disruptive planetesimal collisions depend on the formation locations and migration histories of the giant planets \citep{Carter_2020}.

To gain some intuition for the conditions required to reach impact vaporization, without relying on an $N$-body simulation of a specific solar system formation scenario, we estimated the perturbation from a single planet on a nearby population of planetesimals. See Appendix \ref{app:stirring} for details. Figure~\ref{fig:criticale} shows the eccentricity of planetesimals required to produce collisions with mean velocities that correspond to that required to vaporize different materials (colored lines). The black and grey lines indicate the mean eccentricities of a population of planetesimals excited by dynamical stirring from a planet of a given mass for two different nebular gas densities. 

\begin{figure}
\centering
\includegraphics[width=0.46\textwidth]{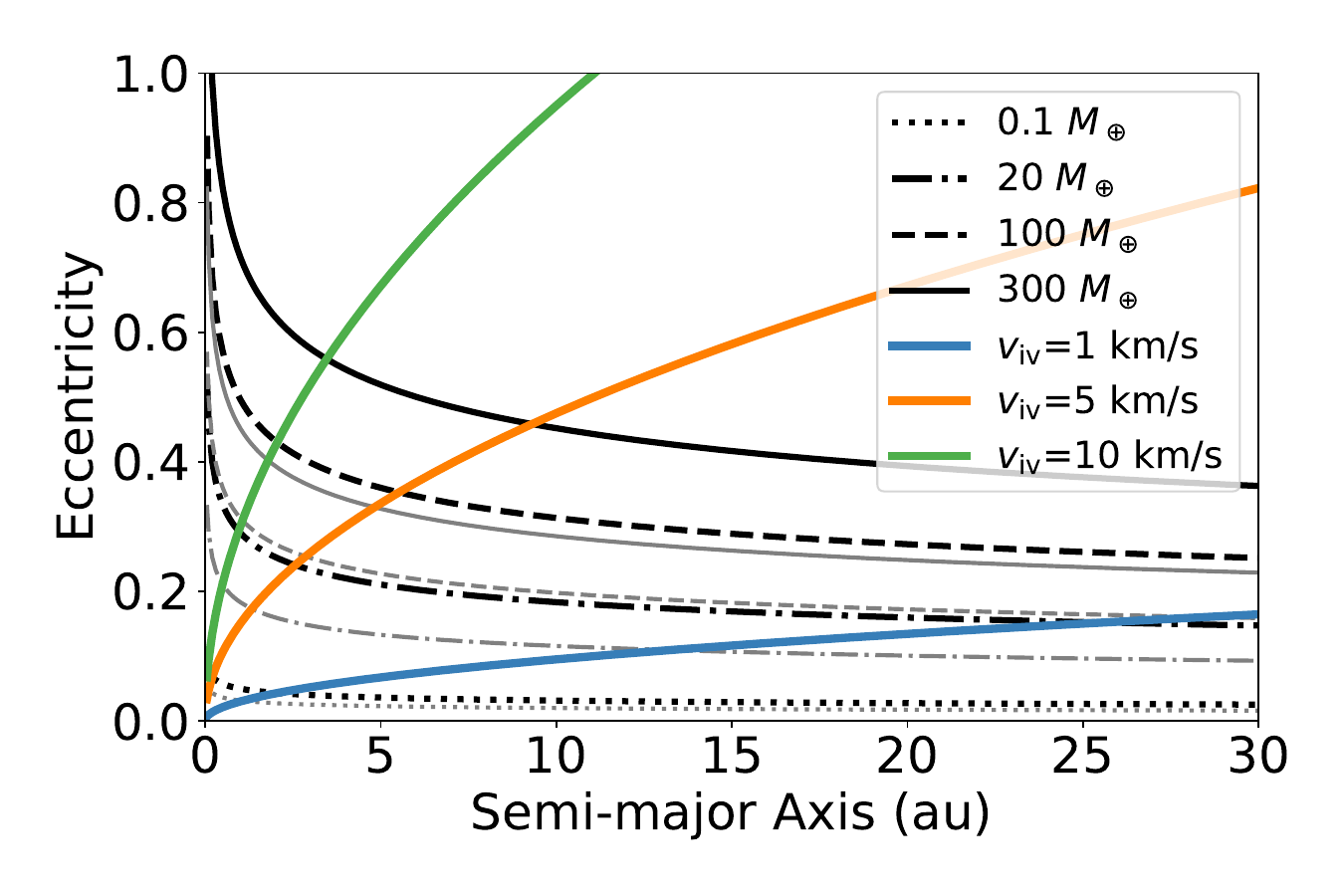}
\caption{{\bf Protoplanets can dynamically excite a population of nearby planetesimals into the vaporization regime.} Critical eccentricities (colored lines) define the threshold where the mean relative velocity between planetesimals is equal to a critical velocity for shock-induced vaporization (Eq.~\ref{eq:eiv}): incipient vaporization for $T_0=150$~K ice (1~km~s$^{-1}$, blue), incipient vaporization for $T_0=300$~K forsterite (10~km~s$^{-1}$, green), and an intermediate velocity (5~km~s$^{-1}$, orange). Black lines denote the mean eccentricities for a population of 100-km radius planetesimals that are dynamically stirred by a single planet of varying mass (line styles) in the oligarchic growth regime with a nebular gas density of $\rho_{neb}=10^{-7}$~kg/m$^3$ (Eq.~\ref{eq:estir}). Grey lines have $\rho_{neb}=10^{-6}$~kg/m$^3$. The magnitude of dynamical stirring is greater for: larger mass planets, planets closer to the sun, and decreasing nebular gas density.  
\label{fig:criticale}}
\end{figure}

We find that stirring from a Mars-mass embryo does not induce vaporizing collisions between planetesimals anywhere except near the Sun (dotted lines). An ice giant planet, or rocky core of a growing gas giant planet, can induce vaporizing collisions between icy bodies ($v_{\rm iv}\gtrsim$1~km~s$^{-1}$) and silicate bodies ($v_{\rm iv}\gtrsim$10~km~s$^{-1}$) within about 25 au and 1 au, respectively. If Jupiter's core grew within about 5 au, then nearby planetesimals would reach the vaporization regime for silicates during its growth. Beyond about 25 au, there would be a limited number of vaporizing collisions between planetesimals when stirred by a growing ice giant planet in the oligarchic growth regime. 

From these analytic estimates and $N$-body simulations \citep{Carter2015,Carter_2020}, we find that typical mutual collision velocities between planetesimals result in disruption with negligible melting or vaporization of silicates and metals \citep[see][]{Leinhardt2009}. However, collisions between the excited planetesimals often exceed the threshold for vaporization of water ice. In addition, the disruptive planetesimal collisions would have generated substantial amounts of dust, generally reflecting the average composition of the original bodies. Since most collisions occur in the midplane of the nebula, planet growth and migration would transiently elevate the midplane dust-to-gas ratio in the excited regions of the solar nebula. Hence, our study is primarily focused on the interactions of water vapor plumes with the surrounding dusty solar nebula under dynamically excited conditions.

\subsection{Planetesimal generations and diversity}

Based on the chronological information in the meteoritic record \citep{Kleine_2009,Connelly_2012,Krot2005}, there must have been multiple generations of planetesimals that formed throughout the lifetime of the solar nebula gas. Since giant planet growth and migration takes place in the gas disk, the gas giant planets instigated erosive and vaporizing collisions between planetesimals over the same time frame. As a result, collisional disruption of planetesimals must have occurred contemporaneously with new planetesimal formation.

The disrupted planetesimals were likely in varying stages of internal evolution, from primitive dust aggregates that survive today as comets and CI-type asteroids to partially or fully differentiated bodies. Differentiated meteorites have been identified in both the CC and NC reservoirs \citep[e.g.][]{Warren2011, Dey_Yin_2022} and such bodies must have formed rapidly and reached sufficient size to melt from internal heating from radioactive decay of $^{26}$Al. However, if a planetesimal experiences a collision that disrupts the body, potentially forming rubble piles, internal heat is released, and internal differentiation would have been more limited or prevented \citep[e.g.,][]{Ren_Hesse_Dygert_Lucas_2024}. Thus, planetesimals that formed at similar times could differentiate or remain undifferentiated depending on their collisional history.

Planetesimals in the proximity of the large cores of the growing gas giant planets would have been dynamically excited into an erosive and disruptive regime, thus limiting their growth and the extent of internal heating and differentiation. Because these disruptive events occur at small absolute velocities (Figure~\ref{fig:disrupt}), the amount of melting of ice (and silicates and metals) is quite limited. Laboratory disruption experiments of ice-dust aggregate bodies demonstrate that most of the debris remains unmelted and collected into variously sized clasts of aggregated dust \citep{Nakamura_2024}. 

Asteroids Bennu and Ryugu are examples of primitive icy planetesimals from the CC reservoir. These bodies have few chondrules and may represent surviving remnants of the early generations of dust aggregate planetesimals that have experienced a history of collisional disruption and aqueous metamorphism. In this work, we focus on primitive icy dust aggregates as the precursor bodies for the formation of water vapor impact plumes. Because vapor plume materials originate from near the surface layers of the colliding bodies, our study is also applicable to partially differentiated bodies that may be in various stages of segregating metal-sulfide liquid into a core but prior to substantial silicate melting. Such bodies may also accrete a surface layer of nebular dust and debris \citep{Weiss_Elkins-Tanton_2013}. Future work will consider the details of vaporizing collisions between fully differentiated bodies. 

\subsection{Unexplored aspects of impact vapor plumes}

Chondrule formation by melting free-floating dust in nebular shock waves has been widely investigated under various driving conditions. Previous studies of nebular shock waves have typically focused on either large-scale features driven by gravitational instabilities in the protoplanetary disk \citep[e.g.,][]{Ciesla_Hood_2002,Morris_2016} or on smaller-scale nebular shocks around eccentric protoplanets and planetesimals \citep[e.g.,][]{Morris2012,Boley_Morris_Desch_2013}. Less attention has focused on shock waves generated in the nebular gas by impact vapor plumes from planetesimal collisions. For application to chondrule formation, recent works have developed a model for the impact-production of molten silicates in stress concentrations that lead to high-velocity jets of material in the impact ejecta field \citep{Johnson2015,Johnson_2018,Cashion_2022}. In the impact jetting chondrule formation model, the molten silicate melt is broken into chondrule-sized fragments by shearing interactions with the nebular gas. The thermodynamic conditions that generate molten jets would also have generated some impact vapor. 

The presence of the nebula dramatically changes the outcomes of vaporizing collisions, compared to vapor expanding into vacuum, due to the interactions of the vapor plume with the nebular gas. One of the early attempts to numerically simulate the interaction between the vapor plume and nebular gas was presented in \citet{Hood2009}. This prescient study explored nebular shocks generated by impact-vaporization of silicates during collisions between planetesimals excited by Jovian resonances. Although the study was hindered by short calculation times and limited equation of state models, which necessitated the use of an overly dense nebular gas that restricted the extent of the vapor plume disturbance, they demonstrated that impact generated nebular shocks could be hot enough to melt free-floating silicate dust. Here, we substantially expand upon this early work to characterize the evolution of impact vapor plumes to late times.

The formation of the chondritic mixture is much less studied than the formation of the molten chondrules alone. These sedimentary rocks must have agglomerated gently, requiring the continued presence of the nebular gas, under physical processes that could collect materials with vastly different thermal histories. The chondritic mixture may be collected by sedimentation of free-floating materials onto the surfaces of pre-existing planetesimals \citep[as suggested by][]{Johnson2015,Johansen_SciAdv_2015}. The interiors of these planetesimals may be in varying stages of differentiation \citep{Weiss_Elkins-Tanton_2013}. In some cases, however, chondrites have physical and chemical features that indicate that the sedimentary rock formed quickly after chondrule formation \citep{Ruzicka2012,Ruzicka_Hugo_Friedrich_Ream_2024}, including the observation of a chemical complementarity between the matrix and chondrules in some meteorites \citep{Palme_Hezel_Ebel_2015}. 

In this work, we present an initial study of the shock hydrodynamics of impact vapor plumes interacting with the surrounding solar nebula gas and dust. These results identify promising pathways to formation of both chondrules and chondritic mixtures.

\section{Overview of the Impact Vapor and Nebular Shocks (IVANS) Model}

The evolution of an impact vapor plume in the solar nebula is qualitatively and quantitatively different from the evolution in the vacuum of space. Impact vaporization experiments on icy bodies in a vacuum, in the laboratory or at the planetary scale of the 2005 Deep Impact mission on comet 9P/Tempel, have produced an expanding gas and dust plume that cools to a quasi-ballistic debris field \citep{Schultz_2007,Richardson_Melosh_Lisse_Carcich_2007,Holsapple_Housen_2007}. We do not have comparable planetary examples for a collision embedded in the nebular gas but there are analog experiments for high-pressure vapor expansion in air and water. 

Explosive production of vapor in air initially expands supersonically, but the outward motion stalls and reverses to an inward flow (which may reshock and expand again in a damped oscillation). The flow reverses because the inertia of the vapor plume allows it to expand to such an extent that it has a lower pressure than the surrounding air, which leads to the formation of reverse shocks that slow and collapse the vapor plume. The inward flow, and subsequent oscillations, restore the hydrostatic equilibrium in the air after the vapor plume perturbation. An example calculation of a spherical explosive charge in air is found in \citet{Brode_1959}. Similarly, a high-pressure gas bubble in water initially expands, collapses, and then repeats this process while decaying in pressure magnitude until the system attains hydrostatic equilibrium \citep[e.g., see Gas Bubble Dynamics in][]{Costanzo_2011}.

\begin{figure*}
\plotone{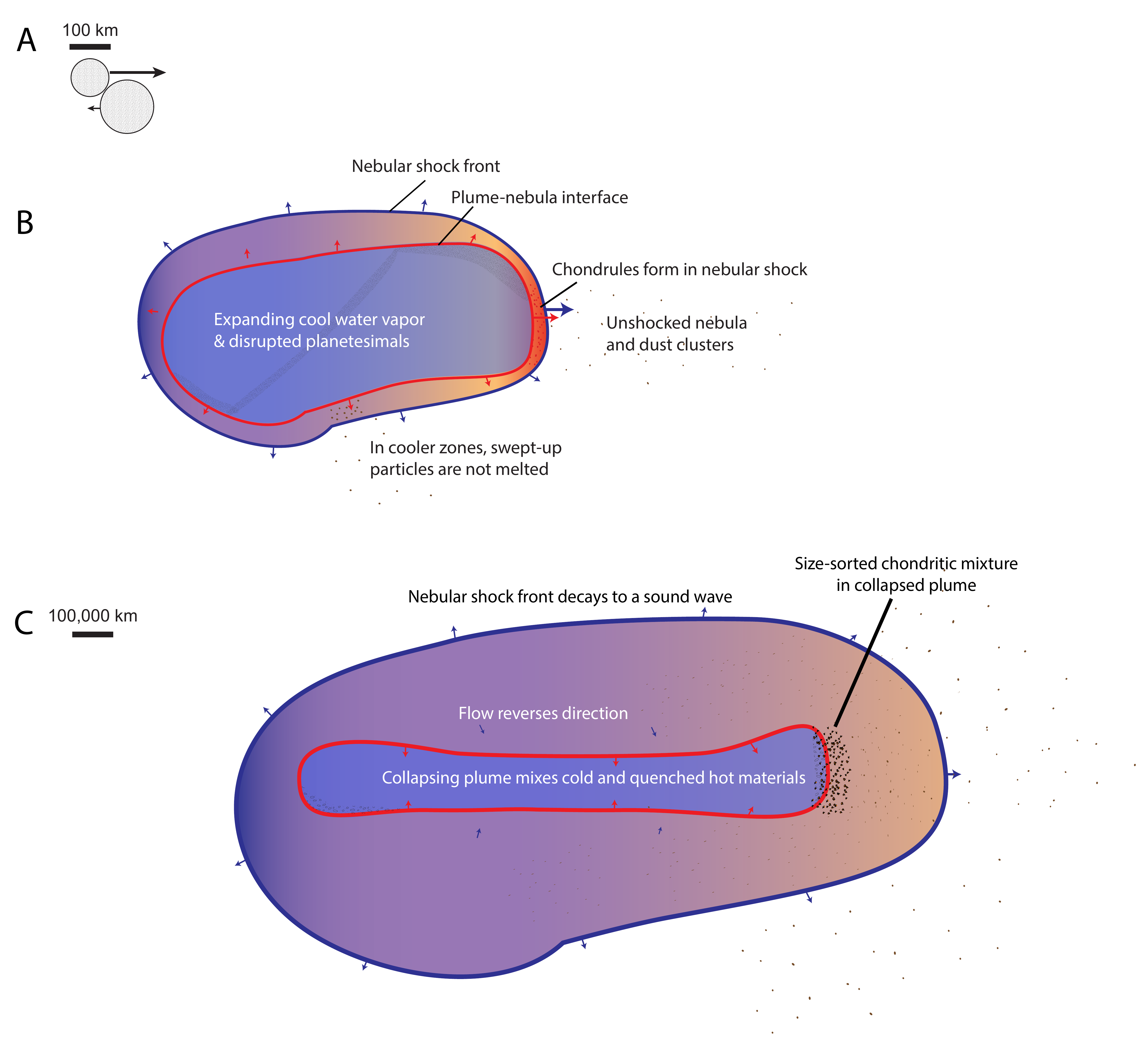}
\caption{{\bf Schematic overview of the hydrodynamics of impact vapor plumes and nebular shocks.} A. Planetesimals comprised of refractory dust and ice collide in a disruptive and vaporizing event surrounded by dusty nebular gas. B. The impact generates an expanding cloud of cooling water vapor and debris from the planetesimal. The vapor plume expands supersonically (inner blue region), driving a warmer shock wave into the solar nebula (outer ring with warm colors). The elongated expanding shell of shocked nebula is hotter in the principal impact direction and cooler in the opposite and lateral dimensions (indicated by the color gradient). Portions of the shocked nebula are warm enough to melt free-floating nebular dust and form chondrules; other regions experience less thermal processing. C. The nebular shock decays to a sound wave as it expands. Expansion of the vapor plume leads to a pressure low in the solar nebula and subsequent hydrodynamic reversal in the flow field. The nebular gas, dust, and chondrules flow into the low pressure region, mixing materials that were processed in different regions. The collapsed mixture has the characteristics observed in chondritic mixtures: quenched chondrules mixed with dust and ice. The mixed region is orders of magnitude larger in scale than the original planetesimals (e.g., Mm-scale mixed regions). The timescale of the collapse is order 10s hours.  \label{fig:cartoon}}
\end{figure*}

In Figure~\ref{fig:cartoon}, we present a schematic of impact vapor plume hydrodynamics in the solar nebula. As in the case of explosive products in air, the supersonic expansion of the vapor plume drives a shock wave into the solar nebula (dark blue outer line, panel B). For typical vaporizing collisions, the nebular shock propagates distances orders of magnitude greater than the size of the planetesimals before decaying to a sound wave (panel C). At early times, the shocked nebula region is typically an elongated shell with asymmetric shock pressures and temperatures (panel B). The contact surface between the vapor plume and nebula (red line) initially expands and then reverses flow (panel C) due to the changing pressure gradients. Initially, the vapor plume is more dense than the surrounding nebula but as the system evolves, the density of the plume front drops below that in the nebular shock and turbulent eddies will form at the interface, mixing the vapor plume materials with the surrounding nebular gas and dust. Depending on the magnitude of the event, the collapsing mixture will form secondary shocks that oscillate and damp until hydrostatic equilibrium is attained.

At present, shock physics codes do not capture the full range of phenomena in a vaporizing planetesimal collision within the solar nebula. The primary challenges include (i) wide ranging equations of state for the constituent materials; (ii) adaptive mesh capability to resolve both the planetesimal scale and the nebular scale of the disturbance; and (iii) multi-phase flow physics for the solids and gas in the system. Previous works have attempted to model impact vapor plume expansion without calculating the shock wave interactions with the nebula \citep[e.g.,][]{dullemond_forming_2014,Dullemond2016,Choksi_2021}. As we will demonstrate in this work, the shock wave interactions are a critical component of the vapor plume phenomena and cannot be neglected.

In this work, we calculate the shock processing of nebular materials generated by expansion and collapse of water and silicate vapor plumes in both a dust-free and dusty-gas nebula. We examine the problem using shock hydrocodes in one, two and three dimensions. Although these are highly three-dimensional (3D) phenomena, it is computationally expensive and difficult to develop scaling laws from 3D calculations. We use a one-dimensional (1D) open source hydrocode to develop scaling laws for the properties of the nebular disturbance from a vaporizing collision. We also present a novel configuration in two dimensions (2D) that captures the essential components of a three dimensional vapor plume and allows for more computationally efficient explorations of the problem.

Schematically, as indicated in Figure~\ref{fig:cartoon}, new chondrules are generated in the shocked nebula region from free-floating nebular dust, generally at early times in the principal direction of the vapor plume. The chondrules cool during the expansion phase. Using a numerical model of the gas drag between the solids and gas, we examine the coupling of dust and chondrules to the inward flow in the collapsing plume. Plume collapse mixes the freshly quenched chondrules with other materials (e.g., ice, dust, previously formed free-floating chondrules not reprocessed by the event). We assess whether or not there could be rapid assembly of this mixture into gravitationally-bound bodies. Finally, we discuss the potential for the impact vapor plume and nebular shock model to explain the origin of chondrules and chondrites and directions for future work.

\section{Shock Hydrocode Calculations}

To investigate the shock interactions between the impact vapor plume and the surrounding solar nebula, we directly modeled the vapor expansion and its resulting effects on the nebula. Our simulations are set in the reference frame of the solar nebula gas; this choice is justified in Appendix~\ref{app:nbody}. The gas reference frame is orbiting the Sun at a sub-Keplerian angular velocity due to the radial pressure gradient. Because the scale of an individual impact is much smaller than the scale height and radius of the disk, the background gas is modeled with a constant pressure and temperature. As planetesimal collisions are statistically more likely to occur near the midplane of the disk \citep{Carter_Stewart_2022}, these pressures and temperatures can be linked to the midplane conditions. The gravitational forces from the disrupting planetesimals were neglected as these terms are much smaller than the pressure gradient forces generated by the shock waves. Similarly, the shear strength of the bodies was neglected and the calculations are purely hydrodynamic.

We modeled the hydrodynamics of vapor production, expansion and collapse using the CTH code (v11.1) developed by Sandia National Laboratories. CTH is a multi-dimensional, multi-material, finite-volume Eulerian shock physics code with adaptive mesh capabilities \citep{McGlaun_Thompson_Elrick_1990,Crawford1999}. We also used the 1D Lagrangian finite difference hydrocode pyKO (v0.8.3), an open source python implementation of the KO shock physics code \citep{Wilkins_1999,Stewart_pyKO_2023}. For the solar nebula, we used an ideal gas, ideal dust-gas mixture (Appendix \ref{sec:dig}), and a hydrogen tabular equation of state \citep[SESAME table 5250,][]{Kerley2003}. The shock compression and expansion of water and silicates is poorly described by an ideal gas equation, so we used tabulated equations of state based on the ANEOS code package \citep{MANEOSv1} for water (H$_2$O, \citet{Stewart_2024_water}) and fused silica (SiO$_2$, \citet{Stewart_Amodeo_2024_silica}) (Table~\ref{tab:aneos_params}).  The vaporization of silicates is initially dominated by the production of SiO gas, so silica provides a general representation for vaporization of major silicate minerals in planetary bodies. The choice of fused silica ($\rho_0=2.2$~g~cm$^{-3}$) instead of crystalline quartz ($\rho_0=2.65$~g~cm$^{-3}$) is motivated by the likely porous structure of the planetesimals. In this manner, we approximated the specific impact energies of the event without the use of a particular porosity model.

\begin{figure}
\centering
\includegraphics[width=.45\textwidth]{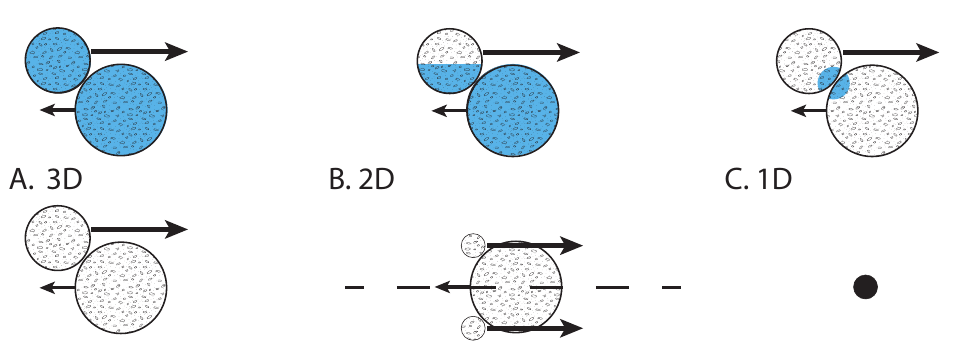} 
\caption{{\bf Schematic of the different dimensionalities for impact vapor plume calculations.} The simulated regions are indicated in blue. A. Full 3D hydrocode calculations of colliding spherical bodies. The direction of the larger velocity vector in the nebular frame of reference is the principal direction of the vapor plume. B. 2D, cylindrically-symmetric calculation with the symmetry axis denoted by the dashed line. The full target is on the center line, and the projectile is represented by an torus with mass equal to the interacting portion of the projectile. C. The regions shocked to the highest pressure states in the 3D event (shaded blue zones) are approximated by a 1D sphere initialized in a shocked state on the principal Hugoniot. In all cases, the initial bodies are surrounded by ambient nebular gas.  \label{fig:dimensions}}
\end{figure}

Figure~\ref{fig:dimensions} presents a schematic of the different types of calculations used to investigate the nebular shocks and hydrodynamic evolution of the vapor plume generated by the impact event. In 2D and 3D, the colliding planetesimals were ice-silica mixtures. The 3D simulations are computationally expensive due to the dynamic range of the length and time scales in the problem. We used a simplified geometry in 2D cylindrical symmetry that focuses on generating a similar vapor plume as in the equivalent 3D event. To accomplish this simplification, the projectile is converted into a torus with a mass equal to the {\it interacting mass} of the projectile (Figure~\ref{fig:dimensions}B). The interacting mass is a concept introduced in catastrophic disruption scaling laws to indicate the mass fraction of the projectile that has an overlapping cross section with the target \citep{Leinhardt2012}. The impacting torus creates a high-pressure region in the target that is similar to the full 3D event. In this 2D case, the details of the disruption of the target are not correct; however the properties of the vapor plume and its interaction with the nebula in the principal plume direction are similar to the full 3D event in Figure~\ref{fig:dimensions}A. In 1D (Figure~\ref{fig:dimensions}C), the vapor plume is generated by decompression and expansion of a high-pressure sphere embedded in the nebula. The high-pressure sphere was pure water or pure silica as the Lagrangian code (pyKO) does not have a mixed material cell capability. The high pressure sphere represents the most highly shocked regions of the two bodies and the source of most of the vapor plume gas. Different initial shock pressures, e.g., corresponding to the principal direction and opposite direction of the vapor plume (shown schematically in Figure~\ref{fig:cartoon}), can be used to analyze different aspects of a 3D event with 1D calculations.
See Appendix~\ref{sec:hydro-methods} for further details about the hydrocode calculations.

\subsection{1D Calculations: Vapor plume expansion and reversal}

\begin{figure*}
\centering
\includegraphics[width=1\textwidth]{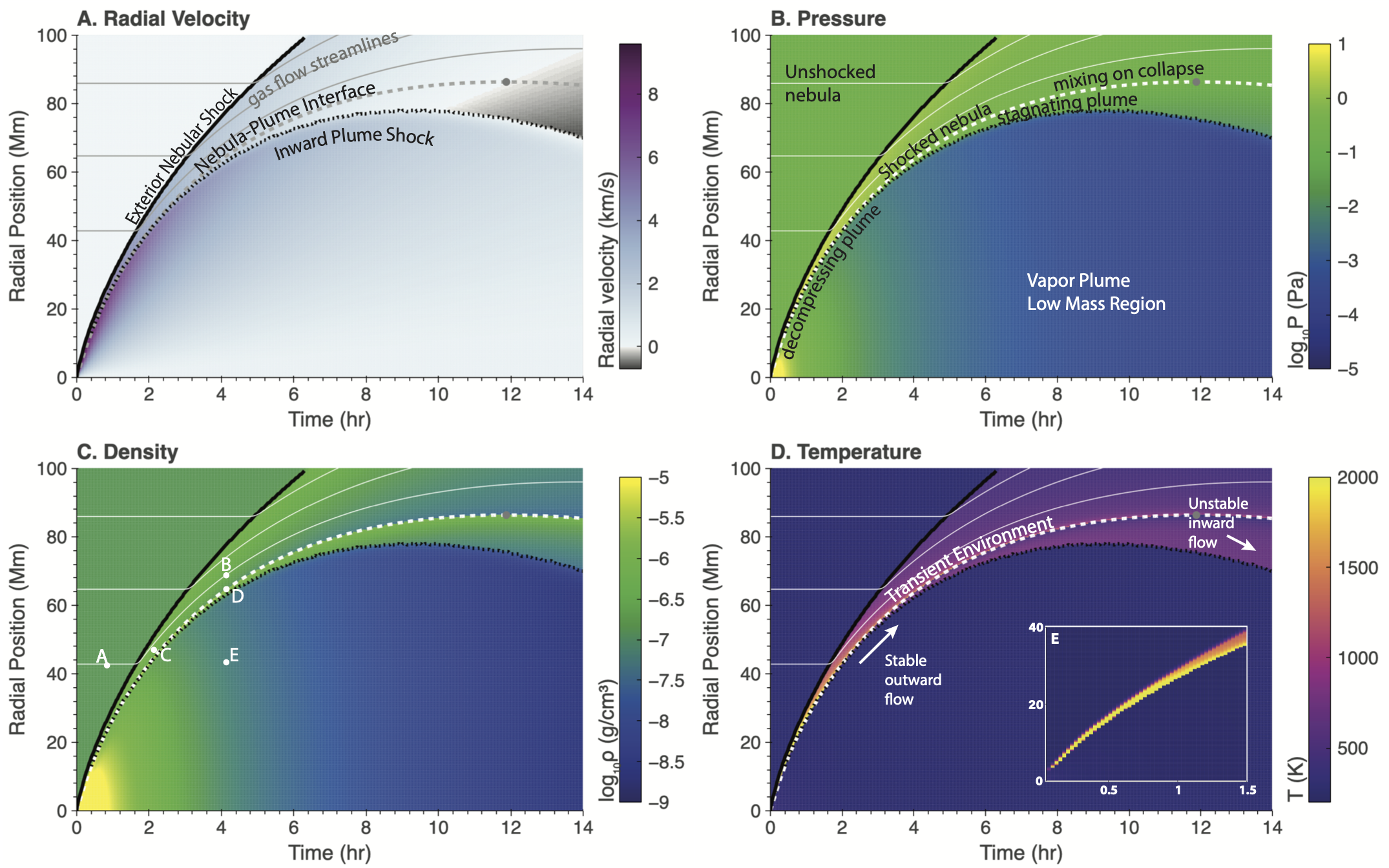} 
\caption{{\bf 1D spherical vapor plume expansion and flow reversal.} Example vapor plume evolution starting with an 80 GPa water sphere of radius 25~km, expanding into a dust-free nebular gas at $10^{-6}$ bars. In panels A-D, the thick dashed line represents the material boundary between the inner water vapor plume and outer nebular gas. The color shading presents radial velocity (A), pressure (B), density (C), and temperature (D) versus time. The streamlines (thin grey in A; thin white in B-D) indicate the flow history of the gas and any fully coupled dust. The nebular shock generated by the expanding vapor plume creates a transiently hot nebular environment that expands and cools on the time scale of hours. The inset (E) is a close-in view of the initial hot nebular shock front. In panel C, points A-E are used to illustrate size sorting in the outward flow in section \ref{sec:sizesorting}. We propose chondrules form and cool in the transiently shocked (dusty) nebular gas. The \href{https://chondrules.net/ivans/}{supplemental materials} contain an interactive version of this figure.  \label{fig:1devolution}}
\end{figure*}

During a high-velocity planetesimal collision, small portions of the bodies are shocked to high pressure and vaporize upon release. The primary evolution of the vapor plume involves an outward supersonic flow that stalls and reverses to an inward collapse of the plume. The key features of the impact phenomena can be explored in 1D simulations that capture the vapor plume interactions with the nebula, as shown in Figure~\ref{fig:1devolution}. This example calculation begins with a 25~km-radius sphere of shocked water, initialized in a state on the ice Hugoniot (80 GPa, see Table~\ref{tab:1dsims}). 
The sphere is surrounded by a dust-free ideal gas representing the hydrogen-helium mixture (nominally 200~K and 0.13 Pa, see Appendix~\ref{sec:dig}). 

This example illustrates the overall gas dynamics between a water vapor plume and the nebular gas. 
In Figure~\ref{fig:1devolution}, the panels present the radial velocity (A), pressure (B), density (C), and temperature (D) versus time. The rapid decompression and expansion of the shocked water drives a strong shock into the nebular gas, indicated by the exterior nebular shock front (solid black line) in panel A. The shock wave imparts both internal and kinetic energy, compressing the nebular gas into a hot, outward flow. The pressure of this leading shock front decays as it propagates into a larger volume with time, eventually decaying to a sound wave in the nebula. 

This 1D planetary vapor plume calculation is comparable to the explosive gas interactions with the atmosphere presented in \citet{Brode_1959}. An analytic model for a spherical plume expansion into an ambient gas is presented in \citet{Arnold_Gruber_Heitz_1999}. However, note that the vapor plume expansion does not follow a perfect Sedov strong shock solution (see Appendix~\ref{sec:hydro-methods}).

Initially, in the first 1-2 hours for this example, the pressure and density of the expanding vapor plume exceeds the pressure and density of the shocked nebular gas. As the pressure drops inside the vapor plume, a reverse shock forms (Fig.~\ref{fig:1devolution}, dotted black lines) that travels inward from the nebula-plume interface (Fig.~\ref{fig:1devolution}, thick white or grey dashed lines). The inward shock slows the plume expansion and heats and compresses the outer shell of the plume. Eventually, the nebula-plume interface stalls and begins to flow inward. The grey dot in each panel indicates this stall point. Nebular gas near the impact event can be shocked to temperatures that exceed the melting point of silicates. The thin shocked region of nebular gas, visible in panels D and E, cools as the zone expands. The streamlines indicate the pressure-temperature paths for a parcel of gas at different initial distances from the impact point. 

These shock fronts generate two distinct transient environments. The first environment is shocked nebular gas (and dust) exterior to the nebula-plume interface (thick dashed line) and interior to the exterior nebular shock (thick solid line). The hot material in the inset (panel E) is shocked nebular gas. The second environment is shocked plume material (water vapor and dust) between the nebula-plume interface (thick dashed white line) and inward plume shock (thick dotted black line), which develops at later times in panel (D). Each region initially has a distinct oxygen fugacity determined by their different compositions. Water vapor plumes are cool compared to the shocked nebular gas; typical collisions vaporize ice but not the silicate dust (as shown in Figure~\ref{fig:criticale}). The outer portions of the water plume are heated by the inward plume shock. The temperature evolution will be discussed in more detail below.

Initially the outward flow is a dense plume gas driving a shock into a less dense nebular gas. Under these conditions, the nebula-plume interface is stable against Rayleigh-Taylor instabilities. However, during the plume evolution, the relative densities reverse, and the interface becomes unstable to turbulent mixing. At this point, the two distinct chemical environments will begin to mix in turbulent eddies. This mixing process is not captured by the 1D Lagrangian calculation and will be explored more below when discussing the 2D and 3D simulations. 

At later times, the 1D plume collapses to the center point, the converging flow reshocks the plume, and the plume expands again. With each outward and inward oscillation, the shock pressure amplitude decays. Because turbulent mixing would modify the later evolution, we truncate the time axis for the presentation in Figure~\ref{fig:1devolution}.  Next, we will consider a 2D model where the planetesimals are ice-dust mixtures.

\subsection{2D Calculations: Turbulent mixing and collapse}

In 2D simulations using the CTH code, we simulated the asymmetric expansion of the vapor plume using the simplification of a torus-shaped projectile with a mass that corresponds to the interacting mass in the 3D case (Figure~\ref{fig:dimensions}). Here, we focus the additional information revealed in a 2D simulation.

\begin{figure*}
\centering
\includegraphics[width=1\textwidth]{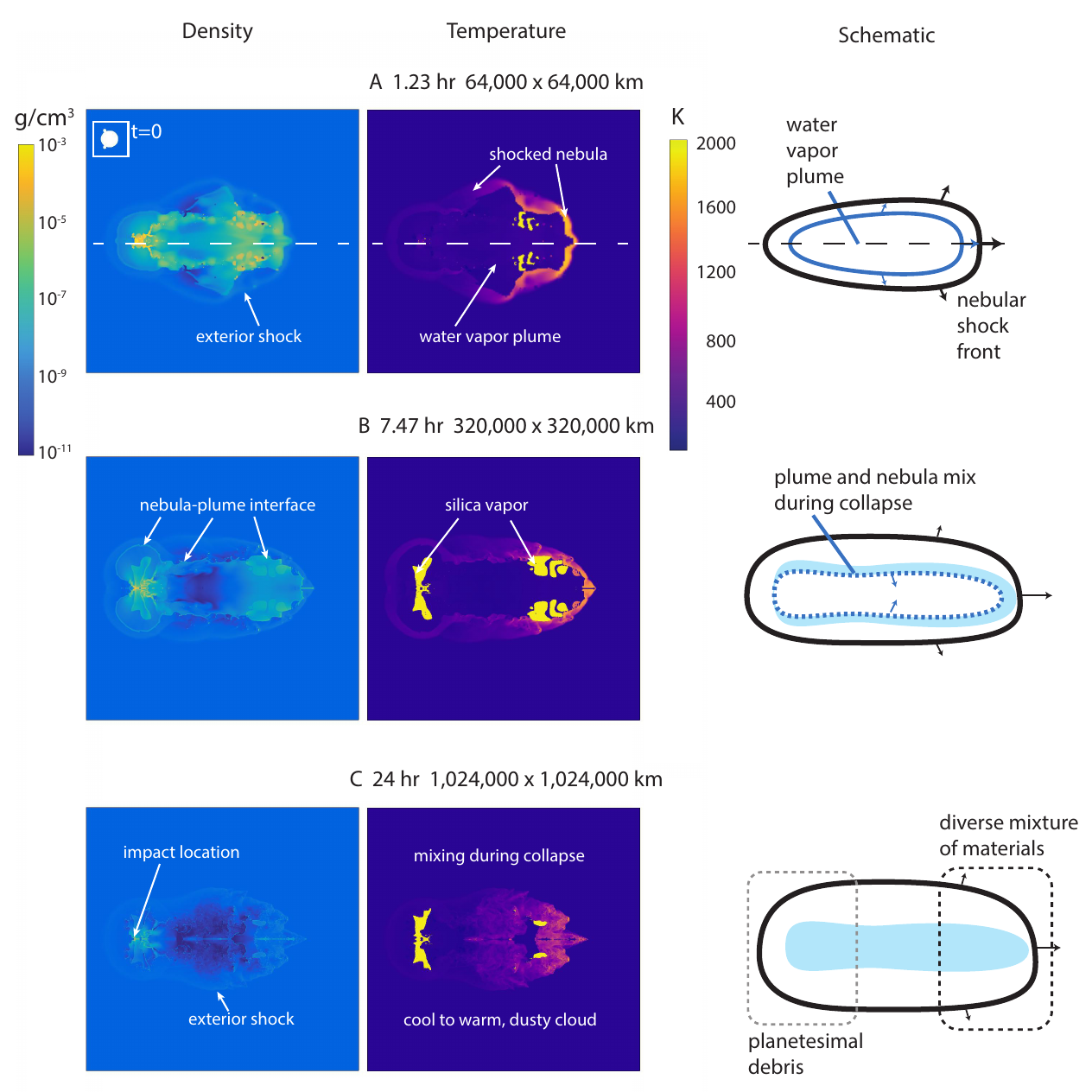} 
\caption{{\bf 2D vapor plume expansion and collapse.} Example 2D cylindrically symmetric vapor plume evolution starting with ice-silica bodies colliding at 6 km~s$^{-1}$. The surrounding nebular gas is dust-free at 0.1 Pa and 200~K. A. Initially the vapor plume is denser than the surrounding shocked nebula, indicated by the shell of hot gas bounded by the exterior nebular shock and the nebula-plume interface. The water vapor plume is cool with small amounts of hot silica gas. The principal direction of the vapor plume generates a much hotter nebular shock than the opposite direction. In the schematic, the region within the blue line contains water vapor and condensed silica. The ring between the blue and black lines contains shocked nebular (H-He) gas. Beyond the black line is the undisturbed ambient nebular gas. B. As the pressure and density drops in the vapor plume, turbulent eddies form at the nebula-plume interface. C. The inward turbulent flow mixes plume and nebular materials in a warm cloud about $10^4$ times greater in size than the original planetesimals. The schematic light blue zone is a mixture of plume materials and nebular materials. Note the change in scale at each time step. Animations of this impact are available in \href{https://chondrules.net/ivans/}{supplemental materials}. \label{fig:2devolution}}
\end{figure*}

In this example, the target has no relative velocity to the surrounding nebular gas to simplify the presentation; however, the overall phenomena is similar when both bodies have a relative velocity to the gas. The choice of no relative velocity between the gas and target is based on the results from $N$-body simulations. During planet migration, many planetesimal collisions occur between an excited body on an eccentric orbit and a body on a more circular orbit in the midplane \citep[see Appendix~\ref{app:nbody} and][]{Carter_2020,Carter_Stewart_2022}. In Figure~\ref{fig:2devolution}, the target is modeled as a 200-km radius sphere and the projectile as a 34-km radius torus. The torus mass is comparable to the interacting mass in a 100-km radius spherical projectile in 3D. The planetesimals were initialized by placing alternating layers of water ice and fused silica in 1-km radial increments, giving a total water ice mass fraction of 14 wt\%. 
The torus impacts the target at 6~km~s$^{-1}$. The surrounding nebular gas is at 200~K and $10^{-6}$ bar and modeled using the SESAME 5250 hydrogen equation of state. The code advects the mixture of materials through the computational mesh but does not include a multi-phase flow feature where the solids and gas can have separate velocity fields.

The plume has a principal expansion direction determined by the velocity vector of the projectile. The strongest nebular shock is in the initial direction of the projectile motion (also referred to as the downrange direction). The magnitude of the nebular shock decreases with increasing azimuthal angles from the principal plume direction. In 2D cylindrical symmetry, the vapor plume has an elongated cigar shape (note that the shape is somewhat modified in 3D). Turbulent eddies first form on the lateral surfaces of the plume as the density and pressure of the plume drops and the interface becomes unstable (B, 7.47 hr). The last region of the plume to collapse is the downrange direction with the strongest nebular shock and greatest outward velocity in the vapor plume (C, 24 hr).

The expanding plume materials are generally cooler than the shell of shocked nebula surrounding the plume. The small regions of hot material in the vapor plume are small amounts of shock-vaporized silica. The silicate vapor has a very low density and should chemically interact with the surrounding water vapor and quench; however, the code does not include these processes. Most of the silicate and ice from the bodies remains in the solid phase and spatially near the impact point. Only a portion of the dust in the bodies is swept downrange with the expanding vapor plume. The collapsed plume has a lower density than the surrounding unperturbed nebula due to the shock-heating of the nebular gas. 
The plume is optically thick \citep[as was observed in the Deep Impact mission,][]{Richardson_Melosh_Lisse_Carcich_2007} so we neglect radiative heat transfer on the 10s of hours timescales of plume evolution for this initial study (see also Appendix~\ref{sec:opacity}). Over time, the materials vaporized by the impact or nebular shock would recondense as the system cools to the temperature of the background nebula.

The gas-dust mixture in the collapsed plume eventually establishes hydrostatic equilibrium in the frame of reference of the orbiting nebular gas. The warm cloud cools through radiative transfer and convection within the mixture. Note that the 100-km scale planetesimals generated a nebular perturbation comparable to the orbit of Io around Jupiter (a factor of $10^4$ larger than the size of the original bodies). The warm, dusty cloud is a mixture of materials that were heated to different degrees along the asymmetric vapor plume. Some regions were heated to the melting point of silicates but others were not heated above the melting temperature of ice. 

Because the hydrocode cannot model the two-phase flow directly, we separately calculate the interactions between different size dust particles and the nebular shock and plume front, which is presented below (Section~\ref{sec:sizesorting}).

\subsection{3D Calculations: Planetesimal disruption within the shocked nebular region}

Our 3D simulations illustrate the separation between the large slow ejecta and fast vapor plume front in vaporizing, disruptive collisions. The colliding bodies are modeled as spherical mixtures of ice and silica. In Figure~\ref{fig:3devolution}, each frame presents the equatorial plane of the two colliding bodies. In this example, a 100-km radius body strikes a 200-km radius body at 6~km~s$^{-1}$. The initial bodies are about 38 wt\% ice. The asymmetric nebular shock front is similar to the 2D case. Here, the disrupted bodies are clearly visible as high density regions in panel A and B. Unlike in collisions in a vacuum, the strewn fragments are slowed by the surrounding gas. The least shocked fragments of the planetesimal bodies are located opposite the hottest regions of the nebular shock (opposite the principal direction of the vapor plume). In our nominal planetesimal model, the solid fragments from the bodies are ice-dust aggregates and the shear from the gas would act to break them into small pieces (Appendix~\ref{sec:coupling}). The size distribution of these solid fragments will be investigated in future work. The turbulent collapse of the vapor plume leads to mixing of material with very different thermal histories. The different regions will be discussed in more detail below.

\begin{figure*}%
\plotone{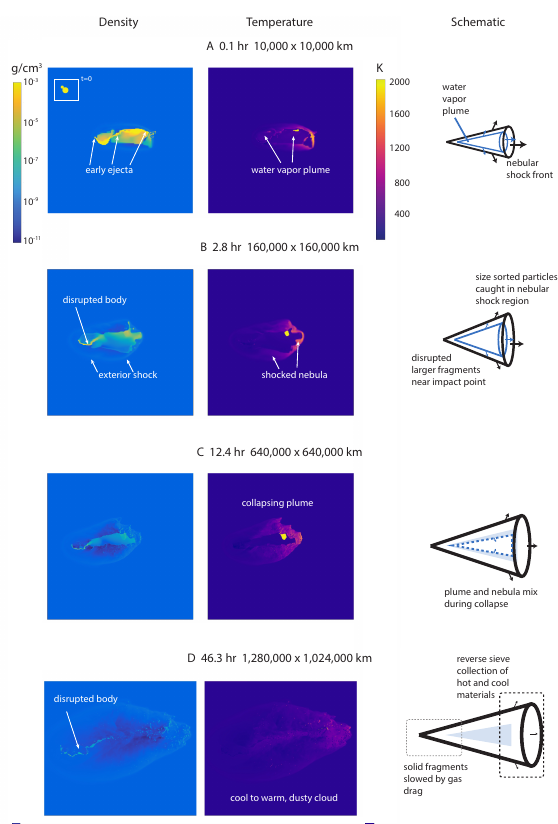}
\caption{{\it Caption on next page.}}
\end{figure*}

\addtocounter{figure}{-1}

\begin{figure*}
\caption{{\it On previous page.} {\bf 3D vapor plume expansion and collapse.} An example 3D vapor plume evolution starting with ice-silica bodies colliding at 6 km~s$^{-1}$. The surrounding nebular gas is dust-free at $10^{-6}$ bars and 200~K. A. The principal direction of the vapor plume expands more rapidly than the solid fragments from the disrupted bodies. In the schematic, the region within the blue cone contains water vapor and condensed silica. The region between the blue and black cones contains shocked nebular (H-He) gas. Beyond the black cone is the undisturbed ambient nebular gas. B. The cool solid fragments from the disrupted bodies have slower velocities compared to the fast impact vapor in the principal direction of the plume. C. The inward turbulent flow in the principal direction mixed shocked nebula and plume debris far downrange from the largest fragments from the disrupted bodies. D. The material collected at the leading face of the plume is sorted by the reverse sieve effect described in section~\ref{sec:sizesorting}. The schematic light blue zone is a mixture of plume materials and nebular materials. Note the change in scale at each time step. Animations of this impact are available in \href{https://chondrules.net/ivans/}{supplemental materials}.  \label{fig:3devolution}}
\end{figure*}

\subsection{Modeling a dusty nebula}

In the example calculations presented above, the nebula is modeled as a dust-free gas. However, the formation of environment for chondrules requires a dust-to-gas ratio that is elevated compared to the initial solar nebula \citep[e.g.,][]{Alexander2008,Shimizu_Alexander_Hauri_Sarafian_Nittler_Wang_Jacobsen_Mendybaev_2021}. During dynamical excitement from giant planets, planetesimal collisions will generate substantial amounts of debris. If a large fraction of the proximal planetesimals have not differentiated, then the debris is likely dominated by small dust particles as found in comets. To explore the effect of high dust-to-gas ratios on the dynamics of an impact vapor plume in the nebula, we implemented an ideal dust-gas mixture equation of state in our 1D simulations following the methods in \citet{Steiner_Hirschler_2002}. In an ideal mixture of dust and gas, the dust is considered incompressible and small enough to be perfectly coupled to the gas. The temperature of both components is the same and the internal energy is partitioned between the components by mass. The ideal dusty gas model is described in detail in Appendix~\ref{sec:dig}.

We are interested in the pressure-temperature conditions in the nebular shock. The presence of dust substantially modifies the shock response. For strong shocks in gases, the density of the shocked gas quickly asymptotes to a value equal to 
\begin{equation}
    \rho_{\rm s}=\rho_{\rm neb} \left ( \frac{\gamma +1}{\gamma -1} \right ),
\end{equation}
where $\rho_{\rm s}$ is the density of the shocked gas, $\rho_{\rm neb}$ is the initial density of the nebular gas, and $\gamma$ is the ratio of isobaric to isochoric specific heat capacities. For hydrogen gas, $\rho_{\rm s} \sim 6 \rho_{\rm neb}$. See Table~\ref{tab:nebula} for the initial states of the nebula. The compressibility of the gas is limited by the thermal pressure and the shock temperature rises quickly with increasing shock pressure. 

\begin{figure*}
\plottwo{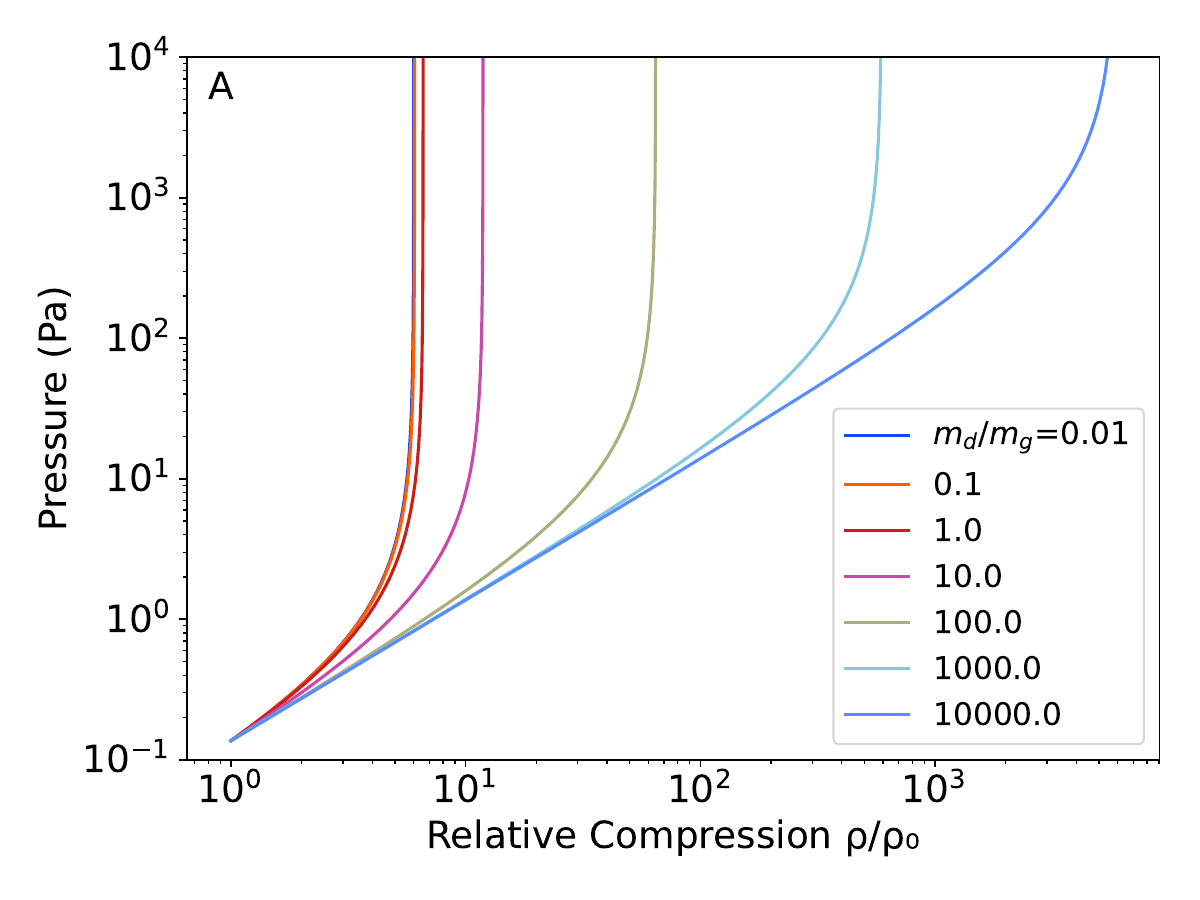}{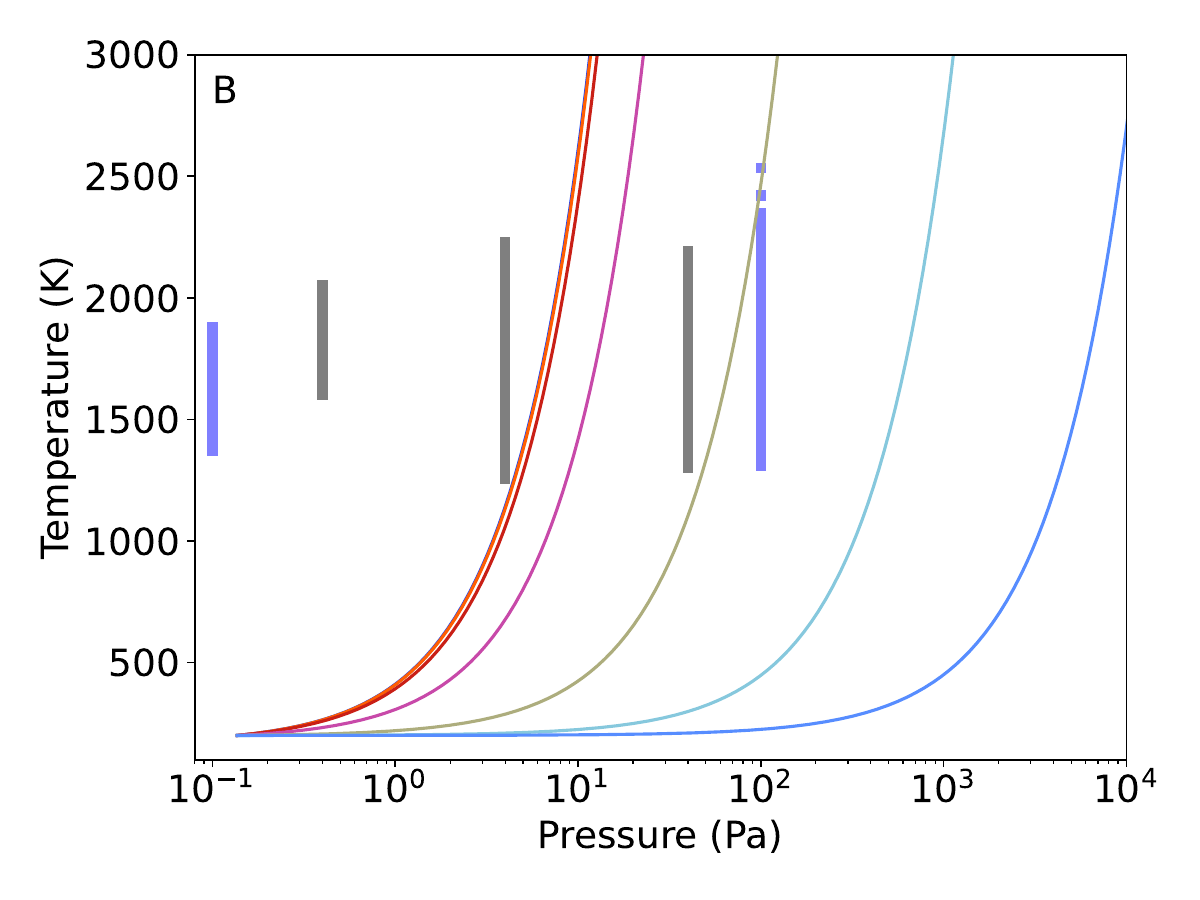} 
\caption{{\bf Ideal dust-gas mixture Hugoniots.} The relative compressibility of a dusty gas is much greater than for a pure ideal gas.  A. Shock Hugoniots for an ideal dusty gas with varying dust-to-gas mass ratios ($m_d/m_g$), with density normalized to the initial density. Each Hugoniot originates at 200~K and 0.137 Pa (see Table~\ref{tab:nebula}). Note that the $m_d/m_g$=0.01 and 0.1 curves are indistinguishable on these scales. B. At a fixed shock temperature, the corresponding shock pressure is greater as the dust-to-gas mass fraction increases. A wide range of pressure-temperature conditions are possible in a shocked dusty gas, as discussed below. Bars show stability range for liquids in various dust-enriched nebular compositions from this study (grey, see Section~\ref{sec:melt_stab} and Table~\ref{tab:meltcalcs}) and \citet{Ebel_Grossman_2000} (blue).
 \label{fig:dig}}
\end{figure*}

Adding dust to the gas increases the bulk compressibility of the system and lowers the shock temperature for a given shock pressure, as shown in Figure~\ref{fig:dig}. The ideal dust-gas mixture temperatures in our model do not account for the latent heat of melting or vaporization, which would buffer the temperature of the system. Adding dust to the nebular gas also reduces the sound speed (Appendix~\ref{sec:dig}). Thus, in dusty regions, it is easier to generate a strong shock compared to a dust-free gas. 

\begin{table*}
\begin{tabular}{lrrrrrrrr}
\hline
Shock Pressures (GPa) & 10 & 15 & 20 & 40 & 80 & 120 & 160 & 200 \\
Shock Density (g/cm$^3$) & 1.90 & 2.01 & 2.09 & 2.33 & 2.63 & 2.84 & 3.00 & 3.12 \\
Shock Temperature (K) &  479 & 682 & 868 & 1865 & 3979 & 6071 & 8342 & 10908 \\
Shock Sp. Energy (MJ/kg) & 1.55 &  2.44 & 3.39 & 7.59 & 16.99 & 27.07 & 37.49 & 48.10\\
Shock Sp. Entropies (kJ/K/kg) & 3.82 & 4.77 & 5.57 & 7.69 & 10.00 & 11.47 & 12.58 & 13.45 \\
Equivalent Planar $V_i$ (km/s) & 3.31 & 4.25 & 5.06 & 7.70 & 11.59 & 14.66 & 17.27 & 19.58 \\
\hline
\end{tabular}
\caption{Summary of initial shock states for 1D simulations of compressed water plumes expanding into dusty and dust-free nebular gas with varying nebular pressures. Initial shock states are on the principal Hugoniot for 150~K ice using the ANEOS equation of state.
The Hugoniot defines the locus of possible shock states that can be reached from a given initial state. 
}
\label{tab:1dsims}
\end{table*}

\begin{table*}
\centering
\begin{tabular}{lrrrrrrrrrrr}
\hline
Pressure (Pa) & 0.137 & 1.37 & 13.7 & 1.37 & 1.37 & 1.37 & 1.37 & 1.37 & 1.37 & 1.37  \\
Dust-to-gas mass ratio, $m_d/m_g$ & 0 & 0 & 0 & 0.01 & 0.1 & 1 & 10 & 100 & 1000 & 10000 \\
Bulk density, $\rho_{\rm neb}$ (kg/m$^3$) & 2E-8 & 2E-7 & 2E-6 & 2.02E-7 & 2.2E-7 & 4E-7 & 2.2E-6  & 2.02E-5 & 2.002E-4 & 2.0002E-3   \\
Temperature (K) &  200 & 200 & 200 & 200 & 200 & 200 & 200 & 200 & 200 & 200 & \\
\hline
\end{tabular}
\caption{Summary of initial states for the dust-free or dusty-gas nebula surrounding the vapor plume. More details are provided in Appendix~\ref{sec:dig}.
}
\label{tab:nebula}
\end{table*}

\section{Plume stall radii and timescales}
\label{sec:scaling}

\begin{figure}
\centering
\includegraphics[width=.475\textwidth]{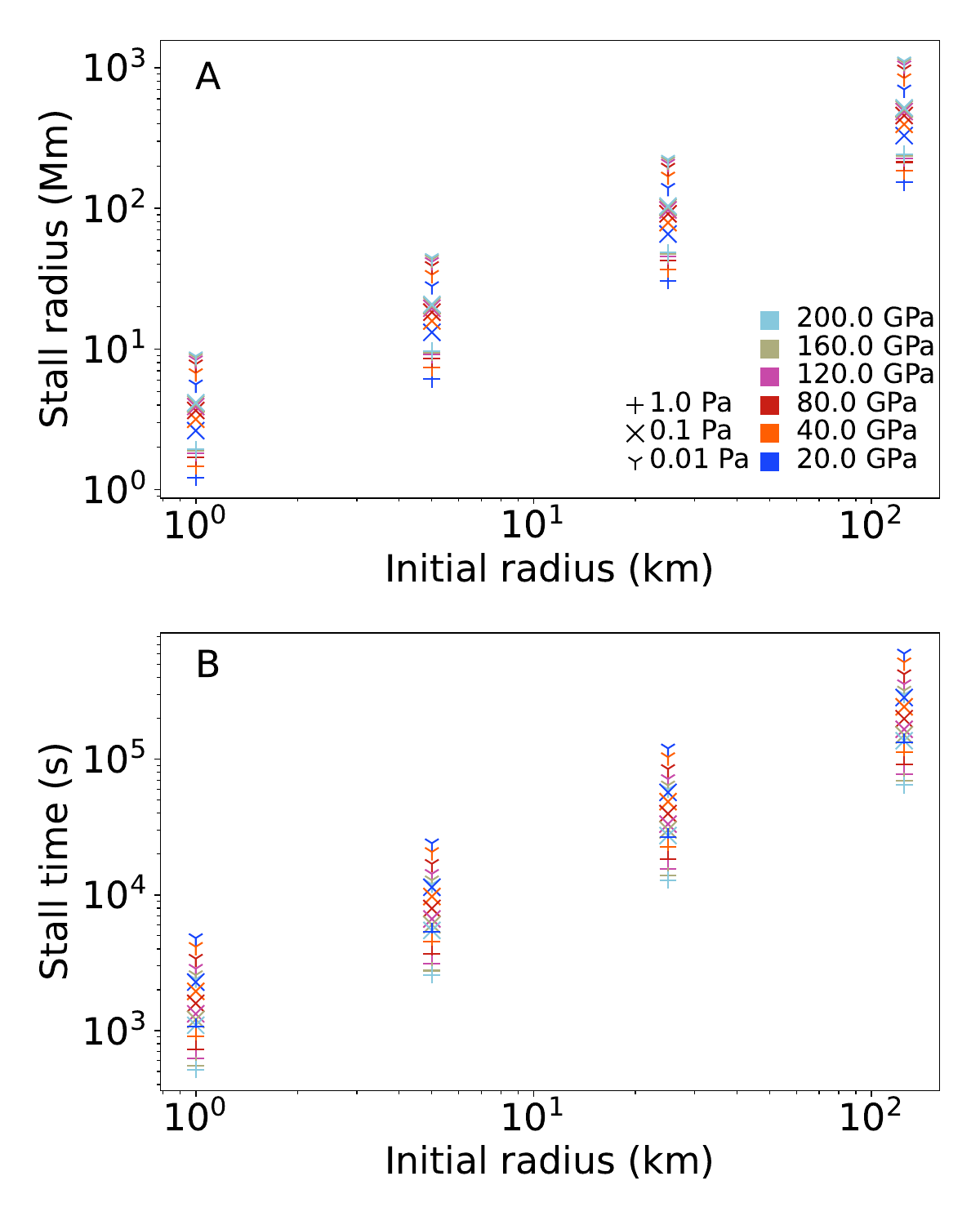} 
\caption{{\bf 1D water vapor plume expansion results.} Results from dust-free nebula simulations demonstrating the systematic dependence for the plume stall radius and time on the initial radius of the plume source, initial shock pressure (colors), and the pressure (or density) of the nebular gas (symbols). 
 \label{fig:waterplumes}}
\end{figure}

The evolution of the expanding plume depends on just a few parameters of the event. Three of the key variables are the composition of the planetesimals, the impact velocity, and the impact angle. Together, these variables determine the peak pressure in the colliding bodies and the amount of vapor produced in a single event. For scaling in the 1D simulations, we account for the combined effect of these parameters by initializing a specified mass of water at a homogeneous initial shock pressure. The expanding vapor does work against the surrounding nebular gas; so the denser the nebular gas, the smaller the expansion distance. The expansion distance is proportional to the original energy of the shocked plume source. Hence, the key variables that control the evolution of the vapor plume are the density (or pressure) of the nebular gas, the masses (or alternatively length scales) of the planetesimals, and the shock pressures attained in the collision.

Using pyKO, we calculated the 1D spherical expansion of a shocked sphere of water vapor surrounded by nebular gas (and dust). We considered a wide range of initial conditions, presented in Tables~\ref{tab:1dsims} and \ref{tab:nebula}. The initial shock pressures were varied with values corresponding to planar impact velocities between about 3 to 20 km\,s$^{-1}$, spanning partial to full vaporization of water ice when decompressed to the triple point pressure (6~mb) \citep{Stewart2008}.  To develop robust scaling laws, we also simulated fused silica vapor plumes (Table~\ref{tab:1dsims-silica} in Appendix~\ref{sec:hydro-methods}). We modeled cases with a pure H-He ideal gas and a dusty gas with varying amounts of dust. The dust-to-gas mass ratios of the surrounding nebular gas were 0, 1, 100, or 10000. We considered nebular gas pressures, which scales with density for an ideal gas, of 1, 0.1, and 0.01 Pa. These values are typical of the range encountered in the inner to outer solar system in an evolved protoplanetary disk \citep[e.g.,][]{Dodson-Robinson_2009}. Our nominal initial gas temperature was 200 K to represent regions where water ice is a component of the nebular dust.

Figure~\ref{fig:waterplumes} shows the systematic variations of the stall radius and time with shock pressure, background gas pressure (or density), and the size of the plume source. The properties of the vapor plume can be non-dimensionalized via the Buckingham Pi theorem to produce general scaling laws for a wide variety of impact conditions.

\begin{figure*}
\plotone{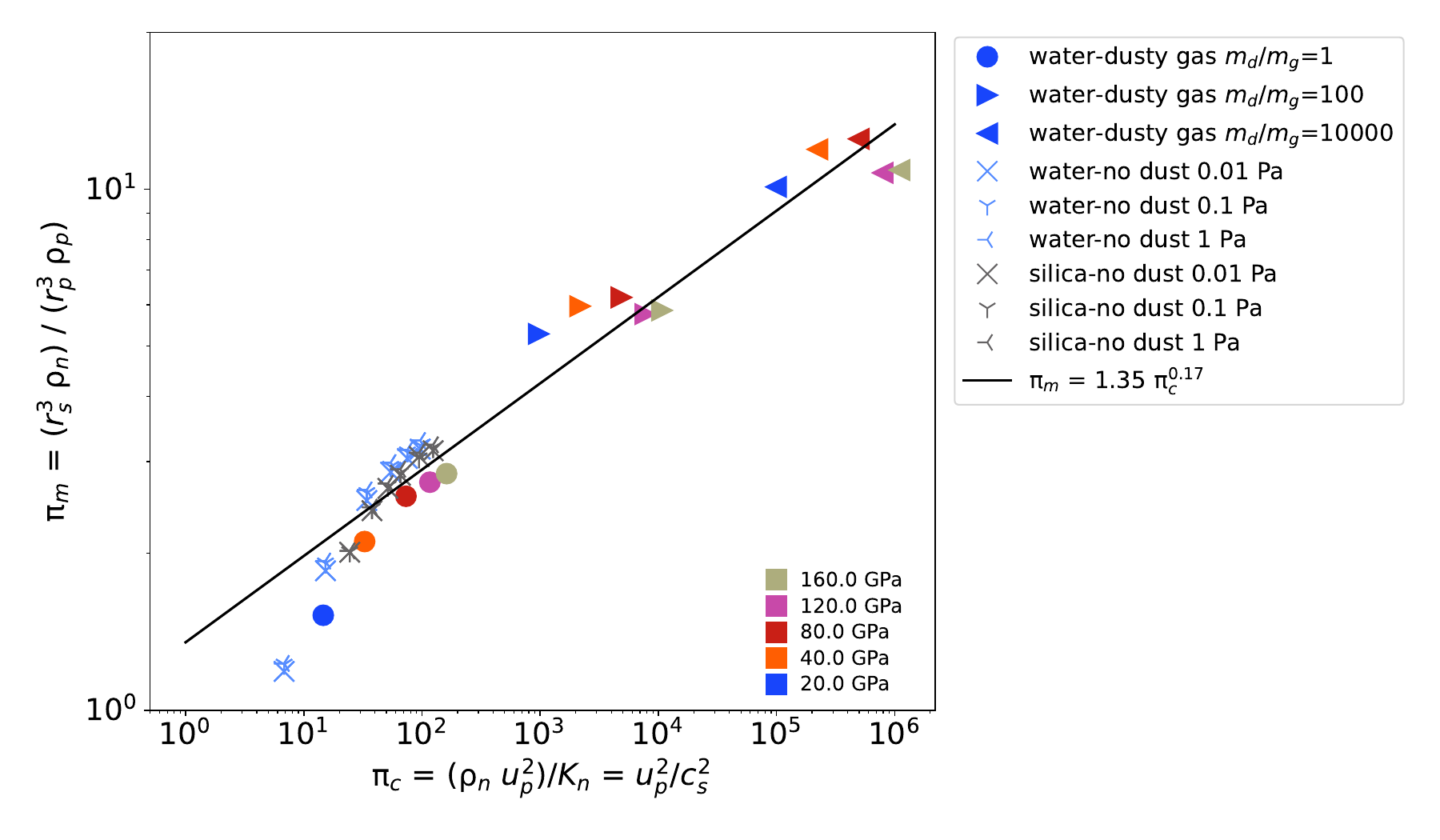}
\caption{{\bf Plume stall radius scaling law.} A power law (black line) relates the mass displaced by the plume expansion to the impact energy for a wide range of impact conditions. Results for water vapor plumes expanding into dusty gas (filled symbols indicating the dust-to-gas ratio). The dusty gas pressure was 0.1 Pa and the initial pressure of the plume is indicated by the color of the symbols. Other symbols show the results for expansion of water (light blue) or silica (grey) into dust-free gas. The lowest energy events do not generate strong shock waves or substantial amounts of vapor and fall off the general power law. See Tables~\ref{tab:1dsims} and \ref{tab:1dsims-silica} for full details of the simulation parameters.
 \label{fig:pimass}}
\end{figure*}

\begin{figure*}
\plotone{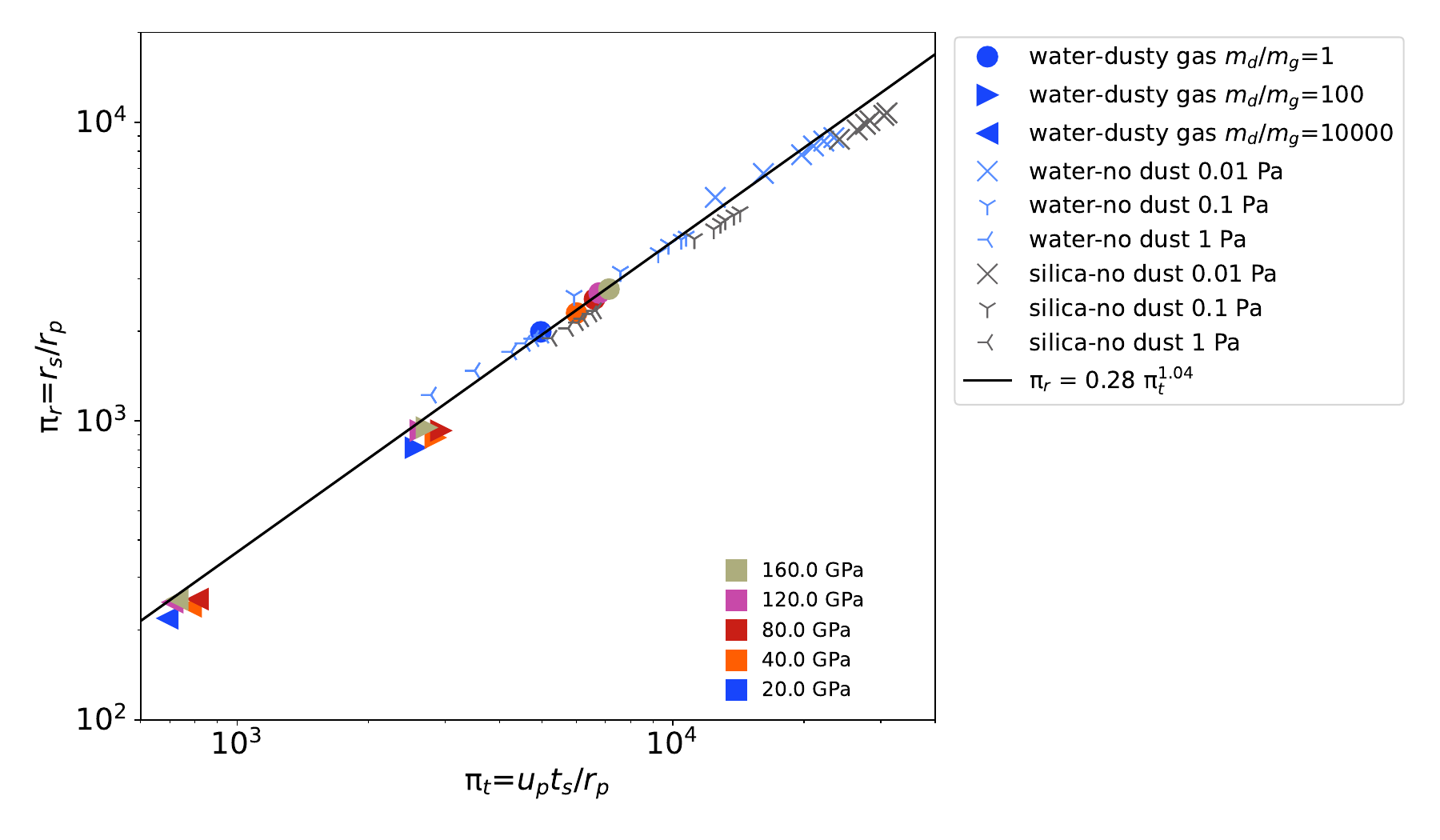} 
\caption{{\bf Plume stall time scaling law.} The power law (black line) relates the plume stall time and radius for a wide range of impact conditions. The dusty gas pressure was 0.1 Pa. Color legend for initial pressure in water-dusty gas cases only; see Tables~\ref{tab:1dsims} and \ref{tab:1dsims-silica}.
 \label{fig:pitime}}
\end{figure*}

The first non-dimensional term that can be constructed is the Cauchy number which is a representation of the strength of the shock wave driving into the nebula,
\begin{equation}
    \pi_c = \frac{\rho_n u_p^2}{K_n} = u_p^2/c_s^2,
\end{equation}
where $\rho_n$ and $K_n$ are the density and incompressibility of the nebula and $c_s$ is the sound speed in the unshocked nebula. $u_p$ is the the shock state particle velocity that represents the energy of the plume. For our idealized initial shock states, the specific kinetic energy is equal to the specific internal energy in the shock state. Thus, the shock state particle velocity is a representative velocity of the event that is independent of the surrounding gas:
\begin{equation}
    u_p = \sqrt{2E_p},
\end{equation}
where $E_p$ is the specific internal energy of the shocked plume source. 

The total nebular mass within the stall radius, $r_s$, is non-dimensionalized by the total mass in the plume source:
\begin{equation}
    \pi_m = \frac{r_s^3 \rho_n}{r_p^3 \rho_p},
\end{equation}
where $r_p$ and $\rho_p$ are the initial radius and density of the plume, respectively. The dependence of $\pi_m$ on $\pi_c$ is shown in Figure~\ref{fig:pimass}, which combines the results from water vapor and silica vapor plumes and different nebular environments. The nebular pressure for the dust-gas mixtures was 0.1 Pa. At low impact energies, the plume size is lower than predicted by the general power law as these events produce only small amounts of vapor. A log-linear fit to the data is $\log(\pi_m)=b+a\log(\pi_c)$, where $[a,b]=[0.16568046, 0.12964904]$ and the covariance matrix, $\Sigma$, is
\begin{center}    
$
\Sigma = \begin{bmatrix}
 0.00012852 & -0.00049172\\
-0.00049172 & 0.00217774\\
\end{bmatrix}.
$
\end{center}

We use the stall time, $t_s$ as a representative time scale for plume expansion. The non-dimensionalized stall radius,
\begin{equation}
    \pi_r = \frac{r_s}{r_p},
\end{equation}
is related to the non-dimensional stall time,
\begin{equation}
    \pi_r = \frac{u_p t_s}{r_p} ,
\end{equation}
as shown in Figure~\ref{fig:pitime}. For a wide range of plausible impact parameters and nebular properties, the stall radius is about 10$^3$ to 10$^4$ times larger than the plume source radius.  A log-linear fit to the data is $\log(\pi_r)=b+a\log(\pi_t)$, where $[a,b]=[1.04023665, -0.55931397] $ and the covariance matrix, $\Sigma$, is 
\begin{center}    
$
\Sigma = \begin{bmatrix}
9.10405834\times10^{-5} & -3.49911677\times10^{-4}\\
-3.49911677\times10^{-4} &  1.35759165\times10^{-3}\\
\end{bmatrix}.
$
\end{center}

These two power laws can be used to estimate the volume of nebular materials that are displaced by the vapor plume from any given impact. The enclosed mass provides an estimate of the amount of nebular material involved in the inward flow during plume collapse. Note that the overall shocked nebular region is larger than the volume estimated from our scaling laws, as the extent of the nebular shock is not defined by the stall radius of the vapor plume but by the distance to the point where the exterior nebular shock decays to a sound wave. Future work will develop relationships that account for changes to these scaling laws that arise from the relative velocity between the center of mass frame of the impact event and the reference frame of the nebular gas.

\section{Nebular shock conditions}
\label{sec:nebconditions}
Next, we examine the properties of the nebular shock and the conditions experienced by materials in the shocked nebula.

\subsection{Shock Pressures, Temperatures and Velocities}
\label{sec:nebpt}

\begin{figure*}
\centering
\includegraphics[width=1\textwidth]{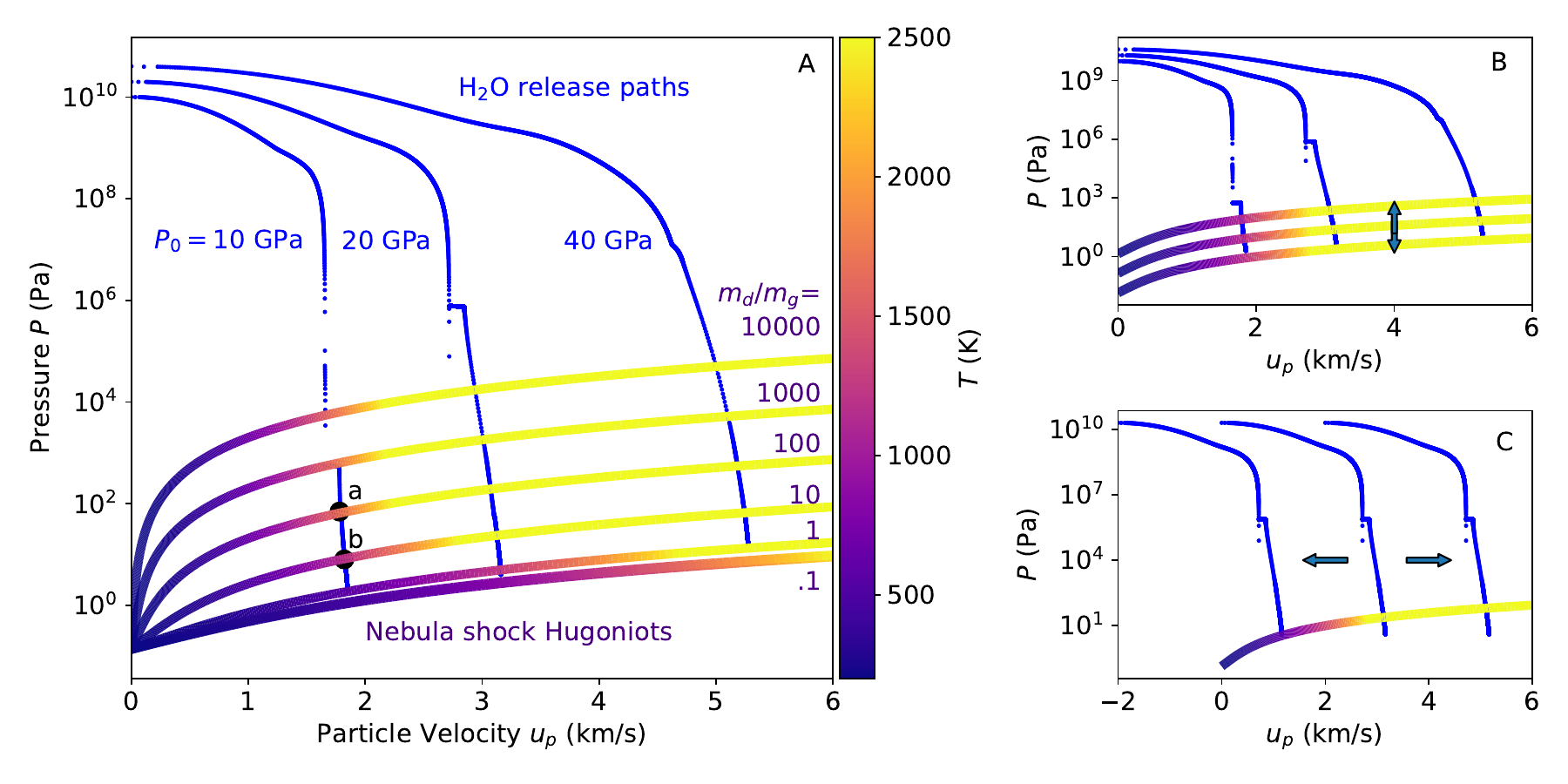} 
\caption{{\bf Shock impedance match solution for water plume and different nebular dust-gas mixtures.} A. Spherical decompression paths (blue curves) for 25-km radius water ice sphere initially shocked to pressures ($P_0$) of 10, 20 and 40 GPa in the rest frame of the nebula ($u_{p,0}=0$), calculated using the 1D pyKO code. The pressure and particle velocity of the initial shock in the ambient nebula is defined by the intersection between the water decompression path and the shock Hugoniot (color gradient lines) of the nebular gas. Here, the nebula initial state is 0.1~Pa and 200~K and the lines with temperature color gradients correspond to varying dust-to-gas mass ratios ($m_d/m_g$) as labelled. Increasing the dust mass fraction in the nebular gas leads to greater initial pressures in the nebular shock (e.g., point a vs.\ b). In addition, adding dust lowers the shock pressure required to reach silicate melting temperatures. The kinks in the H$_2$O decompression paths correspond to the intersection with the water vapor curve. B. Varying initial pressure in the nebula shifts the nebular shock Hugoniot up or down. C. Varying relative velocity of plume to the nebula shifts the plume decompression path left or right. 
 \label{fig:imtemps}}
\end{figure*}

The initial properties of the nebular shock front can be estimated using an impedance match solution, which is the solution that attains continuity in pressure and particle velocity between the expanding plume and shocked nebular gas. This solution determines the initial magnitude of the exterior nebular shock. The intersection is defined by the release isentrope from the shock state of the plume with the shock Hugoniot of the surrounding gas \citep[e.g., chapter 3 in][]{Forbes_2012}. As the plume decompresses, the reduction in pressure and density requires an increase in particle velocity to expand the plume. The time evolution of the particle velocity is found by solving the Reimann integral,
\begin{equation}
    du_p = -\frac{dP}{\rho_p c_s} \bigg |_S,
\end{equation}
where $u_p$ is the particle velocity, $P$ is pressure, $c_s$ is the sound speed, and $\rho_p$ is density in the isentropic (subscript $S$) thermodynamic path through the plume mixture equation of state.

In 1D planar flow, the compression and decompression occurs under uniaxial strain. Therefore, the acceleration required for the first material to match the velocity and pressure of the second material is independent of the length scale of the problem. However, in non-planar geometries, the acceleration of the shocked material depends on the length scale because of the dependence on density. 
The solution to the Reimann integral is commonly presented for planar geometries \citep[e.g.,][]{rice1958compression} but rarely for non-planar geometries. Thus, we present a derivation of the particle acceleration for spherical decompression in Appendix~\ref{app:Riemann}. The acceleration is given by
\begin{equation}
    \Delta u_p = \mp \frac{c_s}{\rho_p}\Delta\rho_p \mp \frac{2u_p c_s}{r}\Delta t,
\end{equation}
where $r$ is radius and $t$ is time. This equation must be solved numerically with the equation of state. We calculated the decompression path for a shocked water sphere expanding into the nebular gas using the 1D pyKO code. In general, because of the faster drop in density, the pressure decays more rapidly with distance in spherical expansion than in planar expansion. As a result, the impedance match particle velocities are also smaller in spherical than in planar geometry.

Example decompression paths are presented in Figure~\ref{fig:imtemps}A (blue lines) for different initial states (pressures) on the water ice shock Hugoniot and different background dust-gas mixtures (Appendix \ref{sec:dig}). Note that the greater the dust fraction, the more compressible the mixture and the impedance match solution is at higher pressures (e.g., point {\bf a} vs.\ {\bf b}). The nebular Hugoniots in Figure~\ref{fig:imtemps}A all start at 200~K and 0.137~Pa. Since these are ideal gas mixture models, changing the background pressure shifts the nebular Hugoniots up or down, leading to higher or lower pressure impedance match solutions, respectively (Figure~\ref{fig:imtemps}B). To attain the impedance match solution, the surrounding nebular gas must be accelerated to the impedance match particle velocity. This is accomplished by generating a shock wave in the nebula, which travels faster than the particle velocity (compare to nebular sound speeds in Figure~\ref{fig:digmodel}) and attains the impedance match pressure. 

The colors on the nebular Hugoniots in Figure~\ref{fig:imtemps} indicate the shock temperature of the nebular gas, where the colors saturate to indicate that silicate melting temperatures are attained. Melting is not included in the model equation of state and so the simple dust-gas mixture equation of state is not robust above the solidus. In addition, vaporization of ice dust and silicate dust would increase the pressure behind the shock front. Thus, the shock pressures could be even greater in the transient environment of the nebular shock than calculated here. We have also neglected radiative transfer due to the high opacity of the vapor plume and high dust-to-gas ratios in the nebula (Appendix~\ref{sec:opacity}). Radiative heat transfer will be considered in future work. In general, planetesimal collisions can generate nebular shock temperatures that exceed the solidus temperature for free-floating silicate dust and generate a substantial increase in pressure.

The impedance match solutions shown in Figure~\ref{fig:imtemps}A and B are presented for a plume with no initial relative velocity to the background gas. As shown in Figures~\ref{fig:2devolution} and \ref{fig:3devolution}, the nebular shock is stronger in the principal plume direction and weaker in the opposite direction. In the acceleration calculation, the initial particle velocity is the relative velocity between the plume source and the surrounding gas, which may be positive or negative with respect to the expansion direction. An initial relative velocity in the direction of the shock increases the intersection pressure and conversely the pressure is decreased in the opposite direction, as shown in Figure~\ref{fig:imtemps}C. Thus, the nebular shock pressures are larger in the principal direction of the vapor plume where the relative velocity is added to the decompression particle acceleration, and smaller in the opposite direction where the relative velocity is subtracted.  

The acceleration associated with the decompression of the vapor plume enables substantial spatial separation of the leading edge of the plume from the solid fragments of the disrupted planetesimals (Figure~\ref{fig:3devolution}). The solid fragments are ejected from the collision point at velocities relative to the center of mass of the collision. In laboratory experiments of catastrophic disruption (where the projectile to target mass ratio is small and more like a point source), most solid fragments have velocities $<1$\% of the impact velocity \citep{Martelli_Ryan_Nakamura_Giblin_1994} and so will remain close to the center of mass of the collision. The dispersed fragments will be further slowed by gas drag within the nebula. In contrast, the vapor plume acceleration increases the principal plume particle velocity. The plume acceleration provides a means of separating solid impactite products, such as partial melts, from the nebular shock products.

\begin{figure}
\centering
\includegraphics[width=.48\textwidth]{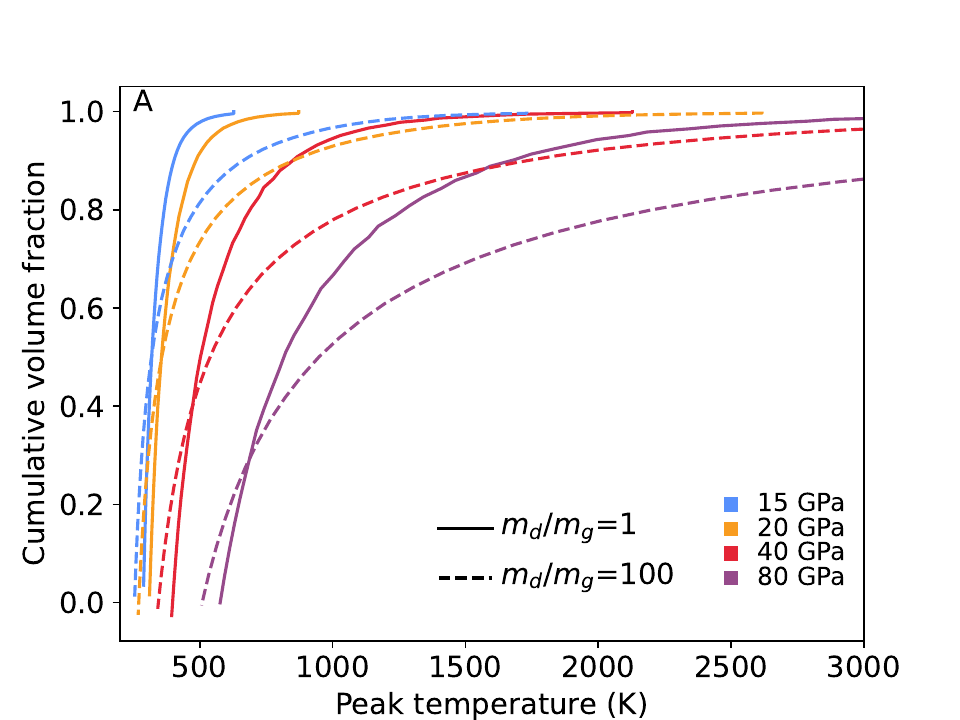} 
\includegraphics[width=.48\textwidth]{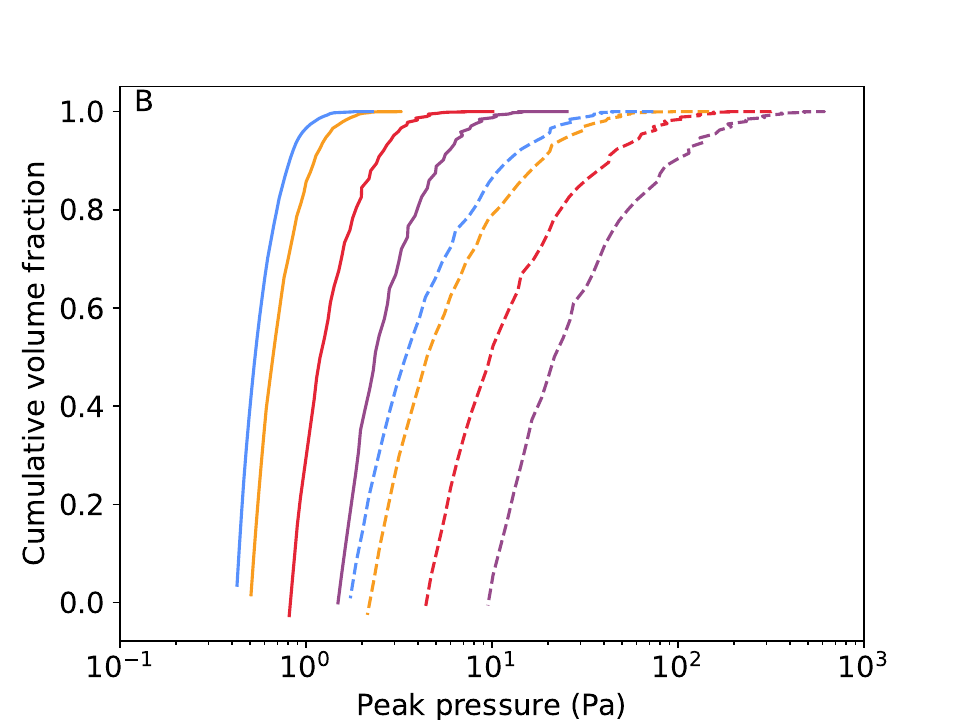} 
\caption{{\bf Nebular temperature and pressure distributions within the stall radius.} Cumulative volume (or equivalently mass) fractions below each temperature and pressure within the plume stall radius. Results for nebular shock states driven by spherical 25-km radius water vapor plume expansion from initial pressures of 15, 20, 40 and 80~GPa. The initial nebular states were 200~K and 0.137~Pa with dust-to-gas mass ratios of 1 and 100.
 \label{fig:vol-shock-pt}}
\end{figure}

In this work, we present scaling results for when the collision occurs in the rest frame of the nebula. This frame of reference is most comparable to the intermediate nebular shock pressures (e.g., lateral directions) in the asymmetric vapor plume. The impedance match solution defines the initial shock states in the exterior nebular shock front but the shock pressure decays with propagation into an ever-increasing volume of the nebula. In Figure~\ref{fig:vol-shock-pt}, we present cumulative volume fractions for the peak pressures and temperatures within the stall radius of the spherical 1D plume. These data reflect the steep decay of the shock pressure with distance from the impact point and the effect of the cubic dependence on the larger spherical radius. Note again that larger dust-to-gas mass ratios lead to increased nebular shock pressures. Each impact event generates a different range of thermal processing in the surrounding nebula.

\begin{figure*}
\centering
\includegraphics[width=1\textwidth]{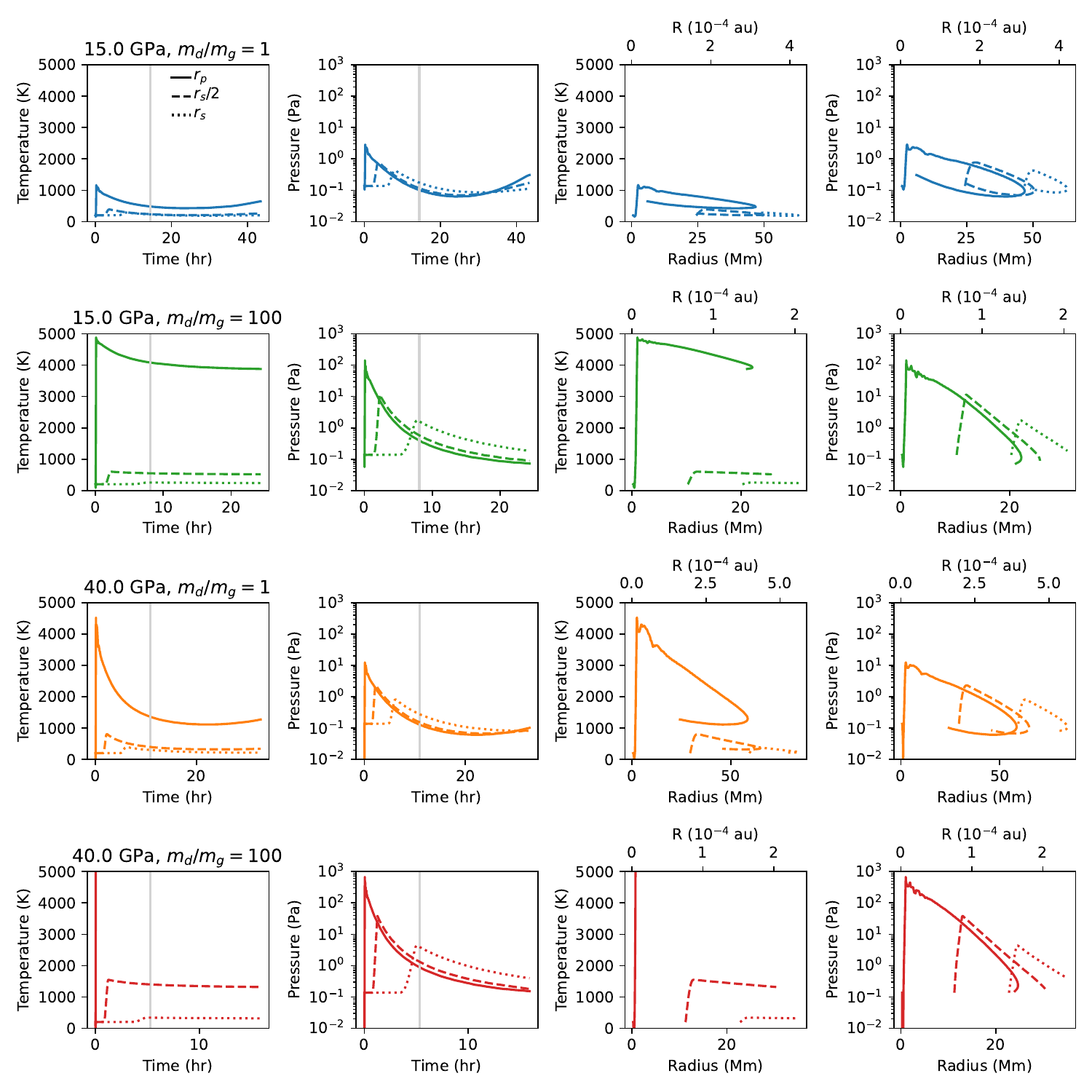}
\caption{{\bf Pressure-temperature histories in expanding nebular shocks.} Example pressure-temperature paths for the shocked nebula with different dust-to-gas ratios. Each row presents the history for material originating at the nebula-plume interface ($r_p$, solid lines), at halfway to the stall radius ($r_s/2$, dashed lines), and at the stall radius ($r_s$, dotted lines) for four events (25-km radius water plume) with dust-to-gas mass ratios of 1 and 100. The vertical grey line indicates the stall time of the nebula-plume interface. The initial nebula was at 200~K and 0.137~Pa.
 \label{fig:tempcool}}
\end{figure*}

\begin{figure*}
\epsscale{.9}
\plotone{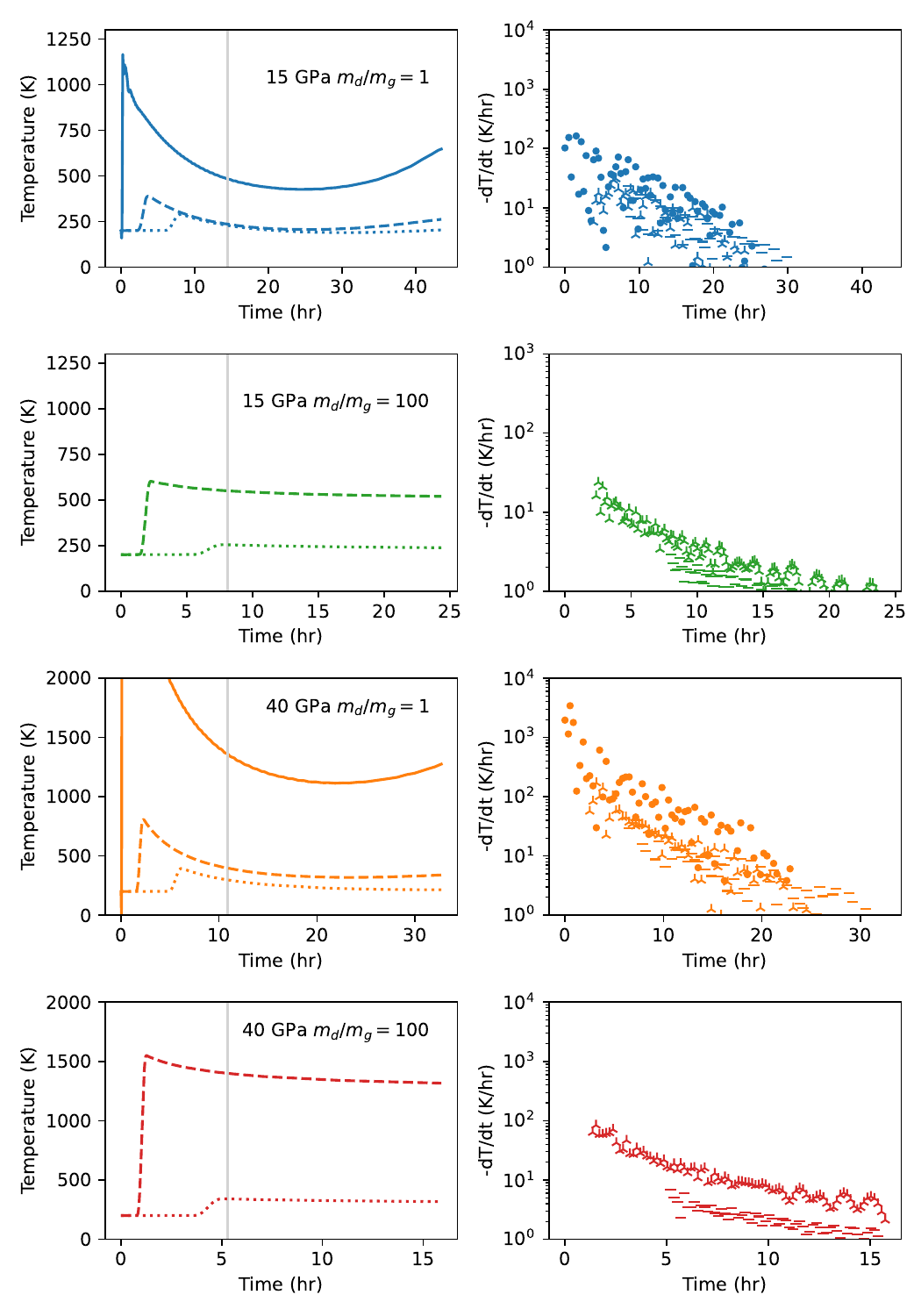} 
\caption{{\bf Temperature histories and cooling rates in expanding nebular shocks.} A. Temperature as a function of time for different parts of the nebular shock as shown in the first column of Figure~\ref{fig:tempcool}. Each row is for a different initial plume pressure and nebular gas-to-dust ratio. The vertical grey line indicates the stall time of the nebula-plume interface. B. Cooling rates for shocked nebular dust-gas mixtures at different times for parcels of nebular material at the nebula-plume interface ($\bullet$), at halfway to the stall radius ($\wye$), and at the stall radius ($-$), excluding the highest temperature examples as the model equation of state does not include melting or vaporization.
 \label{fig:tempcoolrate}}
\end{figure*}

For the perfectly coupled dust-gas mixture, we can also examine the pressure and temperature evolution of nebular material over time. In Figures~\ref{fig:tempcool} and \ref{fig:tempcoolrate}, we present the temperature and time histories for parcels of dust and gas that originated at the initial nebula-plume interface ($r_p$, solid lines), halfway to the stall radius ($r_s/2$, dashed lines), and at the stall radius ($r_s$, dotted lines). There are different cooling rates with different initial plume shock pressures and dust-to-gas mass ratios. For the case with $m_d/m_g=1$, the plume fully collapses within 3 times the stall time, leading to some recompression and heating. This spherically symmetric calculation does not fully capture the complexity of a natural event but provides some insight into the general phenomena associated with vaporizing collisions in the solar nebula.

\subsection{Melt Stability in Nebular Shocks}
\label{sec:melt_stab}

We emphasize that the amount of solids inferred to be in chondrule formation environments is substantial and needs to be included in calculating the properties of the nebular shock. Note that the dust-to-gas mass ratio must be converted to a molar ratio for comparison to geochemical calculations of molten chondrules under nebular conditions. In Table~\ref{tab:meltcalcs}, we present some example calculations of melt stability over a range of compositions and nebular pressures. Following \citet{Wood_Hashimoto_1993}, the nebula was divided into four chemical components: refractory dust, tar (carbonaceous matter), ice, and hydrogen gas. We convert the atomic ratios to weight percent and then calculate the total mass ratio of solids to gas ($m_{\rm s}/m_\mathrm{g}$).  These example melt stability calculations span solids-to-gas mass ratios of order 1 to $10^4$, values that overlap with the those used in the hydrocode calculations presented in Section~\ref{sec:scaling}. 

The melt stability range calculations used an updated version of the GRAINS code \citep{Petaev_2009,Lock_2018, MacPherson_Petaev_2025} extended to 50 elements: H, He, C, N, O, Na, Mg, Al, Si, P, S, Cl, K, Ca, Ti, Cr, Mn, Fe, Co, Ni, Cu, Ga, Ge, Mo, Ru, Pd, Hf, W, Re, Os, Ir, Pt, Au, Zn, Zr, Cd, In, Sn, Sb, Pb, La, Ce, Nd, Eu, Gd, Ho, Er, Tm, Yb,  and Lu. It has two models for calculating stability of silicate melts considering ideal or non-ideal mixing of major and trace element oxides. The non-ideal model was based on \citet{Berman_1983} internally consistent CMAS (CaO-MgO-Al$_2$O$_3$-SiO$_2$) thermodynamic database with other oxides, except FeO, being treated as ideal solutes. The solution properties of FeO were assumed to be identical to those of MgO. We added another non-ideal melt model based on mixing parameters of melt components from  \citet{Fegley_Lodders_Jacobson_2023}. All three models yield similar ranges of melt stability within $\sim$100 K for the upper limit, with the ideal and ‘Fegley’ model typically having higher temperature values. As to the lower limit, the latter two models sometimes extend melt stability to unrealistically low temperatures. Therefore, the melt stability ranges reported in Table~\ref{tab:meltcalcs} are calculated using the CMAS melt model. It has to be noted that the highest temperature melt is enriched in very refractory elements such as Zr, Hf, rare earth elements and Al. Upon cooling, other elements condense into this refractory melt gradually changing its composition to a Ca, Mg, Al, Si-rich (CMAS) chondrule-like melt. Here we have chosen condensation temperature of 1 atomic\% Al and concentration of 10 wt\% SiO$_2$ in melt as the upper stability limits of the Al-rich and CMAS melts, respectively.

The results of our calculations are presented in Table~\ref{tab:meltcalcs} and Figure~\ref{fig:dig}. The calculated melt stability ranges (Figure~\ref{fig:dig}) are very similar to those of \citet{Ebel_Grossman_2000} who used a more advanced thermodynamic model for silicate melts. Progressively increasing the ratio of oxidizing (dust and ice) to reducing (tar and gas) components in the shocked portion of the nebula leads to more oxidizing conditions (higher $f$O$_2$) compared to the classic, unfractionated solar nebula. As a result, subsequent cooling of the modeled `chondrule' melts will crystallize FeO-bearing olivines (Table \ref{tab:meltcalcs}) that are compositionally similar to those in Type I (FeO $\le 10$ wt\%) and Type II (FeO $>10$ wt\%) chondrules. Our calculations demonstrate that it is possible to produce molten silicates under a plausible range of pressures, temperatures, and solids-to-gas mass ratios in impact-generated nebular shock fronts.

\begin{table*}
\centering
    \begin{tabular}{cccccccccccc}
    \hline

$P$ & D & T & I & G & D & T & I & G & $T_{\rm liq}$ & FeO in Ol & $m_{\rm s}/m_{\rm g}$ \\
Pa & at & at & at & at & wt\% & wt\% & wt\% & wt\% & K & wt\% & - \\
\hline
0.4 & 1000000 & 1 & 50000 & 1 & 9.46E+01 & 4.97E-05 & 5.40E+00 & 1.74E-02 & 2054-1600 &  & 5741 \\ 
4 & 1000000 & 1 & 50000 & 1 & 9.46E+01 & 4.97E-05 & 5.40E+00 & 1.74E-02 & 2210-1700 &  & 5741 \\ 
4 & 1000000 & 1000000 & 1000000 & 1 & 3.75E+01 & 1.97E+01 & 4.28E+01 & 6.90E-03 & 2050-1700 &  & 14481  \\ 
4 & 1000 & 100 & 100 & 1 & 7.40E+01 & 3.89E+00 & 8.45E+00 & 1.36E+01 & 2010-1599 &  & 6  \\ 
4 & 10000 & 500 & 500 & 1 & 9.08E+01 & 2.38E+00 & 5.18E+00 & 1.67E+00 & 2180-1323 &  & 59  \\ 
4 & 10000 & 500 & 500 & 1 & 9.08E+01 & 2.38E+00 & 5.18E+00 & 1.67E+00 & 2180-1612 &  & 59  \\ 
4 & 10000 & 10000 & 10000 & 1 & 3.72E+01 & 1.96E+01 & 4.25E+01 & 6.86E-01 & 1860-1430 & $\sim$0.2-24.8 & 145  \\ 
4 & 10000 & 10000 & 10000 & 1 & 3.72E+01 & 1.96E+01 & 4.25E+01 & 6.86E-01 & 2040-1700 & $\sim$3.4-4.9 & 145  \\ 
4 & 10000 & 10000 & 10000 & 1 & 3.72E+01 & 1.96E+01 & 4.25E+01 & 6.86E-01 & $\sim$2040-1601 & $\sim$19.9 & 145  \\ 
4 & 10000 & 1 & 1 & 1 & 9.82E+01 & 5.16E-03 & 1.12E-02 & 1.81E+00 & 2200-1293 &  & 54  \\ 
4 & 10000 & 1 & 10000 & 1 & 4.63E+01 & 2.43E-03 & 5.28E+01 & 8.53E-01 & 2130-1300 &  & 116  \\ 
4 & 10000 & 1 & 10000 & 1 & 4.63E+01 & 2.43E-03 & 5.28E+01 & 8.53E-01 & 2130-1635 &  & 116  \\ 
4 & 10000 & 10000 & 10000 & 1 & 3.72E+01 & 1.96E+01 & 4.25E+01 & 6.86E-01 & 2040-1300 &  & 145  \\ 
4 & 100000 & 100000 & 100000 & 1 & 3.75E+01 & 1.97E+01 & 4.28E+01 & 6.90E-02 & 2220-1600 &  & 1448  \\ 
4 & 1000 & 1000 & 1000 & 1 & 3.51E+01 & 1.84E+01 & 4.00E+01 & 6.46E+00 & 1980-1660 &  & 14  \\ 
4 & 1000000 & 1 & 1 & 1 & 1.00E+02 & 5.25E-05 & 1.14E-04 & 1.84E-02 & 2230-1690 & $\sim$23.7 & 5431  \\ 
4 & 1000000 & 1 & 1 & 1 & 1.00E+02 & 5.25E-05 & 1.14E-04 & 1.84E-02 & 2230-1800 & $\sim$6.4-23.7 & 5431  \\ 
4 & 1000000 & 1 & 50000 & 1 & 9.46E+01 & 4.97E-05 & 5.40E+00 & 1.74E-02 & 2215-1256 &  & 5741  \\ 
4 & 1000000 & 1 & 50000 & 1 & 9.46E+01 & 4.97E-05 & 5.40E+00 & 1.74E-02 & 2215-1609 &  & 5741 \\ 
4 & 1000 & 100 & 1 & 1 & 8.08E+01 & 4.24E+00 & 9.22E-02 & 1.49E+01 & 2000-1601 &  & 6  \\ 
4 & 1000 & 1 & 100 & 1 & 7.70E+01 & 4.04E-02 & 8.79E+00 & 1.42E+01 & 2030-1687 & $\sim$3.8 & 6  \\ 
4 & 1000 & 1 & 100 & 1 & 7.70E+01 & 4.04E-02 & 8.79E+00 & 1.42E+01 & 2030-1505 & $\sim$8.2 & 6  \\ 
4 & 100 & 1 & 100 & 1 & 2.51E+01 & 1.32E-01 & 2.86E+01 & 4.62E+01 & 1850-1446 & $\sim$6.2 & 1  \\ 
4 & 100 & 1 & 25 & 1 & 3.19E+01 & 1.68E-01 & 9.11E+00 & 5.88E+01 & $\sim$1820-1386 & $\sim$1.9 & 1  \\ 
4 & 1000 & 1 & 50 & 1 & 8.05E+01 & 4.23E-02 & 4.59E+00 & 1.48E+01 & $\sim$2030-1480 & $\sim$5.7 & 6  \\ 
4 & 1000 & 1 & 1 & 1 & 8.43E+01 & 4.43E-02 & 9.62E-02 & 1.55E+01 & $\sim$2020-1461 & $\sim$3.1 & 5  \\ 
40 & 10000 & 1 & 1 & 1 & 9.82E+01 & 5.16E-03 & 1.12E-02 & 1.81E+00 & 2160-1721 &  & 54  \\ 
40 & 10000 & 10000 & 10000 & 1 & 3.72E+01 & 1.96E+01 & 4.25E+01 & 6.86E-01 & 2190-1300 &  & 145  \\ 
40 & 1000 & 1000 & 1000 & 1 & 3.51E+01 & 1.84E+01 & 4.00E+01 & 6.46E+00 & 2125-1364 &  & 14  \\ 
\hline
    \end{tabular}
\caption{{\bf Summary of melt stability calculations.} At each pressure ($P$), we calculated the range of temperatures where CMAS melts are stable ($T_{\rm liq}$) and the wt\% of FeO in olivine (Ol) for varying atomic ratios of dust (D), tar (T), ice (I) and nebular gas (G) (compositions are defined in \citet{Wood_Hashimoto_1993}). The atomic proportions are converted to weight percent and the total ratio of solids (D+T+I) to gas by mass ($m_{\rm s}/m_{\rm g}$). We find that molten chondrules are thermodynamically stable under plausible formation conditions by impact-generated nebular shocks.}
\label{tab:meltcalcs}
\end{table*}

\section{Size sorting and coupling to nebular shock region}
\label{sec:sizesorting}
Up until this point, we considered the dust-gas mixture to be perfectly coupled. However, the protoplanetary nebula is likely to have a size distribution of individual dust grains, previous generations of chondrules, and different sized grain clusters. Previous studies have examined how dust particles respond to a nebular shock wave, but in these works, the nebular shocks are driven by different nebular processes, such as spiral density waves.

\begin{figure}
\includegraphics[width=.49\textwidth]{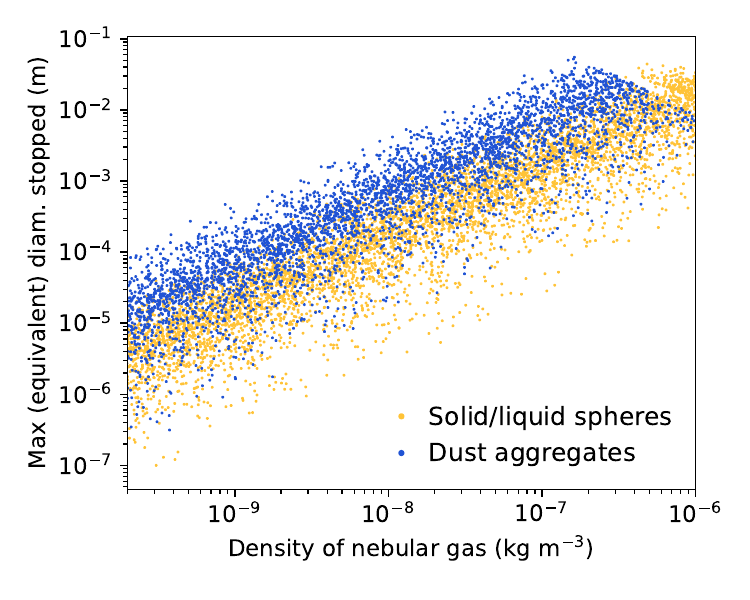} 
\caption{{\bf Particle-gas coupling in the expanding plume and nebular shock.} The maximum size of particles coupled to the outward flow in a Monte-Carlo suite of simulations that explored a wide range of plume and nebular properties. The upper right edges of the point cloud reflect the end ranges of parameters considered in the Monte-Carlo model. The size of coupled particles (droplets or aggregates) depends primarily on the density of pre-shock gas density for a dust-free H$_2$-He mixture. The size of aggregates is given as the equivalent size of the aggregate after it has been melted to form a chondrule. 
 \label{fig:coupling}}
\end{figure}

Here, we consider the response of particles subject to two possible disturbances: (i) the nebular shock followed by (ii) the nebula-plume interface, where there is a change in the composition and density of the gas. Again, we consider the frame of reference to be the background nebular gas, assuming that the particle or cluster is initially coupled to the gas. At the nebula-plume interface the two materials are at the same velocity, but the particle velocity increases rapidly in the plume behind the interface (Figure~\ref{fig:1devolution} and ~\ref{fig:imtemps}).  From the perspective of an individual grain or cluster, the surrounding gas instantaneously jumps in relative velocity and so we additionally consider the interface as a velocity discontinuity. 

We calculated the drag force on individual dust grains and aggregates to determine which size particles are returned to being coupled to the gas over the time and length scales of the impact-driven nebular shocks. Using a Monte-Carlo approach, we varied the densities of the nebular gas, plume gas, nebular shock velocity, nebular shock length scale, and plume length scale over a large range. Under the expected range of conditions in the solar nebula, the drag forces span multiple dynamical regimes, which are described in Appendix~\ref{sec:coupling} and \citet{Lock_coupling_IVANS_prep}. In our coupling calculations, we treated particles as isolated particles in an ideal gas and did not consider a dusty gas, where particles may collide with each other, leaving this problem for future work.

We found that the dominant parameter that determines the maximum size particle that is coupled to the nebular shock or plume front is the density of the nebular gas. The maximum size of particles coupled to the gas as a function of initial nebular gas density is presented in Figure~\ref{fig:coupling}, which displays the results of the Monte-Carlo simulations for both dust grains or (partially) molten droplets (orange) and dust aggregates (blue). The size of aggregates is given as the equivalent size if the aggregate melted to form a chondrule.

To illustrate the interaction of different size particles with the nebular shock and plume front, consider a particle located at point A in Figure~\ref{fig:1devolution}C. The exterior nebular shock arrives about 1.5 hr after the impact, and the particle experiences the shock environment with the denser, moving gas. If the particle is small enough to remain perfectly coupled to the shocked nebular gas, the particle will follow a gas streamline, e.g. moving to point B at 4 hours after the impact. However, if the particle takes some time to become coupled to nebular shock flow, its motion will lag behind a streamline, e.g., at point C, before coupling to the flow, e.g., at point D. Free-floating debris that are too large to be coupled to the nebular shock would not be perturbed by the flow, e.g., remaining at the same distance from the impact event at later times (point E). In this manner, the expanding nebular shock collects a relatively narrow size range of particles near the nebula-plume interface that depends primarily on the initial density of the nebular gas. Solar nebula midplane gas densities range from $10^{-6}$ to $10^{-8}$~kg~m$^{-3}$ from about 1 to 20 au \citep[corresponding to surface densities from][]{Dodson-Robinson_2009}. For likely values of the nebular density in the terrestrial and gas giant planet forming regions, the expected size of the stopped population is on the order 0.1-1~mm, consistent with the typical size range of chondrules. This size-sorted collection of material is physically separated from the impact location of the event by the accelerating plume.

Under the conditions of impact-driven nebular shocks, the maximum size of coupled particles agrees well with the maximum size range of chondrules for the plausible range of nebular gas densities (Figure~\ref{fig:coupling}). Chondrules typically span 0.1 to a few mm in size \citep[e.g.,][]{Metzler_2018}. Different chondrite groups have different maxima and mean size chondrules \citep{Jones2012}, which is interpreted to mean that they formed in distinct nebular environments. Within the framework of our model, the variation in the nebula pressures/densities at the time/place the chondrule forming impact occurred is responsible for the difference in chondrule sizes.

\subsection{A reverse sieve in the nebula}

The principal direction of the vapor plumes acts like a reverse sieve in the solar nebula: large particles pass through while small particles are retained. Consider the plume front in Figure~\ref{fig:3devolution}; the decompressing plume is schematically a cone expanding away from the impact point, shown in the third column. The leading edge is the exterior plume shock (black cone) followed by the nebula-plume interface (blue cone). The size of the cone and the area of the leading face increases by orders of magnitude with time. The cone sweeps through the nebula and collects particles that are small enough to couple to the front. These coupled particles are transported distances that are 3-4 orders of magnitude larger than the scale of the plume source (Figure~\ref{fig:pitime}). The collapse of the plume mixes materials coupled to the inflowing gas. In \href{https://chondrules.net/ivans/}{Supplemental Movies S2 and S8}, Lagrangian tracer particles placed in the nebular gas approximate the flow of perfectly coupled particles. These animation illustrate the sweeping effect of the vapor plumes. In Figure~\ref{fig:3devolution}D, the particles collected by the reverse sieve are predominantly found in the black dashed region. In contrast, the large solid fragments from the colliding planetesimals are closer to the impact location (grey dotted region).

\begin{figure*}
\centering
\includegraphics[width=1\textwidth]{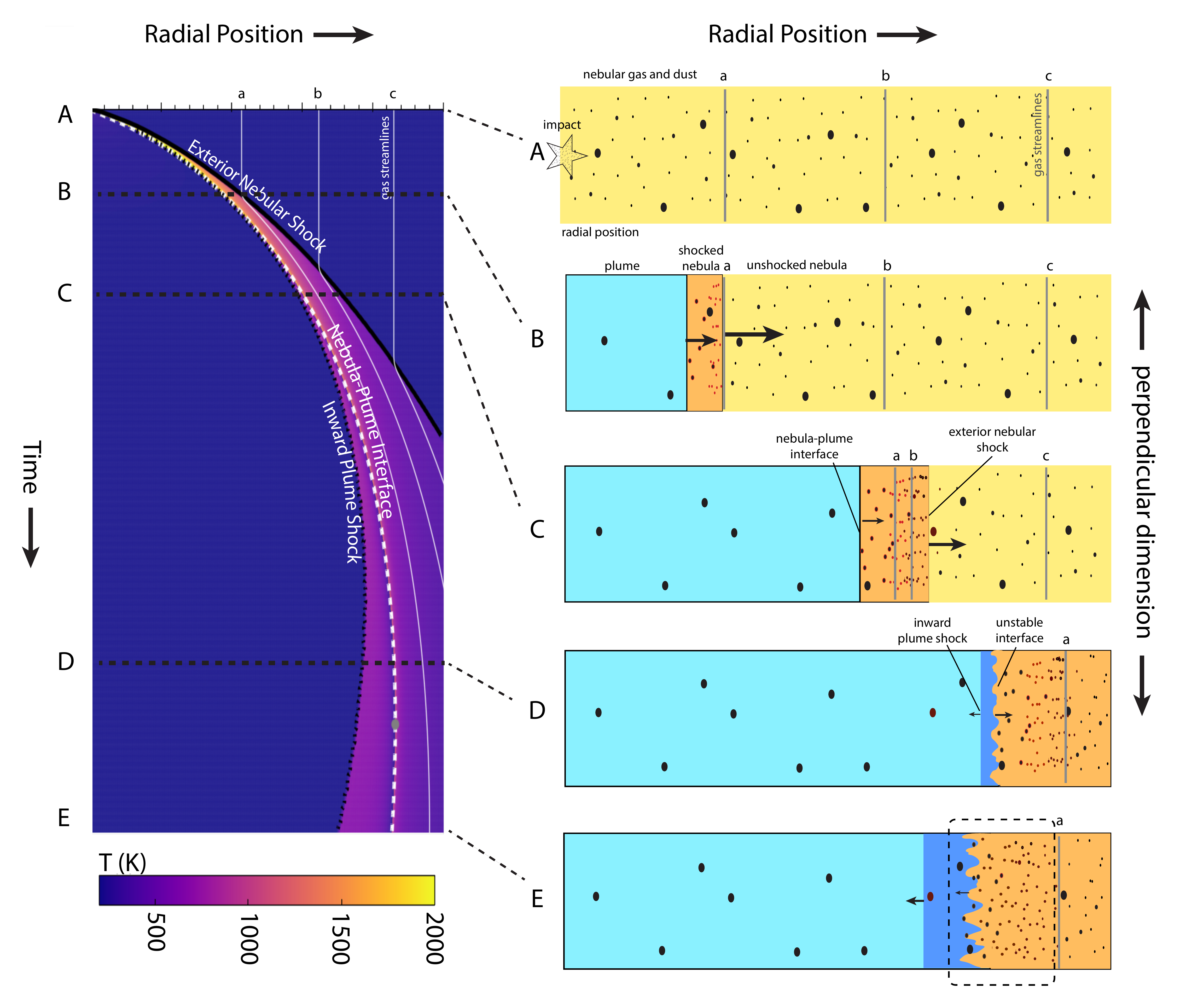} 
\caption{{\bf Schematic of particles coupled in nebular shock front during expansion and collapse of a spherical vapor plume.} The nebular shock front expands in size and decays in strength as it propagates into the nebula. Panels A-E are not-to-scale representations of particle thermal processing and size sorting at different times during plume evolution shown on the left (example shown in  Figure~\ref{fig:1devolution}. Gas flow streamlines are shown with thin grey lines labeled a, b, c, which illustrate the compression from the plume expansion and nebular shock. Panel A-E background colors represent the water vapor transient environment (blues) and the shocked nebula transient environment (orange). The plume region heated by the inward propagating shock is dark blue (D, E). The shock front acts like a reverse filter, collecting smaller particles and flowing past larger particles. The nebula-plume interface becomes unstable to turbulent mixing as the density in the plume decreases (D). Size-sorted particles with different thermal histories are collected and mixed near the nebula-plume interface during plume collapse (dashed line region in E). Note that the 1D code does not capture the turbulence at the nebula-plume interface nor the plume collapse from the lateral directions, as shown in Figures~\ref{fig:cartoon} and \ref{fig:3devolution}.
 \label{fig:couplingcartoon}}
\end{figure*}

The schematic in Figure~\ref{fig:couplingcartoon} illustrates the reverse sieve effect of the expanding nebular shock in the principal direction of the plume. The shocked region collects the smallest size fraction that couples to the moving gas. The shock front passes by particles too large to become coupled to the gas. The particles collected at early times experience higher temperatures and pressures, in which the free-floating dust can melt (Figure~\ref{fig:couplingcartoon}B, orange region), compared to later times when the nebula shock is weaker and colder (Figure~\ref{fig:couplingcartoon}D). As the shock front propagates, the environment surrounding the molten particles evolves. The dust-to-gas ratio increases, which can buffer the volatile elements in the melt within the transient environment and inhibit substantial isotopic fractionation. The nebular shock and vapor plume are initially separate redox environments (orange and blue regions). The redox of the nebular shock region evolves as the dust-to-gas mass ratio increases. As the nebula-plume interface becomes unstable and turbulently mixes (Figure~\ref{fig:couplingcartoon}D, E), the redox (and chemical/thermal) environments for particles change again. 

The size-sorted materials are gathered from the swept-up volume of the cone and concentrated in the principal direction.
The size-sorted particles in the dashed region in Figure~\ref{fig:couplingcartoon}E have experienced a wide range of shock temperatures in the decaying nebular shock. In 3D, these particles in the principal plume direction are mixed with material collapsing from the lateral directions which experience lower shock pressures and temperatures (Figure~\ref{fig:3devolution}).
Based on the geometry of the system, this process can enhance the mass of the small size fraction by several times on top of the background dust-to-gas ratio. Thus, the dust-to-gas ratio in the chondrule-forming environment changes with time during the evolution of the event. 

We have demonstrated that vaporizing planetesimal collisions can generate nebular shock waves that process and collect particles in the chondrule size range and smaller. In this manner, a vaporizing collision can generate a chondritic mixture of variously processed and size-sorted material.

\section{Self-gravitating regions}

The meteoritic record contains evidence for rapid warm accretion of some chondrules \citep{Metzler_2012,Ruzicka_Hugo_Friedrich_Ream_2024} and some chondrite groups have a chemical complementarity between the matrix and chondrules \citep[e.g.,][]{Hezel2018}. These data indicate that some chondrites formed quickly after the chondrule formation event and that both the matrix and chondrule components experienced the same local transient environment. These features are not universal, indicating a diversity in the formation processes for chondrites.

Here, we examine whether or not a collapsing vapor plume could become gravitationally unstable and quickly form a new planetesimal from the processed nebular particles. The self-gravitational collapse criteria is given by Equation~15 from \citet{Choksi_2021}:
\begin{equation}
R_{\rm Jeans}\sim \frac{c_{s, \rm mix} \left( 1+ \rho_{\rm particles}/\rho_{\rm g} \right)^{-1/2}}{\sqrt{G (\rho_{\rm g}+\rho_{\rm particles}) }},
\end{equation}
where $c_{s,\rm mix}$ is the sound speed of the dust-gas mixture, $\rho_\mathrm{g}$ is the density of the gas and $\rho_{\rm particles}$ denotes the bulk density of solids in the cloud. Here, we consider the number density of particles a free parameter, given in terms of the total solids-to-gas mass ratio, $\rho_{\rm particles}/\rho_{\rm gas}=m_{\rm s}/m_\mathrm{g}$. We use a reference gas density of $\rho_{\rm g}=10^{-7}$ kg\,m$^{-3}$, dusty gas sound speed of $c_{s, \rm mix}\sim500$ m\,s$^{-1}$, and $G$ is the gravitational constant. To convert the cloud mass to a new planetesimal radius, we use a dust material density of $\rho_{\rm d}=3000$ kg\,m$^{-3}$.

The particles dominate the bulk mass, and the free fall time of the cloud is of order
\begin{equation}
t_{\rm ff} \sim \frac{1}{\sqrt{G \rho_{\rm particles}}} \; .
\end{equation}
If the cloud collapses into a single planetesimal, the radius of the new aggregate would be 
\begin{equation}
R_{\rm new} = \left( \frac{R_{\rm Jeans}^3 \rho_{\rm     
particles}} {\rho_{\rm d}} \right)^{1/3}.
\end{equation}

\begin{figure}
\plotone{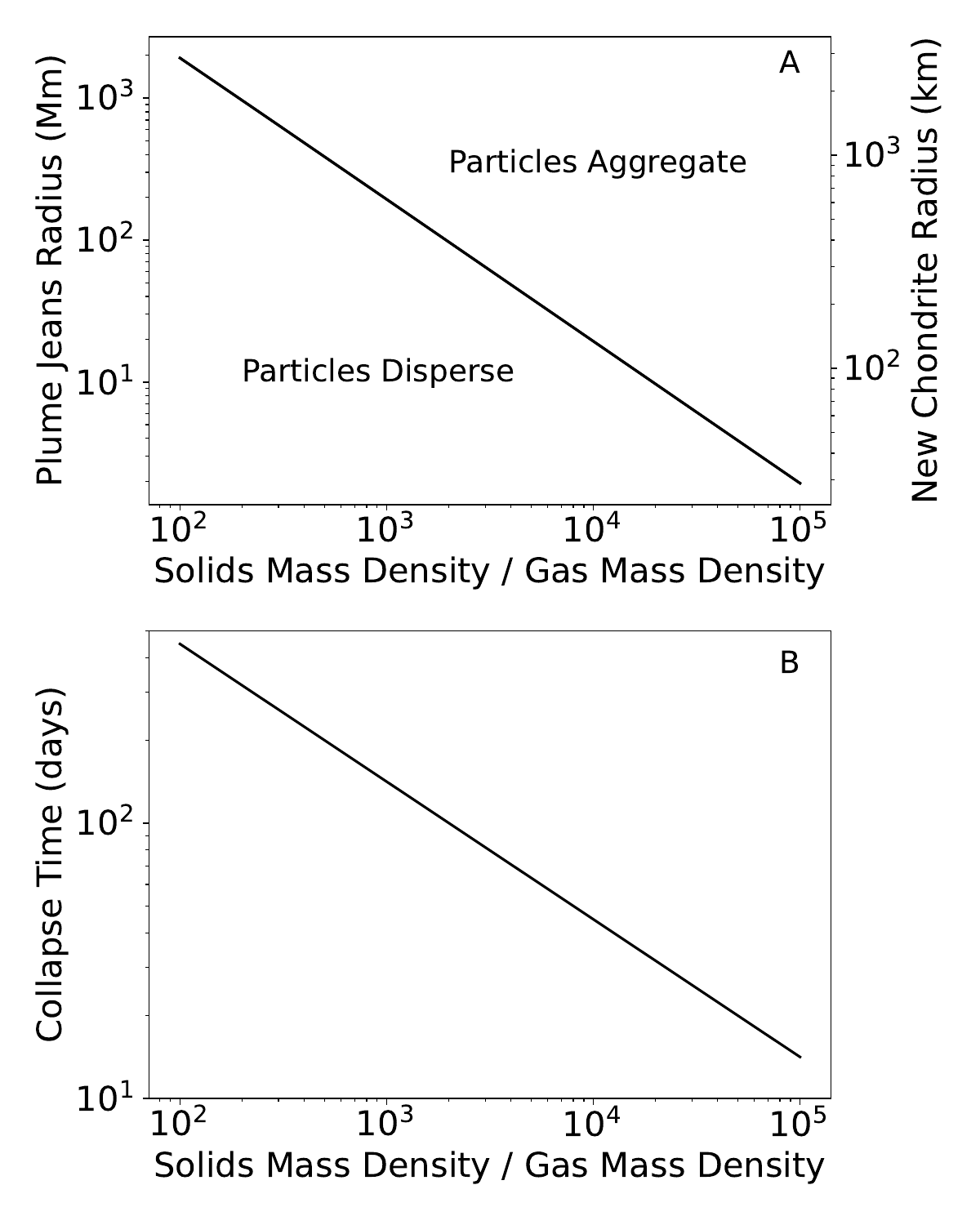}
\caption{{\bf Self-gravitating cloud collapse criteria.} A. With the high solids-to-gas mass ratio implied by chondrule compositions (100s to 10000s), the plumes generated by vaporizing collisions must be 100s to 10s Mm  in size to quickly form a new planetesimal. B. In such cases, the collapse time would be order 10s to 100s days. 
 \label{fig:selfgrav}}
\end{figure}

The results of this calculation are shown in Figure~\ref{fig:selfgrav} for a range of solid to gas mass ratios. For example, if the number of chondrules per cubic meter is 100, with a mean radius of 0.5 mm and a solid density of 3000 kg\,m$^{-3}$, the equivalent $\rho_{\rm particles}=1.57\times10^{-4}$ kg\,m$^{-3}$, and $\rho_{\rm particles}/\rho_{\rm gas}=1570$ and the required cloud radius is $>123$ Mm. In general, for $m_{\rm s}/m_{\rm g}\sim$ 100s to 1000s, the required cloud size for gravitational instability is 100s of Mm. 

For 3D vapor plumes, the region of interest is the mixture of materials in the principal plume direction. Our hydrocode modeling indicates that this size range is possible but borderline for the scale of plumes under the range of nebular and impact conditions explored here. Alternatively, other physical mechanisms may enable more efficient planetesimal formation or growth after the collapse of the vapor plume: e.g., streaming instabilities \citep{Youdin_Goodman_2005,Johansen_2007}, cavitating bubbles \citep{Chiang_2024}, or sedimentation onto other planetesimals \citep[e.g.][]{Johnson2015,Matsumoto_Oshino_Hasegawa_Wakita_2017}.

We infer that it is possible for some plume products to become gravitationally unstable but that most events, which are at low impact velocities, would deliver their impact products to the nebula with small relative velocities. The free-floating chondrules and dust would then be assembled into new planetesimals by other physical processes.

\section{Discussion}

In this work, we examined the physical phenomena associated with nebular shock waves generated by supersonic impact vapor plumes. Vaporizing collisions between planetesimals occurred during planet formation over large regions of the solar nebula \citep{Carter_2020, Carter_Stewart_2022}. The outcomes of these collisions must be reflected in the meteoritic record and preserved in the surviving planetesimals that are stored in the asteroid belt. 

We propose that chondrules are one of the major products of impact vapor plume-driven nebular shock waves. The precursor materials were free-floating nebular dust and dust aggregates around the impact event. The principal driver of impact vapor plumes was shocked water ice and other major volatiles that were present in the first generation of planetesimals. This model is called the impact vapor and nebular shocks (IVANS) model for chondrule formation to distinguish it from other collision models and sources of nebular shocks.

Here we focus our discussion on the most common types of chondrites, ordinary and carbonaceous, and aspects of vaporizing collisions that are consistent with their major characteristics. The diversity of more rare types of meteorites are beyond the scope of this work and will be discussed briefly along with avenues for development of future physical and chemical models.

\subsection{Complexity in the precursor planetesimals}

We motivated our calculations by the existence of a first generation of planetesimals that were composed of ice, organic compounds, and refractory dust, such as CI chondrites and primitive asteroids, like Bennu and Ryugu. We expect water to be present in primitive planetesimals in both the inner and outer solar system as the ice line migrates inward during the evolution of the protoplanetary disk \citep[e.g.,][]{Dodson-Robinson_2009}. Ordinary chondrites, which are presumed to have formed in the asteroid belt region, contained some water at the time of their formation \citep{Alexander_2019,Sutton_2017}.

As shown in Figure~\ref{fig:3devolution}, only a small portion of the colliding bodies is vaporized. In a vapor plume composed of a mixture of silicates and water, most of the silicate fraction remains solid (Figure~\ref{fig:disrupt} and Appendix \ref{sec:imp-prob}). Our calculations are consistent with the vapor plume produced in the NASA Deep Impact mission, where a copper-rich projectile impacted a comet at 10.3 km~s$^{-1}$. The observed dust temperatures in the ejecta plume were only about 375~K \citep{Gicquel_2012}. 
The diverse products in the principal direction of the vapor plume are physically separated from the bulk of the solid fragments of the planetesimals. These fragments, which experience shock temperatures lower than the vapor plume, would remain dust aggregates as in the original bodies. Although some intermediate impactite materials may form (e.g., partial melts), these would be a small volume fraction of the original bodies.

Some fraction of planetesimal collisions likely involved bodies in various stages of differentiation. For example, iron meteorites are derived from the cores of disrupted differentiated bodies and some of these disruptive collisions occurred during the lifetime of the solar nebula. The rare CB and CH group meteorites, which contain a high mass fraction of metals, are thought to be products of an impact with a differentiated planetesimal within the lifetime of the nebula \citep[e.g.,][]{Krot_2022}. If the collision generated a vapor plume, then a nebular shock would develop similar to those as examined in this work. 

Modeling such complex mixtures of different materials in hydrocodes presents a number of challenges. Our 2D and 3D hydrocode simulations used two pure phases, water ice and fused silica, to represent the planetesimal materials. When ice is mixed with a higher impedance material like silicates, the shock pressures are greater for a given impact velocity compared to a collision between two pure ice bodies (see Appendix~\ref{sec:imp-prob}). Conversely, porosity in the planetesimals would reduce the peak shock pressure.  Our hydrocode simulations do not include any chemical reactions between the silicate and ice or the surrounding nebula, which leads to temperature artifacts in the simulations (e.g., the zones of hot silicate in the vapor plumes in Figures~\ref{fig:2devolution} and \ref{fig:3devolution}). However, the overall hydrodynamics of vaporizing collisions and the shock interactions between the plume and nebula are captured by the current numerical models.

\subsection{Key parameters for the IVANS regime}

The IVANS model requires the impact-production of a gas phase that expands supersonically into the surrounding nebular gas. Some condensed volatiles (e.g., hydrogen, carbon, and sulfur compounds) are present in all chondrites. We found that a small amount of vapor can generate a large disturbance in the nebula (section~\ref{sec:scaling}). These scaling laws relate the portion of the colliding bodies that reaches partial vaporization to the plume effects on the nebula. Future work will investigate the expected variations in the compositions and internal structures planetesimals to relate the impact conditions to the amount of vapor produced.

Because the formation of a nebular shock depends on several impact parameters, there is not a single impact velocity threshold to enter the regime of the IVANS model. First, the collision must generate a sufficiently strong shock pressure in the volatile component(s) to produce a vapor plume. This depends on the impact velocity, impact angle, and composition and internal structure of the colliding bodies (Appendix~\ref{sec:imp-prob}). From our 1D models (section~\ref{sec:scaling}), we find that vapor plumes from shocked non-porous water ice follow a single scaling law when shocked to pressures of 40~GPa or greater. This pressure can be achieved over a wide range of shock velocities depending on the composition (e.g., 3.1 to 9.9 km~s$^{-1}$ in Figure~\ref{fig:impact-pres-prob}). Lower shock pressures can generate vapor plumes but the corresponding nebular shocks would process a smaller volume of the nebula compared to the scaling laws presented in this work. 

Second, the magnitude of the nebular shock depends on the relative velocity of the vapor plume source to the surrounding gas. The impact velocity required to reach silicate-melting temperatures in the nebular shock depends on the impedance match solution between the expanding vapor plume and the nebular gas, as illustrated in Figure~\ref{fig:imtemps}. Every collision between planetesimals has a different relative velocity to the nebular gas rest frame. Our calculations explored the case where one of the planetesimals had no relative velocity to the background gas. When the principal direction of the vapor plume is aligned with the planetesimal relative velocity to the gas, the nebular shock may process a larger volume compared to our scaling laws and vice versa. Future work will investigate the effects of varying planetesimal compositions and impact configurations to extend these scaling laws to include more complexity. 

Since the magnitude and volume of the nebular shock depends on these various parameters, the volume of nebula that exceeds the melting temperature of silicates to form chondrules is also sensitive to these parameters. Because chondrites contain chondrules that formed at different times, the proportion of chondrules to matrix dust is not directly related to the parameters of a single impact event.

In the IVANS model, we propose that processing of nebular dust occurs via many collisions within a population of planetesimals. Thus, increasing the chondrule to dust ratio in the nebula would require a larger volume of nebula to be processed by shock waves. In this scenario, a larger chondrule fraction implies a larger number of collisions leading to a larger fraction of nebular dust processed through collisions. Studies of planetesimal collisions from migrating giant planets have found substantial variations in the fraction of planetesimals that experience mutual collisions \citep{Carter_2020,Carter_Stewart_2022}. For example, in the Grand Tack model, a much larger fraction of planetesimals in the asteroid belt experienced collisions compared to planetesimals near Jupiter or beyond Jupiter's orbit (see Figure 7 in \citet{Carter_2020}).

\subsection{Characteristics of chondrules and chondrites}

We have investigated the main features of nebular shocks generated by impact vapor plumes. The characteristics of these types of nebular shocks can be compared to the inferred environments for chondrule formation based on evidence gathered from the meteorite record. Here, we summarize the major connections between impact-generated nebular shocks and meteoritic data.

{\it Timing of chondrule formation.} The ages of individual chondrules appear to span the duration of the lifetime of the solar nebula with some clustered around 2 Myr \citep{Connelly_2012,Krot2005}. Based on Figure~\ref{fig:criticale}, the oligarchic growth of the cores of the gas giant planets would have initiated vaporizing conditions and the formation of chondrules. Migration of the giant planets would have generated intense periods with an increased number of collisions \citep{Carter_2020} with the potential for many chondrule formation events happening over a shorter period of time. Chondrule-forming collisions would become rare later in the gas disk phase when the population of planetesimals has been depleted and excited by giant planets.  As chondrules form in nebular shocks, their production would cease entirely as the nebular gas dissipated.

{\it The CAI storage problem.} One of the challenges of chondrite formation is preserving CAIs in an accessible reservoir to be combined with chondrules that form millions of years later. Free-floating dust spirals into the Sun on kyr time scales. Hence, CAIs must be stored in a reservoir for millions of years, e.g., within primitive planetary bodies or in long-lived structures such as nebular pressure bumps created by the giant planets \citep[e.g.,][]{Desch_Kalyaan_Alexander_2018}. In our model for chondrule formation, the first generation of planetesimals, comprised of undifferentiated mixtures of dust and ice, may be a storage reservoir. For every vaporizing collision, there are many more disruptive collisions (Figure~\ref{fig:disrupt}) that would continuously disperse a portion of the dust in these planetesimals to become free-floating dust or dust clusters in the nebula. The released CAI grains could then be incorporated into chondritic mixtures with younger chondrules. Further work is required to determine if the physical and isotopic properties of CAIs can be preserved through this process.

{\it Chondrules have distinct heating and cooling rates.} Chondrules are heated on the time scale of minutes and cooled over hours. These timescales have been difficult to match with other nebular processes. We find that impact vapor-plume driven nebular shocks, with plausible energies and length scales from planetesimal impacts, overlap with the inferred heating and cooling time scales for chondrules (Figure~\ref{fig:tempcoolrate}).

{\it Chondrules have a characteristic size.} The shearing forces between a liquid droplet or dust aggregates and a surrounding gas can break down larger particles. Drag forces due to the relative velocity between a particle and a gas can also couple a particle to the gas flow. We calculated the size of particles that are efficiently coupled to the nebular shock and plume front (Figure~\ref{fig:coupling}). 
For the expected range of nebular gas densities at the time of chondrule formation \citep{Dodson-Robinson_2009}, our results are consistent with the observed size range of chondrules \citep{Scott2014}.

{\it High dust-to-gas environment.} Chondrules contain volatiles, which implies a high local partial pressure of sodium, water, and carbon monoxide during chondrule formation \cite[e.g.,][]{Alexander2008, Shimizu_Alexander_Hauri_Sarafian_Nittler_Wang_Jacobsen_Mendybaev_2021}. In addition, the oxygen fugacities of chondrules are much greater than the nebular gas. Cosmochemical studies for the local environment around molten chondrules favor a high dust-to-gas ratio and/or elevated pressures \citep[e.g.,][]{Ebel_Grossman_2000}. As argued in the introduction, vaporizing collisions would have occurred in an environment with many more lower-energy impacts that would temporarily raise the local dust-to-gas ratio. In addition, the shock pressures in a dusty nebular gas can be greater than in a dust-free gas. We find that the nebular shocks driven by impact vapor plumes have characteristics compatible with the required physicochemical environment for chondrules (Section~\ref{sec:nebconditions}). The transiently shocked nebular environment is not a perfectly closed nor a perfectly open system: the moving shock front incorporates additional dust, like a reverse sieve, as it expands and cools (Figure~\ref{fig:couplingcartoon}). Because the nebular shock front cools as it expands, the potential for isotopic fractionation of the early time melts is limited compared to an open melt system in contact with the background nebula. Future work on possible isotopic and chemical fractionations should consider the pressure-temperature-dust content paths in the IVANS model.

{\it Warm chondrules and matrix complementarity.} Some chondrites show evidence for rapid accretion of warm chondrules \citep{Ruzicka_Hugo_Friedrich_Ream_2024}. In addition, some but not all chondrites have complementary elemental abundances between the matrix and chondrules \citep{Palme_Hezel_Ebel_2015}. We suggest that the collapsed plume environment would be a transient local environment that could lead to complementarity and rapid aggregation of some chondrule clusters. However, our initial analyses suggest that the impact products would often be distributed into the nebula and would need to be aggregated into a chondrite parent body by other processes or rapidly accreted onto other pre-existing bodies. This diversity of outcomes is compatible with the diversity observed in the meteoritic record.

{\it Type I and II chondrules.} The plume front, shocked nebula and varying dustiness leads to a variety of redox environments. There is evidence that chondrules were reprocessed in different redox environments \citep[e.g.,][]{Libourel_Nagashima_Portail_Krot_2023}. In this model, some of the free-floating nebular solids would have been previously formed chondrules and these chondrules could be reprocessed and/or incorporated into new chondrites.

{\it Dusty rims, primitive matrix.} The collapsing plume mixes together materials that experience a wide variety of thermal processing, from essentially none to complete vaporization. These materials are brought together rapidly enough to create dusty rims around chondrules and matrix materials. This environment is consistent with the complexity of chondrule rims \citep[e.g.,][]{Ruzicka_Floss_Hutson_2012}.

{\it Varying chondrule volume fraction.} The frequency of vaporizing collisions would be controlled by the history of the giant planets \citep{Carter_2020}. The sweeping resonances through the asteroid belt region from inward planet migration may result in a larger fraction of shock-processed dust compared to the number of collisions in the outer solar system. More collisional processing in the asteroid belt region may explain the smaller matrix fraction and greater occurrence of broken chondrules in ordinary chondrites compared to carbonaceous chondrites \citep[see Figure~7 in][]{Carter_2020}. Finally, vaporizing collisions are less likely beyond about 25 au (Figure~\ref{fig:criticale}) and thus some planetesimals in the outermost solar system would remain free of chondrules. However, those planetesimals would need to be scattered into the asteroid belt to be sampled in the meteoritic record.

{\it Planets are different from chondrites.} Earth and Mars have distinct isotopic signatures from the main chondrite groups. Chondrites appear to have formed primarily from materials that were not processed through planetary differentiation and only a small fraction of chondrules carry any signatures of silicate differentiation \citep{Libourel_Krot_2007,Sheikh_2021,Sheikh_Humayun_2021}. An origin of chondrules and chondrites from small planetesimals that were dynamically limited in their growth would be consistent with these observations.

{\it Multiple heating events.} In this work, we have focused on a single vapor plume interacting with the nebula. However, during planet migration, many collisions in close proximity could generate overlapping nebular shock waves. In addition, our work has not explored the effects of turbulence and partial particle coupling in generating more complex thermal histories than shown in Figures~\ref{fig:tempcool} and \ref{fig:tempcoolrate}. Other physical processes, such as lightning, may also be present in the nebular shocks and vapor plumes. Lightning is observed in volcanic eruptions, and electrical discharges observed in shock tube experiments, due to charge build up on dust grains in the supersonic gas flows \citep[e.g.,][]{Cimarelli_Genareau_2022}. These unresolved processes must be considered when attempting detailed comparisons to observations in meteorites.

\subsection{A planetary context for chondrites}

Our development of the IVANS model for the formation of chondritic mixtures offers a new physical link between the meteoritic record and the overall process for planet formation. Collisions were an ubiquitous process during planet formation, and they must leave some mark in the meteorite record. As a result, collisions have been invoked as a possible origin for chondrules \citep[e.g.,][]{asphaug_chondrule_2011,sanders_origin_2012,Johnson2015,lichtenberg_impact_2018}
but the physical constraints from chondrules did not match the properties of direct impact ejecta \citep{Desch2012}. The body of data appears to favor nebular shocks \citep{Connolly_Jones_2016,Marrocchi_2024}. We propose the main mechanism for chondrule formation are the nebular shocks generated by impact vapor plumes. 

Based on the large body of meteoritic data, the chondrule forming environment must have had a high dust-to-gas ratio and many nebular shock waves \citep[e.g., as synthesized in][]{Ruzicka2012}. In previous work, nebular shock waves were generated by separate processes from collisions: e.g., spiral density waves in the disk or bow shocks around planetary embryos or planetesimals. Here, we demonstrate that mutual collisions between planetesimals can generate both elevated dust-to-gas ratios in the disk and nebular shocks in localized environments in the nebula. Astronomical observations of evolved protoplanetary disks, 1-3 Myr old, have found dust-to-gas ratios of order 1 \citep{Pinte_2015,Ansdell_2016}. Sweeping resonances could generate local regions with even greater dust masses. However, the chemical requirements for the amount of dust required during chondrule formation needs to be re-evaluated in the context of the IVANS model.

Our calculations demonstrate that impact-generated nebular shocks can process volumes of the nebula that are many orders of magnitude larger than the volume of the initial planetesimals (Figures~\ref{fig:pimass}, \ref{fig:pitime}, \ref{fig:tempcool}). The nebular shocks generated by impact vapor plumes have size scales between those generated by eccentric planetesimal bow shocks and from spiral density waves in the protoplanetary disk. Most planetesimal collisions occur in the midplane and most of the dust mass is in the midplane; thus, impact-generated nebular shocks can process substantial amounts of free-floating dust in the nebula. Note that the number of collisions in $N$-body simulations of planet formation is limited by the smallest size bodies that are resolved in the calculation (e.g., 100s km in Fig~\ref{fig:disrupt}). Even so, order $10^5$ collisions were resolved in recent works \citep[e.g.,][]{Carter_Stewart_2022}, and each resolved collision that exceeded the ice vaporization threshold could have processed order $10^{-4}$-au-sized regions of the nebula (Figure~\ref{fig:tempcool}). Many more collisions between smaller (unresolved) planetesimals would also have also processed the dust in the solar nebula. 

The growing major planets, especially the gas giant planets, are the main perturbers for small bodies \citep[e.g.,][]{Raymond2017a,Oshino_Hasegawa_Wakita_Matsumoto_2019,
Carter_2020}. Thus, the orbital dynamics of the gas giants were the ultimate source of the energies required for chondrule formation. Most of the planetesimals that survived planet formation originated from between the (proto)planets, which is consistent with the different isotopic signatures and bulk compositions between chondrite groups and the final planets \citep{Marrocchi_2024}.

In the inner solar system, most of the impact vapor plume materials from hypervelocity collisions onto the terrestrial protoplanets themselves, and the products of their associated nebular shocks, were likely accreted onto the growing rocky planets or onto the Sun. It is possible that small chondrite-like bodies formed during the evolution of materials being accreted onto the growing planets, but little of this material was implanted into the asteroid belt. Future work will need to consider new planetesimal formation and their processed compositions in addition to tracing the composition of ejecta from collisions \citep[see composition tracking in][]{Carter_Stewart_2022}. 

In the outermost solar system, the mutual collision velocities are smaller, eventually falling below the threshold for melting nebular dust (Figure~\ref{fig:criticale}). Thus, our model provides a natural explanation for why some primitive bodies have no chondrules even though collisions occurred throughout the solar system. Chondrule formation also ends when the nebula is dispersed or when planetesimal scattering reduces the number of collisions.

We propose an evolved view of the genetic relationship between planets and chondritic planetesimals: (i) The gas giant planets provided the gravitational perturbations that formed chondrites by processing planetesimals and debris while rocky protoplanets were growing; and (ii) the final terrestrial planets accreted a substantial mass fraction of processed material, some as debris and some as chondritic planetesimals. 

Chondrites are precious time capsules of primitive nebular materials. The different meteorite groups have experienced different degrees of thermal processing through nebular shocks, reflecting the different collisional histories in different locations and at different times in the solar nebula, as investigated by \citet{Carter_2020}. We expect these processes to occur universally during planet formation in our solar system and in exoplanetary systems.

\section{Conclusions and Future Directions}

Previous works have emphasized a distinction between chondrule formation mechanisms that are derived from nebular processes versus planetary (including impact) processes \cite[e.g.,][]{Connolly_Jones_2016}. Most studies of impact processes have neglected the presence of the solar nebula or only considered the early time evolution of the impact vapor plume in the nebular gas \citep{Hood2009}. In this work, we presented a wide-ranging study of effects of impact vapor plumes on the gas and dust in the solar nebula to late times. We have identified new physical processes related to planetary collisions, including the generation of strong nebular shocks and hydrodynamically collapsing impact vapor plumes. 

Based on this work, we introduce the impact vapor and nebular shocks (IVANS) model for chondrule and chondrite formation. Our model includes favorable aspects of both nebular and planetary models. The transient plume environments have pressure-temperature histories that are distinct from other physical processes occurring in the solar nebula \citep{Desch2012}.

Our work affirms the meteoritical view that chondrites contain a direct collection of materials from the solar nebula. We find that thermal processing of free-floating nebular dust by impact-generated nebular shocks is consistent with the thermophysical properties of chondrules. Size sorting during coupling of particles to the expanding nebula shock provides a reverse sieve effect that concentrates a size range of particles consistent with that observed for chondrules. The IVANS model also explains how proximal materials in the solar nebula may experience a wide range of thermal processing and yet be rapidly assembled into a chondritic parent body.

Our work presents a physical link between chondritic meteorites and a nebular environment dominated by mutual collisions between planetesimals that were excited by the growing planets. Future work must link the detailed chemistry of chondrules and chondrites to the transient environments generated in the IVANS model. Ultimately, the rich record contained within meteorites could be used to help constrain the unknown formation and migration histories of the gas giant planets.


\appendix

\section{Planetesimals excited by embryos}
\label{app:stirring}

The mean impact velocity in a population of dynamically excited planetesimals is given by 
\begin{equation}
    v^2_{\rm i} = v^2_{\rm esc}+v^2_{\rm rand},
    \label{eq:vimp}
\end{equation}
where $v_{\rm esc}$ is the mutual escape velocity and $v_{\rm rand}$ is the random velocity between a population of bodies. When the escape velocities of the bodies are negligible, as in the case of planetesimals, the impact velocities are dominated by the random velocity, 
\begin{equation}
v^2_{\rm rand} \simeq v_{\rm K}^2 \left ( \langle e \rangle^2 + \langle i \rangle^2 \right )^{0.5}, 
\label{eq:vrand}
\end{equation}
where $\langle e \rangle$ and $\langle i \rangle$ are the mean eccentricity and inclination of the population and $v_{\rm K}$ is the Keplerian orbital velocity. In Figure~\ref{fig:criticale}, we converted the impact velocities required for the onset of vaporization upon shock and release (commonly called the criteria for incipient vaporization, $v_{\rm iv}$) to a mean eccentricity of a population of planetesimals ($e_{\rm iv}$) at semi-major axis $a$. We neglected the escape velocity from the planetesimals for generality and because $v_{\rm esc} \ll v_{\rm iv}$ in most cases. Using Eq.~\ref{eq:vrand} and setting $v_{\rm iv} \simeq v_{\rm rand}$, the critical eccentricities are
\begin{equation}
\langle e_{\rm iv} \rangle = \left ( \frac{ \left( v_{\rm iv}/v_{\rm K} \right)^2 }{1+\beta^2} \right )^{1/2} = \left ( \frac{ v_{\rm iv}^2 a  }{(1+\beta^2) G M_{\odot}} \right )^{1/2},
\label{eq:eiv}
\end{equation}
where $\beta = \langle i \rangle /  \langle e \rangle$. $\beta\sim0.5$ for a population of dynamically stirred planetesimals \citep{Kokubo1996}, $G$ is the gravitational constant, and $M_{\odot}$ is the mass of the Sun. Because the circular Keplerian velocity, $v_{\rm K}$, decreases with increasing semi-major axis, the critical eccentricity for incipient vaporization increases with semi-major axis (colored lines in Figure~\ref{fig:criticale}). 

The magnitude of the equilibrium dynamical excitement of a population of planetesimals by a single planet in the oligarchic growth stage \citep{Tanaka2013} is estimated by 
\begin{equation}
\langle e_{\rm stir} \rangle =1.75 \left( \frac{4 r \rho_{\rm pl}}{3 \pi C_{\rm D} \rho_{\rm neb} a} \right)^{1/5} r_h,
\label{eq:estir}
\end{equation}
where $r$ and $\rho_{\rm pl}\sim2000$~kg~m$^{-3}$ are the radius and density of the planetesimals, respectively. $C_{\rm D}$ is the gas drag coefficient of order 1, and $\rho_{\rm neb}=10^{-7}$ (or $10^{-6}$)~kg~m$^{-3}$ is the nebular gas density. $r_h$ is the reduced Hill sphere of the perturbing protoplanet, 
\begin{equation}
    r_h = \left (\frac{M_{\rm pl}/M_{\odot}}{3} \right )^{1/3},
\end{equation}
where $M_{\rm pl}$ is the mass of the protoplanet.
The dynamical stirring from a single planet increases with its mass and proximity to the Sun (black and grey lines in Figure~\ref{fig:criticale}), and so the possibility of vaporizing collisions will vary both spatially and temporally during planet growth.

\section{Collision velocity components}
\label{app:nbody}
\begin{figure*}
\plotone{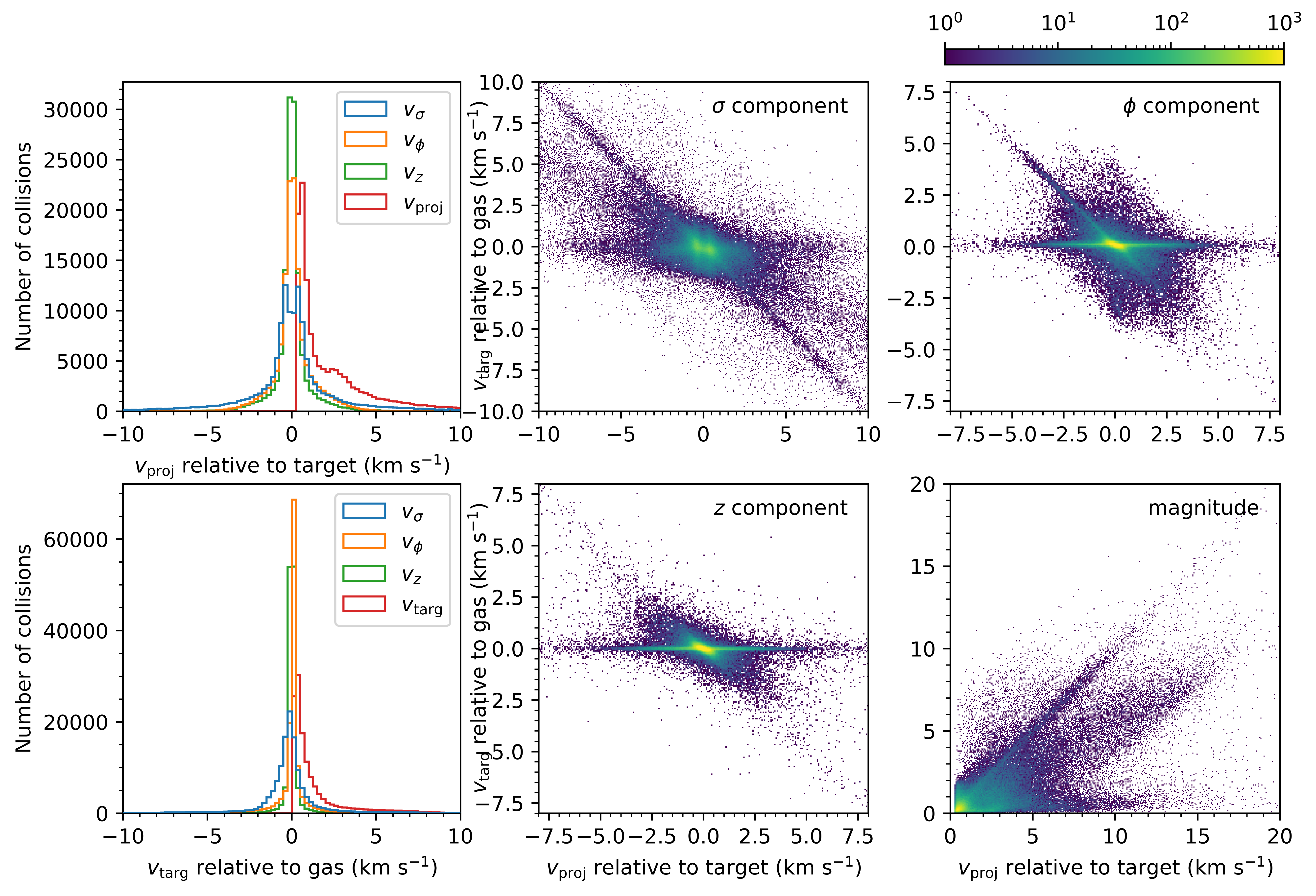}
\caption{{\bf Impact parameter distributions.}
In most collisions one of the impactors is on a low excitation orbit, while the other is excited. Velocity of the target with respect to the nebular gas, and the projectile with respect to the target for collisions in the simulation from Figure~\ref{fig:disrupt} \citep[run 022GTJf6hgas from][]{Carter_Stewart_2022}. The left two panels show histograms for the target and projectile velocities, while the right four panels show 2D histograms for the velocity of target and projectile, where the color scale indicates the number of impacts in each bin. The azimuthal, radial and vertical velocity components are shown, as well as the overall magnitude.  Collision velocity components mostly fall into two distinct bands: the horizontal band indicates that the target is moving on approximately a circular orbit similar to the nebular gas, while the projectile is on an excited orbit; the band with a slope of approximately $-1$ indicates collisions in which the target is excited while the projectile has almost the opposite motion, corresponding to an orbit similar to the gas. Outside of these bands both target and projectile are on excited orbits, with the distributions being determined by the dynamics of the giant planets. In the magnitude panel these bands become horizontal and $+1$ sloped.
\label{fig:veltogas}}

\end{figure*}

We analyzed the velocity components of $N$-body simulations of terrestrial planet accretion in the presence of migrating planets from \citet{Carter_2020}. We found that most of the collisions occurred with one of the bodies near the reference frame of the gas  (Figure~\ref{fig:veltogas}). A low relative velocity to the nebular gas is referred to as a dynamically cool orbit; large relative velocities are considered excited or dynamically hot. Planetesimals excited and captured into sweeping resonances from migrating planets had a high probability of colliding with a dynamically cool planetesimal. In this work, we use the nebular gas as the frame of reference to study nebular shocks generated by impact vapor plumes to simplify the study of the impact outcome. However, once all of the planetesimals are dynamically excited, both bodies have a substantial relative velocity to the gas. Note that the gas interactions between the vapor plume and nebular gas occurs in the reference frame of the gas, regardless of the pre-impact relative velocities, because the restoring force for the pressure low in the expanding vapor plume is sourced from the background nebular pressure field.

\section{Impact Pressure Probabilities}
\label{sec:imp-prob}
In this work, we present the properties of nebular shocks in relation to the shock pressure in the source region of the vapor plume. By focusing on a representative size and shock pressure of the vapor plume source, we simplified the details of the collision parameters and material properties to understand the general phenomena of supersonic vapor plume expansion into the nebular gas. 

\begin{figure}[ht]
    \centering
    \includegraphics[width=0.6\linewidth]{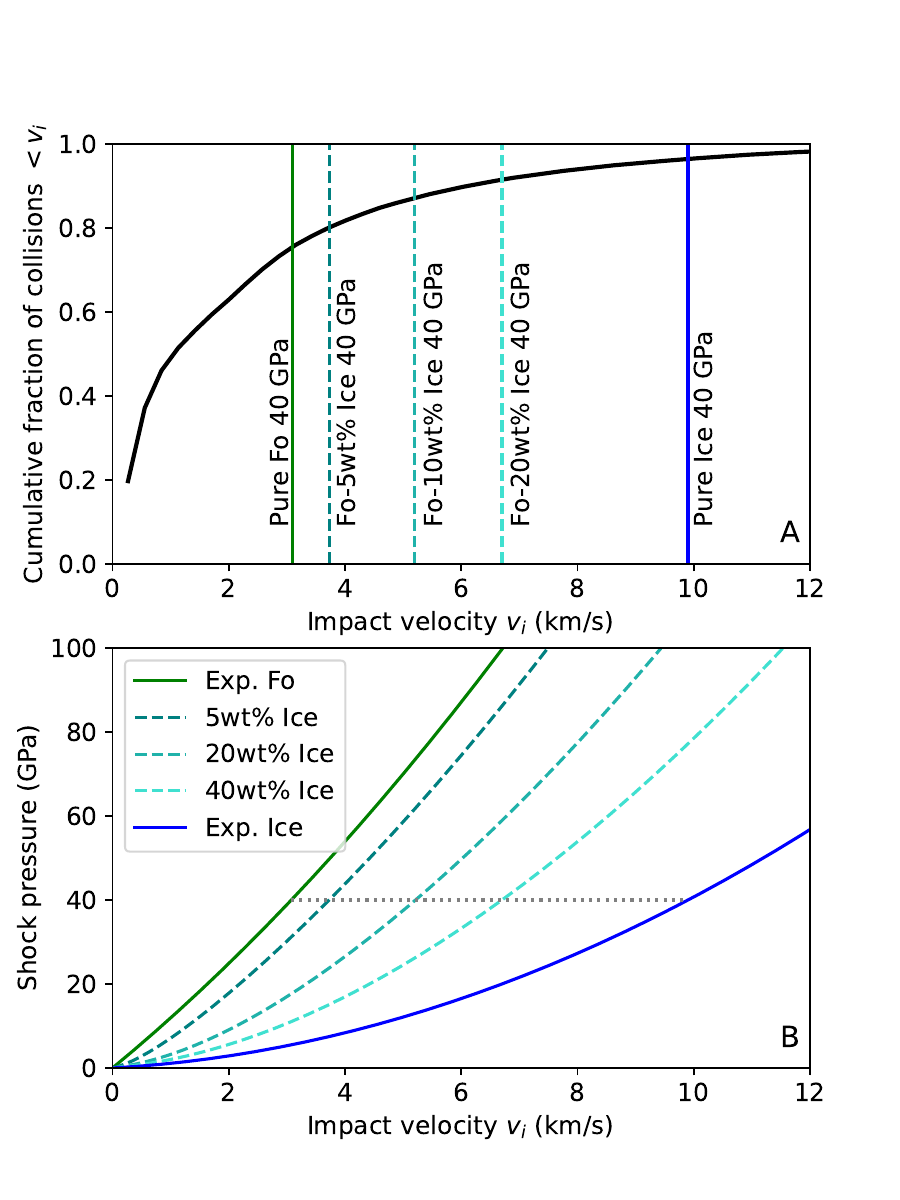}
    \caption{{\bf Impact velocity probabilities and compositional dependence on shock pressure.} A. Example cumulative probability distribution for mutual planetesimal collision velocities (corrected for particle size inflation) from a Grand Tack N-body simulation (022GTJf6hgas) in \citet{Carter_2020}. This is the same distribution as shown in Fig.~\ref{fig:disrupt}. B. Shock Hugoniots for pure forsterite (Fo, green, experimental) and pure H$_2$O ice (blue, experimental) and Fo-ice mixtures (dashed, calculated) in terms of the impact velocity between the same composition bodies. The impact velocity required to reach 40 GPa (grey dotted line) depends on the composition. The probabilities corresponding to 40~GPa events are indicated by vertical lines in panel A. In about half the impacts, (i) the planetesimal is disrupted, (ii) the silicate fraction is not melted  \citep[$P<189$~GPa,][]{Davies2020}, and (iii) the ice fraction is partially vaporized \citep[$P>3.5$~GPa,][]{Stewart2008, Kraus2011}.}
    \label{fig:impact-pres-prob}
\end{figure}

Here, we examine the probability of attaining a given shock pressure in the context of the distribution of mutual impact velocities between planetesimals. Figure~\ref{fig:impact-pres-prob}A presents the cumulative probability distribution of impact velocities that were derived from $N$-body simulations of planet formation (using the same dataset presented in Fig.~\ref{fig:disrupt}). For this Grand Tack example, 50\%, 25\% and 5\% of the collisions were above 1.05, 3.05, and 8.7 km~s$^{-1}$, respectively.

The impact velocity required to attain a given shock pressure in water ice depends on the composition of the planetesimals. We illustrate this effect by considering the range of impact velocities required to reach 40~GPa shock pressure in varying rock-ice mixtures. Figure~\ref{fig:impact-pres-prob}B presents the shock pressures attained in varying mixtures in terms of the impact velocity when both bodies have the same composition. The effect of composition on the relationship between impact velocity and shock pressure is indicated by the dotted grey line at 40~GPa, which spans 3.1 to 9.9 km~s$^{-1}$. The reason for this substantial range of values is that the Hugoniots for varying mixtures have notably different compressibilities. 

The end member composition Hugoniots are based on experimental data for pure forsterite (green line) and pure water ice (blue line). The Hugoniots for ice-rock mixtures were calculated using the method of averaging kinetic energy, which gives good agreement with experimental data on mixtures \citep{Petel_Jetté_2010}. The method assumes that both components reach the same shock pressure, and the mixture particle velocity is given by a mass-weighted mean of the components:
\begin{equation}
    u^2_{\rm mix} = X_{\rm ice} u^2_{\rm ice} + (1-X_{\rm ice})u^2_{\rm fo},
\end{equation}
where $u_{\rm ice}$ and $u_{\rm fo}$ are the particle velocities on the shock Hugoniots for the pure phases and $X_{\rm ice}$ is the mass fraction of ice. To determine the corresponding shock velocity, we use the linear Hugoniot parameters for the liquid region on the ice Hugoniot ($u_p>1.59$ km~s$^{-1}$) defined by \citet{Stewart_Ahrens_2005}: 
\begin{equation}
    U_s = 1.440(\pm0.035) u_p + 1.7(\pm0.130),
\end{equation}
where $U_s$ is the shock velocity and $u_p$ is the particle velocity in km~s$^{-1}$. For forsterite, we fit a linear Hugoniot to the low pressure data ($0.8<u_p<3.25$ km~s$^{-1}$, $P<112.7$~GPa) compiled in Figure 1 of \citet{Mosenfelder_Asimow_Ahrens_2007}. We found 
\begin{equation}
    U_s = 0.645 u_p + 7.082,    
\end{equation}
in km~s$^{-1}$ with a covariance matrix, $\Sigma$,
\begin{center}    
$
\Sigma = \begin{bmatrix}
0.00182436 & -0.00383161\\
-0.00383161 & 0.00911675\\
\end{bmatrix}.
$
\end{center}
At each shock pressure, $P$, a pure phase particle velocity is given by 
\begin{equation}
    u_p = \frac{c}{2s} \left(\sqrt{1 + \frac{4s P}{\rho_0 c^2}} -1 \right ),
\end{equation}
where the pure phase linear Hugoniot has the form, $U_s=s u_p+c$, and $\rho_0$ is the initial volume. Here, we used $\rho_0=3220$ kg~m$^{-3}$ for forsterite and $\rho_0=932$ kg~m$^{-3}$ for cold H$_2$O ice. For our illustrative example of impacts between the same composition bodies, the impact velocity to attain each pressure is $v_i=2u_{\rm mix}$.

Since planetesimals were mixtures that incorporated varying amounts of H$_2$O ice (0-20s wt\%), it is appropriate to consider the impact velocities required to attain a certain peak pressure in an icy mixture. Figure~\ref{fig:impact-pres-prob}B illustrates that ice in a mixture can reach a certain shock pressure at much lower impact velocities compared to pure ice. Experimental measurements of shock temperatures in a SiO$_2$-ice mixture found that the assumption of pressure equilibrium is reasonable for predicting phase changes in the ice fraction \citep{Kraus_Stewart_Seifter_Obst_2010}.

Finally, we note that the impact velocity estimates are made on the assumption of perfect half space collisions and do not include the complexity of real planetesimal shapes nor the effects of porosity. Note that the ANEOS model for water was developed for higher shock pressures and is slightly different in the region presented in Figure~\ref{fig:impact-pres-prob}B. Detailed analysis of collisions between icy mixtures will be addressed in future work.

\section{Hydrocode Methods}
\label{sec:hydro-methods}

The vapor plume hydrocode simulations were purely hydrodynamic (no strength or gravity) because the impact velocities of a system of dynamically excited planetesimals exceed the gravitational escape velocity and the critical velocities for disruption of 100-km scale planetesimals (Figure~\ref{fig:criticale}). We used the no-tension tabulated equations of state (EOS) for water and fused silica using the ANEOS code package (v1.0) (Table~\ref{tab:aneos_params}) in SESAME format \citep{Lyon1992}. See the \href{https://chondrules.net/ivans/}{supplemental materials} for the EOS tables. The nebular gas was modeled with the hydrogen 5250 SESAME table \citep{Kerley2003} or an ideal hydrogen-helium mixture with varying dust content (Appendix \ref{sec:dig}). 

The CTH code is licensed from Sandia National Laboratories (\url{https://www.sandia.gov/cth/}). pyKO is open source and available on GitHub \citep[\url{https://github.com/ImpactsWiki/pyko},][]{Stewart_pyKO_2023}. The ANEOS code is available on GitHub \citep[\url{https://github.com/isale-code/M-ANEOS,}][]{MANEOSv1}.

\begin{figure}
\centering
\includegraphics[width=.6\textwidth]{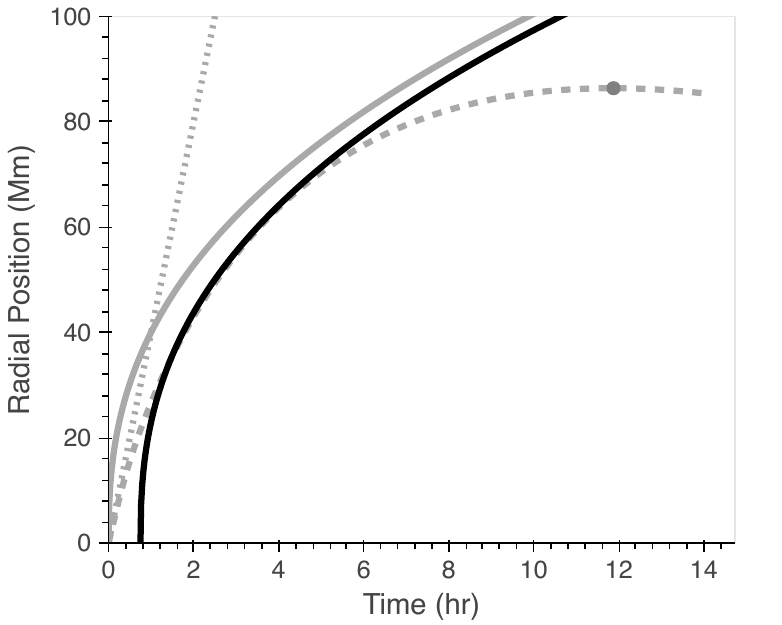} 
\caption{{\bf Plume radius with time.} The 1D plume expansion front (dashed line from Figure~\ref{fig:1devolution}) does not follow a self-similar solution where $R\propto t^{2/5}$ (solid grey line) \citep{sedov2018similarity}. Initially the expansion is more linear (dotted line) and then has an expansion stage with a $t^{2/5}$ curvature (black line is time-shifted grey line) for about a quarter of the time before the stall point ($\bullet$). A similar offset in plume front curvature is seen in laboratory experiments of vapor expanding into a background gas \citep{Pilgram_2024}.
 \label{fig:notsedov}}
\end{figure}

The 1D Lagrangian pyKO simulations initialized a high pressure sphere of water or fused silica on their respective principal Hugoniots. The sphere was surrounded by the solar nebula under different ambient conditions (200~K and varying initial pressure). Example input files and replication data files are in Animations of this impact are available in \href{https://chondrules.net/ivans/}{supplemental materials}. A hydrocode calculation is required for the 1D spherical expansion case because the evolution does not follow an idealized self-similar solution, as shown in Figure~\ref{fig:notsedov}. The initial release isentrope has a faster radial expansion than the Sedov solution with $R \propto t^{2/5}$ \citep{sedov2018similarity}. At later times, the background nebular pressure is no longer negligible and slows the radial expansion compared to the Sedov solution.

The 2D and 3D Eulerian simulations required the adaptive mesh refinement capabilities in CTH \citep{Crawford1999}. Cubic blocks were adjusted by factors of 2 in length according to user-defined criteria. Here, the criteria focused on tracking material interfaces and changes in density to capture the many orders of magnitude increase in length scale between the planetesimal size and the plume expansion. We neglected radiative cooling of the vapor plume because most of the system is optically thick during plume expansion and collapse (Appendix \ref{sec:opacity}). The initial ice-silicate planetesimals were initialized by inserting thin radial shells of ice and fused silica. Massless Lagrangian tracer particles were used to visualize the gas flow streamlines. See example input files and animations in Animations of this impact are available in \href{https://chondrules.net/ivans/}{supplemental materials}.

\begin{table}
\caption{Material parameters for ANEOS tabulated equations of state.}
\begin{tabular}{llll}
Parameter & Definition & Fused Silica & Water \\
V01 $N_{\rm elem}$ & Number of elements & 2 & 2 \\
V02 EOS Type & 4 is solid-liquid-gas with ionization & 4 &4 \\
V03 $\rho_0$ & Reference density, g cm$^{-3}$ & 2.20 & 1.25  \\
V04 $T_0$ & Reference temperature, K & 298 & 150\\
V05 $P_0$ & Reference pressure, dynes cm$^{-2}$ & 1E6 & 1E6 \\
V06 $B_0$ & Bulk modulus, dynes cm$^{-2}$  & 0.4E12 & 2.7E10 \\
V07 $\gamma_0$ & Gr\"uneisen parameter at reference state & 0.57 & 0.60 \\
V08 $\theta_0$ & Debye temperature, K & -1600 & -290 \\
V09 $t$ & Cold curve model parameter $T_\gamma=t-1$  & 1 & 1 \\
V10 3$\gamma_{\rm inf}$ & Gr\"uneisen parameter as $\rho \rightarrow \infty$ & 6 & 6\\
V11 $E_{\rm sep}$ & Zero temperature separation energy, erg g$^{-1}$ & 1.2E11 & 2.9E10\\
V12 $T_{\rm melt}$ & Melting temperature at reference pressure, K & 1996 & 273 \\
V13 $C53$ & Critical point adjustment parameter, erg g$^{-1}$ & -1.E12 & 0 \\
V14 $C54$ & Critical point adjustment parameter  & 0.9 & 0 \\
V15-V16 & Thermal conductivity parameters, not used & 0 & 0 \\
V17 $\rho_{\rm min}$ & Minimum density for solid, default 0.8$\rho_0$ & 0 & 0\\
V18-V22 & High-pressure phase transition parameters, not used & 0 & 0 \\
V23 $H_{\rm fusion}$ & Enthalpy of fusion, erg g$^{-1}$ & 6.6E9 & 1.4E9 \\
V24 $\rho_l/\rho_s$ & Volume change on melting & 0.93 & 0.89 \\
V25 Upper & Upper limit to cold curve extension, default 1 & 0 & 0 \\
V26 Lower & Lower limit to cold curve extension, default 0 & 0 & 0 \\
V27 $\alpha$ &  Liquid model parameter ($0<\alpha<1$, default=0.3) & 0.3 &  0.05 \\
V28 $\beta$ &  Liquid model parameter ($0<\beta<1$, default=0.1) & 0.1 &  0.1 \\
V28 $\gamma$ &  Liquid model parameter ($0<\gamma<1$, default=0.2) & 0.2 &  0.2 \\
V30 $C60$ & Gr\"uneisen model adjustment parameter & 0 & 0 \\
V31 $C61$ & Gr\"uneisen model adjustment parameter ($-1<C61<0$) & -0.851 & -0.4 \\
V32 $C62$ & Critical point adjustment parameter ($0<C62<1$) & 0.5 & 0.5 \\
V33 Flag & Ionization model, 0=Saha, 1=Thomas-Fermi & 0 & 0 \\
V34-V35 & Reactive chemistry model parameters, not used & 0 & 0  \\
V36 $N_{\rm atom}$ & Number of atoms in molecular clusters & 2 & 3 \\
V37 $E_{\rm bind}$ & Molecular cluster binding energy, eV & 5.0 & 4.67 \\
V38 RotDOF & Rotational degrees of freedom & 2 & 3 \\
V39 $R_{\rm bond}$ & Length of molecular bond, cm & 1.5E-8 & 96E-10 \\
V40 VibDOF & Vibrational degrees of freedom & 1 & 3 \\
V41 $T_{\rm Debye}$ & Vibrational Debye temperature, K & 2000 & 2000 \\
V42 LJ Flag & Flag for Lennard-Jones (1) or Morse (0) potential & 1 & 1 \\
V43 $a_{\rm exp}$ & Exponent in Lennard-Jones potential & 1.70 & 1.5 \\
V44 $f_{cv}$ & Thermal model adjustment parameter, max. $c_v=3f_{cv}NkT$ & 1.29 & 1. \\ 
V45 $QCC1$ & Density to transition to ideal gas, g cm$^{-3}$ & 1E-30 & 1E-30 \\
V46 $QCC6$ & $\psi$ value to transition to ideal gas & 1E5 & 1E5 \\
 & Atomic Numbers & 8  & 1 \\
 &                & 14 & 8 \\
 & Atomic Fractions & 0.6667 & 0.6667 \\
 &                  & 0.3333 & 0.3333 \\
 & EOS table version & SLVTv0.2G1 & SLVTv0.3G1 \\
\end{tabular}
\newline  \small This work used ANEOS code package version 1.0 \citep{MANEOSv1} with parameters for water \citep{Stewart_2024_water} and silica \citep{Stewart_Amodeo_2024_silica}. Note input values of 0 sometimes flags the use of a default value. V08: Negative means use full Debye model; positive means use high-temperature approximation; $T_{\rm Debye}$ does not represent the true Debye temperature of the solid when $f_{cv}\neq1$. V36-43: parameters for molecular gas developed by \citep{melosh2007} and extended to triatomic molecules \citep{MANEOSv1}. V44: thermal model adjustment parameter to fit liquid shock temperatures $f_{cv}$ \citep{stewart2020}. V45-V46: user adjustments to ideal gas transition \citep{Stewart2019aneosmanual}.
\label{tab:aneos_params}
\end{table}

\begin{table*}
\centering
\begin{tabular}{lrrrrrr}
\hline
Shock Pressures (GPa) & 100 & 150 & 200 & 250 & 350 & 450 \\
Shock Density (g/cm$^3$) & 4.72 & 4.89 & 5.04 & 5.17 & 5.38 & 5.56   \\
Shock Temperature (K) & 7210 & 11275 & 15330 & 19354 & 27247 & 34936  \\
Shock Sp. Energy (MJ/kg) & 12.18 & 18.80 & 25.65 & 32.67 & 47.08 & 61.84  \\
Shock Sp. Entropies (kJ/K/kg) & 3.85 & 4.47 & 4.90 & 5.23 & 5.72 & 6.08  \\
Equivalent Planar $V_i$ (km/s) & 9.85 & 12.25 & 14.31 & 16.15 & 19.40 & 22.23  \\
\hline
\end{tabular}
\caption{Summary of initial shock states for 1D simulations of compressed fused silica plumes expanding into varying pressures of nebular gas. These results were used to develop scaling laws in section~\ref{sec:scaling}.}
\label{tab:1dsims-silica}
\end{table*}

\section{Ideal Dust-Gas Mixture Equation of State}
\label{sec:dig}

There is an extensive body of work on shocks in dusty gases.  \citet{Capecelatro_Wagner_2024} and \citet{Igra_Ben-Dor_1988} provide introductory overviews of the literature. In this work, the ideal dust-gas mixture equation of state model assumes that the solid particles are distributed evenly through the ideal gas and that the particles are small enough to be in thermal equilibrium with no relative velocity. The dust is assumed to be incompressible with a constant specific heat capacity. The specific internal energy is partitioned between the phases by mass fraction. Our implementation of the dust-gas mixture equation of state follows \citet{Steiner_Hirschler_2002}.

The mass of the mixture, $m_{\rm mix}$, is
comprised of the mass of the dust, $m_{\rm d}$, and the mass of the gas, $m_{\rm _g}$:
\begin{equation}
    m_{\rm mix}=m_{\rm d}+m_{\rm g}.
\end{equation}
The mass fraction of dust is given by
\begin{equation}
    k_{\rm d}= \frac{m_{\rm d}}{m_{\rm mix}}.
\end{equation}
The total volume is the sum of the components:
\begin{equation}
    V_{\rm mix} = V_{\rm d}+V_{\rm g}.
\end{equation}
The dust is approximated as incompressible, with constant $\rho_{\rm d}=3000$~kg~m$^{-3}$, and the volume fraction of the components is given by
\begin{eqnarray}
    V_{\rm d}  & = & \frac{m_{\rm d}}{\rho_{\rm d}}, \\
    V_{\rm g} & = & \frac{m_{\rm g}}{\rho_{\rm g}}.
\end{eqnarray}

The volumetric fraction of the dust in the mixture at a given density and temperature state ($\rho$, $T$) is given by:
\begin{equation}
    Z = \frac{V_{\rm d}}{V_{\rm mix}} = Z_0 \frac{\rho_{\rm mix}}{\rho_{{\rm mix},0}},
    \end{equation}
where the subscript 0 indicates the reference state, here taken as the initial state of the ambient nebula. Then $Z_0$ is defined by:
\begin{equation}
    Z_0= \frac{V_d}{V_{{\rm mix},0}} = \frac{k_{\rm d}}{(1-k_{\rm d})(\rho_{\rm d}/ \rho_{\rm g})+ k_{\rm d}}.
\end{equation}

The pressure and temperature of the mixture satisfies the partitioning of the ideal gas law:
\begin{equation}
    P_{\rm mix} = \left ( \frac{1-k_{\rm d}}{1-Z} \right ) R_{\rm g} \rho_{\rm mix} T_{\rm mix},
\end{equation}
where $R_{\rm g}$ is the specific gas constant of the gas component. 

The specific internal energy of the mixture is
\begin{equation}
    e_{\rm mix} = \left ( \frac{1-Z}{\Gamma -1} \right ) \left( \frac{P_{\rm mix}}{\rho_{\rm mix}} \right ),
\end{equation}
where
\begin{equation}
    \Gamma = \frac{\gamma+\delta \beta}{1+\delta \beta}
\end{equation}
and $\gamma$ is the standard ratio of specific heat capacities in the ideal gas,
\begin{equation}
    \gamma_{\rm g} = \frac{c_{p,{\rm g}}}{c_{v,{\rm g}}}.
\end{equation}
The incompressible dust has $\gamma_{\rm d}=1$, with $c_{p,{\rm d}}=c_{v,{\rm d}}=c_{\rm d}=1000$~J~K$^{-1}$~kg$^{-1}$, and
\begin{eqnarray}
    \beta & = & \frac{c_{\rm d}}{c_{v,{\rm g}}}, \\
    \delta & = & \frac{k_{\rm d}}{1-k_{\rm d}}. 
\end{eqnarray}
The mixture heat capacities are
\begin{eqnarray}
    c_{p,{\rm mix}} & = & k_d c_{\rm d} + (1-k_d) c_{p,{\rm g}} \\
    c_{v,{\rm mix}} & = & c_{p,{\rm mix}}/\Gamma.
\end{eqnarray}
The temperature of the mixture is given by $e_{\rm mix}/c_{v,{\rm mix}}$.

The sound speed of the mixture decreases with increasing dust fraction, given by
\begin{equation}
    c_{s,{\rm mix}} = \sqrt{ \left ( \frac{\Gamma (1-k_{\rm d})} {(1-Z)^2} \right ) \left ( R_{\rm g} T_{\rm mix} \right ) }.
\end{equation}

The molecular weight of the nebular gas is given by
\begin{equation}
    \mu_{\rm g} = \frac{1 f_{\rm H} + 4 f_{\rm He}}{(f_{\rm H}+f_{\rm He})},
\end{equation}
where $f_{\rm H}=12$ and $f_{\rm He}=10.984$ for the proto-Sun from \citet{Lodders_2003}. Then $\mu_{\rm g}=2.43$. For simplicity we assumed a diatomic gas with $\gamma_{\rm g}=1.4$.

\begin{figure*}
\plotone{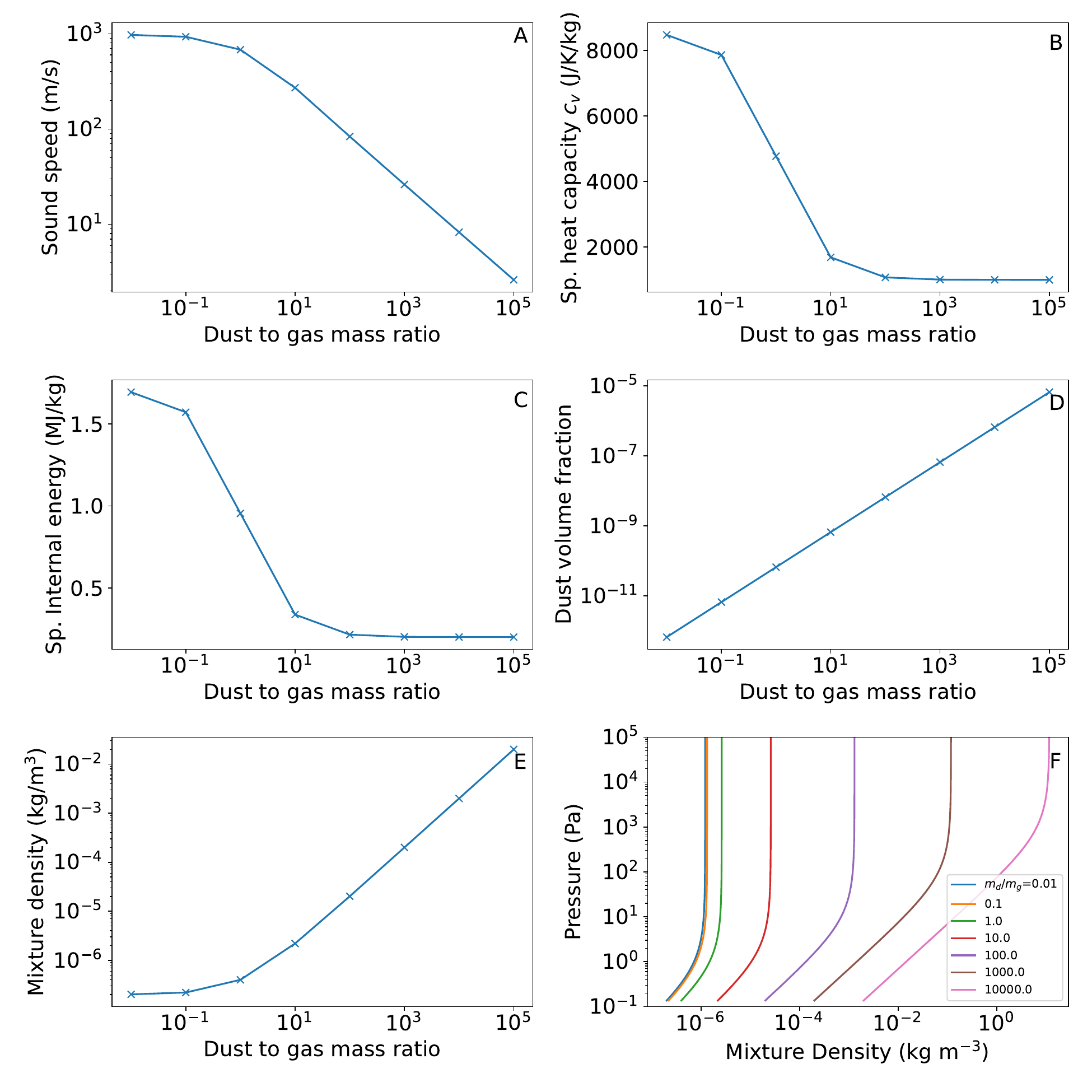} 
\caption{{\bf Properties of an ideal dust-gas mixture equation of state.} Mixture physical properties (A-E) and shock Hugoniots (F) for different initial dust-to-gas mass ratios for an initial state at 200~K and 0.137 Pa. For $m_d/m_g \leq 0.1$, the shock Hugoniots are similar to a pure gas. }
\label{fig:digmodel}
\end{figure*}

The properties of the mixture are presented in Figure~\ref{fig:digmodel} for an initial temperature of 200~K and $\rho_{\rm g}=2\times10^{-7}$~kg~m$^{-3}$. The initial pressure of the mixture is 0.137~Pa.

We also present the shock Hugoniots for a wide range of dust-to-gas mass ratios (Figure~\ref{fig:digmodel}F). Notice that increasing the dust mass fraction increases the compressibility of the mixture.

The range of dust-to-gas mass ratios inferred from the chemical composition of chondrules (Table~\ref{tab:meltcalcs}) would be optically thick. We estimate the mean free path, $L_{\rm MFP}$, for different dust-to-gas mass ratios for a monotonic distribution of spherical particles with radius $r_d$:
\begin{equation}
     L_{\rm MFP}= \left[\pi r_{d}^2 n \right]^{-1},
\end{equation}
where $n$ is the number density of particles, determined from the volume fraction of dust divided by the volume of a single particle. For a wide range of expected parameters, the mean free paths are small than the scale of the stall radius of the impact vapor plumes (100s Mm, Fig.~\ref{fig:waterplumes}). Thus, we neglect radiation into the surrounding unshocked nebula when estimating the cooling rates in the expanding nebular shock. A more sophisticated estimate for the mean free path in the driving vapor plume is described in Appendix~\ref{sec:opacity}.

\begin{figure*}
\centering
\includegraphics[width=.5\textwidth]{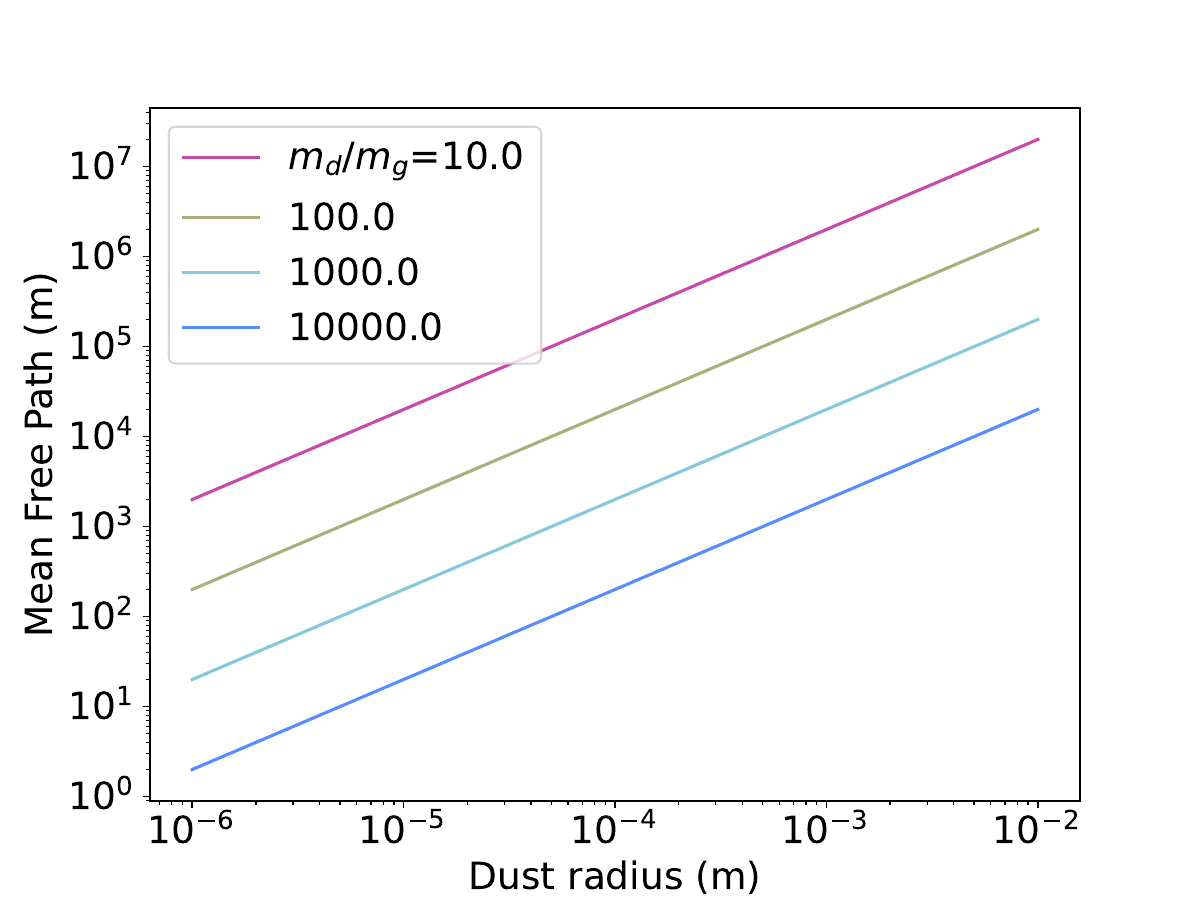} 
\caption{{\bf Mean free path in ideal dust-gas mixtures.} For a wide range of expected dust-to-gas mass ratios ($m_d/m_g$, see Table~\ref{tab:meltcalcs}) and dust sizes, the mean free path is much smaller than the scale of the nebular shocks generated by the expanding impact vapor plume (100s Mm). Thus, the shocked nebular gas and the surrounding ambient nebular gas will be optically thick. }
\label{fig:mfpdigmodel}
\end{figure*}

\section{Spherical Decompression Acceleration}\label{app:Riemann}

For a one dimensional planar shock that is decompressing by rarefaction wave, the particle velocity of the expansion is given by

\begin{equation}
u_{\rm p, exp}=u_{{\rm p},in} - \int_{P_{in}}^{P_{\rm amb}} \frac{dP}{ \sqrt{ \left( - \partial P / \partial V \right)_S } } = u_{{\rm p},in} -  \int_{P_{in}}^{P_{\rm amb}} \frac{V}{c} dP, \label{vap:s:eq:release}
\end{equation}
where the subscript $S$ in the derivative denotes an isentrope, $u_{{\rm p},in}$ is the initial particle velocity given by the Rankine-Hugoniot jump conditions, ${P_{in}}$ is the pressure of the initial shock state, and $P_{\rm amb}$ is the decompressed pressure. This relation was derived by \citet{rice1958compression}. For a this work, equation \ref{vap:s:eq:release} does not accommodate the change in volume appropriate to spherical expansion. Thus, we derive the change in velocity here in a similar manner to \citet{rice1958compression}.

For a fluid, conservation of mass yields

\begin{equation}
    \frac{\partial \rho}{\partial t} + \nabla \cdot (\rho \textbf{u}) = 0
\end{equation}
where u is the mass flux velocity. For a sphere, where there is only velocity in the radial direction, conservation of mass becomes

\begin{equation}
    \frac{\partial \rho}{\partial t} + \frac{1}{r^2} \frac{ \partial ( r ^2 \rho u)}{\partial r} = 0
\end{equation}

which we rewrite as 

\begin{equation}
   \frac{\partial \rho}{\partial t} + \frac{2}{r} \rho u + u \frac{\partial \rho}{\partial r}  + \rho\frac{\partial u}{\partial r} = 0
    \label{S:eqn:conMass}
\end{equation}
where subscripts convey partial derivatives. For a perfect fluid, the force that controls flow is a result of the pressure gradient. In this work, gravity and other body forces are assumed to be negligible and all forces are in the radial dimension. For very large bodies, i.e. planets, gravity must be considered. For a pressure gradient in only the radial direction
\begin{equation}
    \textbf{F}=-dP \cdot d\textbf{A} = dP r^2 \sin\theta d \theta d\phi \hat{r}
\end{equation}
causes a mass $\rho d \textbf{V} = \rho r^2 \sin\theta d \theta d\phi d r \hat{r}$ to experience an acceleration 
\begin{equation}
    \frac{d u}{d t} = \frac{\partial u}{\partial t} + \frac{d r}{d t} \frac{\partial u}{\partial r} = \frac{\partial u}{\partial t} + u \frac{\partial u}{\partial r}.
\end{equation}
We note here that because all motion occurs in the radial direction, unit vectors have been dropped from this point on. From Newton's second law $F=ma$, 
\begin{equation}
-dP r^2 \sin\theta d \theta d\phi  = \rho r^2 \sin\theta d \theta d\phi d r \left[  \frac{\partial u}{\partial t} + u \frac{\partial u}{\partial r} \right]
\end{equation}
which simplifies to
\begin{equation}
-\frac{dP}{dr} = \rho \left[  \frac{\partial u}{\partial t} + u \frac{\partial u}{\partial r} \right].
\label{S:eqn:secondLaw}
\end{equation}

Neglecting heat transport and asserting that the decompression path is isentropic, such that $dE = - PdV$, then 
\begin{equation}
    P=f(\rho, S_0),
    \label{S:eqn:Pfun}
\end{equation}
where $S_0$ is constant entropy. By the chain rule
\begin{equation}
    \frac{dP}{dr} = \frac{\partial \rho}{\partial r} \left( \frac{\partial P} {\partial \rho} \right)_S,
    \label{S;eqn:chain}
\end{equation}
which we substitute into equation \ref{S:eqn:secondLaw}. Here, it is useful to consider sound speed which is defined as
\begin{equation}
    c(\rho) \equiv \left( \frac{\partial P} {\partial \rho} \right)^{1/2}_{S},
\end{equation}
which we then divide the sound speed into both sides of equation \ref{S:eqn:secondLaw} and solve for 0, yielding
\begin{equation}
     c \frac{\partial \rho}{\partial r}+ \frac{\rho}{c}\left[  \frac{\partial u}{\partial t} + u \frac{\partial u}{\partial r} \right] = 0.
    \label{S:eqn:2ndlaw_2}
\end{equation}
Equation \ref{S:eqn:2ndlaw_2} is equal to 0, and thus can be added and subtracted the conservation of mass relation, equation \ref{S:eqn:conMass}. Gathering like terms gives the two relations,
\begin{equation}
    \frac{\partial \rho}{\partial t} + (u + c) \frac{\partial \rho}{\partial r} +  \frac{\rho}{c}\left[  \frac{\partial u}{\partial t} +  \frac{2}{r}cu + (u+c)\frac{\partial u}{\partial r} \right] =0
    \label{S:eqn:positive}
\end{equation}
and
\begin{equation}
    \frac{\partial \rho}{\partial t} + (u - c) \frac{\partial \rho}{\partial r} -  \frac{\rho}{c}\left[  \frac{\partial u}{\partial t} -  \frac{2}{r}cu + (u-c)\frac{\partial u}{\partial r} \right] =0.
    \label{S:eqn:negative}
\end{equation}
For an observer moving at $\frac{dr}{dt}$, it is generally held that for some function $\xi (r,t)$ that $\frac{d \xi}{dt} = \xi_t + \left(\frac{dr}{dt} \right) \xi_r$. Then along $\frac{dr}{dt} = u + c$, equation \ref{S:eqn:positive} becomes

\begin{equation}
    d\rho + \frac{\rho}{c} [du + \frac{2uc}{r} dt]=0,
    \label{S:eqn:posC}
\end{equation}
and along $\frac{dr}{dt} = u - c$, equation \ref{S:eqn:negative} becomes
\begin{equation}
    d\rho - \frac{\rho}{c} [du - \frac{2uc}{r} dt]=0.
    \label{S:eqn:negC}
\end{equation}.

These pairs of relations define characteristic equations. For a more in depth discussion of characteristic equations, please refer to \citet{rice1958compression}. The family of equations that correspond to the +$r$ direction are along the $(u + c)$ characteristic and the -$r$ direction are along the $(u - c)$.

The form of equations \ref{S:eqn:posC} and \ref{S:eqn:negC} are difficult to integrate analytically, but assuming infinitesimal steps, the velocity can be easily calculated in a step wise manner by,
\begin{equation}
    \Delta u = \mp \frac{c}{\rho}\Delta\rho \mp \frac{2uc}{r}\Delta t,
    \label{eq:app:expansion}
\end{equation}
respective to the $(u \pm c)$ characteristic.
This result is quite similar to equation \ref{vap:s:eq:release}, with the addition of a spherical term. The spherical term models the volume expanding into a geometrically increasing volume, compared to a linearly expanding volume as it is in the planar case.

\section{Coupling and breakup of condensed particles}
\label{sec:coupling}

We calculated the maximum size of particles (molten droplets or dusty aggregates) that would be stopped in the nebular shock (Figure~\ref{fig:coupling}). A subsequent paper will describe the model and results at greater length \citep{Lock_coupling_IVANS_prep}, but here we provide a summary of the method. The code used for the calculation in Figure~\ref{fig:coupling} is included in the supplementary materials. The acceleration of a particle entering the nebular shock due to gas drag from the nebula was calculated using the \texttt{scipy.integrate.solve\_ivp} function with an explicit Runge-Kutta method of order 8 (the `DOP853' method). For simplicity we did not consider the reciprocal force on the vapor by the particles. When a particle has a radius $r_{\rm cond} > (9/4)  L^{\rm vap}_{\rm MFP}$, where $L^{\rm vap}_{\rm MFP}$ is the mean-free path of molecules in the shocked nebular calculated using a molecular collisional cross section of $2\times10^{-19}$~m$^{2}$ \citep{Tabata2000_H2_crosssection}, the particle was assumed to be in the viscous regime with a velocity-dependent drag coefficient \citep{brown2003drag}. If $r_{\rm cond} \leq (9/4)  L^{\rm vap}_{\rm MFP}$, the particle is in the Epstein regime with a fluid velocity-independent drag coefficient \citep{Probstein1968Ep_drag}. The viscosity of the nebular gas was taken as a linear molar average of the viscosity of the He and H$_2$ components (with a mole fraction of H$_2$ of 0.818) with the end member viscosities found by fitting power-laws to a compilation of literature data \citep{Guevara1969H_He_visc,Hannley1969H_visc,Flyn1963He_visc,Peterson1970He_visc}. 

For liquid droplets in the viscous regime, the stability of particles was assessed using experimentally-determined criteria \citep{Theofanous2008,Theofanous2011,Theofanous2012}, which are consistent with previous work directly focused on the breakup of chondrules \citep{Kadono2008}. In the Epstein regime, the maximum size of particles was determined by a balance between the surface tension of the droplet and the drag force. The surface tension and viscosity of molten silicates were taken from \citet{Boca2003surface_tension} and \citet{giordano_viscosity_2008}, respectively. For aggregates, the maximum size in both regimes was determined by a balance between drag force and tensile strength \citep{gundlach_tensile_2018} assuming a spherical aggregate. Unstable particles break up on timescales much shorter (a factor of $10^{-5}$--$10^{-2}$) than the coupling timescale. Therefore, particles that became unstable were assumed to break up into small fragments.

To explore the range of particle sizes that could be stopped in the nebular shock, we conducted a Monte Carlo survey of $\sim$~5000 simulations varying the shock velocity (1--5~km~s$^{-1}$), the density of the shocked nebula ($10^{-9}$--$10^{-5}$~kg~m$^{-3}$), the temperature of the shocked nebula and condensates (1000--2500~K), the distance between the front of the nebula shock and the edge of the vapor plume (10$^5$--$10^7$~m), and the velocity at which a particle is considered stopped $v_{\rm stop}$ (1--500~m~s$^{-1}$). The density of the droplets and aggregate monomers was $3\times 10^3$~kg~m$^{-3}$ and the shock was assumed to be an ideal gas in the strong shock regime. We performed calculations for liquid droplets and aggregates with a volume fraction of solids of 0.5. We found that the dominant factor on the particle size stopped in the nebular shock was the density of the shocked gas. The size of droplets caught in the shock is therefore a direct function of the density of the nebula at the time and location of the collision.

\section{Opacity of the vapor plume and nebular shock}
\label{sec:opacity}

For the calculations in this work, we have neglected radiative cooling on the grounds that the material within the nebular shock and vapor plume are optically thick. Here, we justify that assertion. 

To evaluate whether a volume is optically thick, we compare the optical depth with the size of the geometries we are considering. To provide an upper limit on the optical depth, we neglect the absorption of vapor and only consider opacity due to condensates. In a condensate-dominated system, the optical depth is given simply by the mean free path of photons traveling through the condensate field. We describe a condensate population by a number density distribution, $g(r)$, which is normalized such that
\begin{equation}
\int_0^{\infty}g(r') \mathrm{d}r' = 1 \,,
\label{sup:eqn:gr}
\end{equation}
where $r$ is the condensate radius. $g(r)$ is the fraction of particles that have a radius within an increment of radius at $r$ and has units of inverse length. Consider photons traveling a distance $L$ through the condensate field. The average fraction of the area perpendicular to the line of flight that is occupied by condensates is
\begin{equation}
\frac{\mathcal{A}_{\rm cond}}{\mathcal{A}}= L \int_0^{\infty} \pi r'^2 n g(r') \mathrm{d}r' \,,
\label{sup:eqn:rad:Afrac}
\end{equation}
where $n$ is the number density of particles of all sizes, and we have assumed $r \ll L$. 
We define the mean free path as the $L=L_{\rm MFP}$ at which this ratio is unity and from Equation~\ref{sup:eqn:rad:Afrac}
\begin{equation}
L_{\rm MFP} = \left [ \int_0^{\infty} \pi r'^2 n g(r') \mathrm{d}r' \right ]^{-1} \,.
\label{sup:eqn:rad:MFP_n}
\end{equation}
We recast this expression in terms of the mass density of condensates. The mass density for a system of condensates of variable radius $r$, $\xi(r)$, is related to the number density distribution by
\begin{equation}
\Xi \xi(r) = \frac{4\pi}{3} r^3 \rho(r) n g(r) \,,
\label{sup:eqn:rad:fg}
\end{equation}
where $\Xi$ is the mass of all condensates per unit volume, and $\rho(r)$ is the material mass density of condensates which can be a function of particle radius. Using this expression, Equation~\ref{sup:eqn:rad:MFP_n} becomes
\begin{equation}
L_{\rm MFP} = \left [ \frac{3\Xi}{4} \int_0^{\infty} \frac{\xi(r')}{r'\rho} \mathrm{d}r' \right ]^{-1} \,.
\label{sup:eqn:rad:MFP_P}
\end{equation}

\begin{figure}
\centering
\includegraphics[width=3.5in]{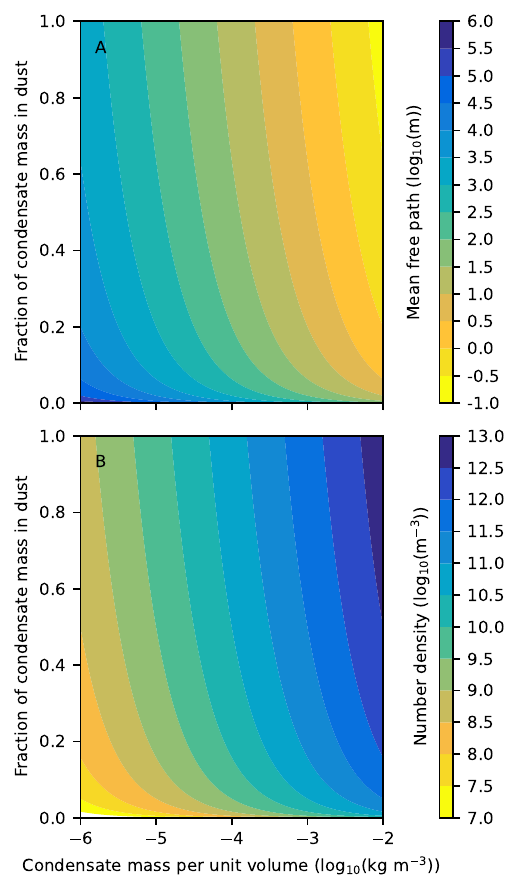}
\caption{{\bf The nebular shock and expanding vapor plume are likely to be optically thick.} (A) The mean free path of photons in a mixture of micron-sized dust and millimeter-sized spherules as a function of the total mass density of condensates and the fraction of the mass that is dust (Equation~\ref{sup:eqn:rad:MFP_P}). (B) The corresponding number density of particles for each total mass density of condensates and dust mass fraction.}
\label{sup:fig:rad}
\end{figure}

The mass density distribution of condensates varies substantially in different regions of the expanding vapor plume and nebular shock. To approximate the range of mean free paths, we calculated the mean free path for a bimodal population of micron-diameter dust and millimeter-diameter spherules. Such a bimodal condensate distribution is described by
\begin{equation}
\xi(r)=\sum_{j=1}^2 \xi_j \delta(r-r_j) \,,
\end{equation}
where $\xi_j$ is the fraction of mass in condensates with radius $r_j$, and $\delta(r)$ is the Dirac delta function. Equation~\ref{sup:eqn:rad:MFP_P} in this case is
\begin{equation}
L_{\rm MFP} = \left [ \frac{3\Xi}{4}\sum_{j=1}^2 \frac{\xi_j}{r_j\rho_j} \right ]^{-1} \,.
\label{sup:eqn:rad:MFP_P_disc}
\end{equation}
Figure~\ref{sup:fig:rad}A shows the mean free path for this bimodal condensate distribution with varying condensate mass density and different fractions of the condensate mass in dust. The corresponding number density of condensates is given in Figure~\ref{sup:fig:rad}B. 

Chondrule number densities of 1 to 100 m$^{-3}$ corresponds to bulk mass densities of about $10^{-7}$ to $10^{-5}$~g~cm$^{-3}$. If 20\% of the condensed mass were dust, the mean free path is less than 100 m, which is much smaller than the scale of the vapor plume (100s Mm). Hence, we neglect radiative cooling of the vapor plume during the first 24 hours of impact plume evolution. 

\newpage
\begin{acknowledgments}
{\bf Acknowledgements:} We thank for Guy Libourel, Alex Ruzicka, Munir Humayun, Daniel Sheikh, and Steve Desch for discussions and feedback that improved this work. We thank the anonymous reviewers for their detailed and helpful comments that improved the presentation of the IVANS model. This work was developed over several years aided by discussions in the Origins Group at UC Davis.
{\bf Funding:} This material is based upon work supported by the Department of Energy, National Nuclear Security Administration under Award Numbers DE-NA0003904 (SJB,STS,MIP), DE-NA0004084 (SBJ,STS,MIP), DE-NA0003842 (STS), and DE-NA0004147 (STS). Additional support from NASA grant NNX15AH54G (STS); NASA grant NNX16AP35H (EJD); UK Science and
Technology Facilities Council grants ST/V000454/1 and
ST/Y002024/1 (PJC); UK National Environmental Research Council grant NE/V014129/1 (SJL); US National Science Foundation awards EAR-1947614 and EAR-1725349 (SJL); and the Division of Geological and Planetary Sciences of the California Institute of Technology (SJL). This work was partially supported by a Benjamin Meaker Visiting Professorship to U. Bristol (STS). The hydrocode simulations were performed at UC Davis and supported by UC Davis research computing technical staff. {\bf Author contributions:} STS, PJC and SJL developed the main physical concepts, conducted key calculations and wrote the text. EJD derived Appendix~\ref{app:Riemann} and vaporization criteria for silicates. SJL performed the fluid dynamics breakup and opacity calculations, contributed to ideas regarding size sorting, and wrote these sections of the text. SBJ and MIP contributed meteoritic expertise and motivating ideas; MIP calculated stability of melts in dusty nebular gas. All authors participated in discussions of the manuscript. {\bf Competing interests:} The authors declare that they have no competing interests. {\bf Software:} This work made use of PKDGRAV \citep{Richardson2000, Stadel_2001}, CTH \citep{McGlaun_Thompson_Elrick_1990}, pyKO \citep{Stewart_pyKO_2023}, ANEOS \citep{MANEOSv1}, NumPy \citep{van2011numpy}, SciPy \citep{2020SciPy-NMeth}, matplotlib \citep{hunter2007matplotlib}, Jupyter \citep{kluyver2016jupyter}, HoloViews \citep{stevens2015holoviews}, Pint (\url{https://github.com/hgrecco/pint}), Accessible Color Tables \citep{Petroff_2024}. {\bf Supplementary materials:} The supplemental materials and replication information for this work is available in \url{https://github.com/ststewart/ivans} (doi:10.5281/zenodo.14969068). This repository includes python scripts and data files to remake the figures, animations from the CTH simulations, and equation of state tables. The repository is compiled as an online Jupyterbook hosted at \url{https://chondrules.net/ivans/}.
\end{acknowledgments}

\bibliography{refs}{}
\bibliographystyle{aasjournal}

\end{document}